\documentclass[acmsmall]{acmart}

\usepackage{graphicx}
\usepackage[utf8]{inputenc}
\usepackage{makecell}

\AtBeginDocument{%
  \providecommand\BibTeX{{%
    \normalfont B\kern-0.5em{\scshape i\kern-0.25em b}\kern-0.8em\TeX}}}

\begin{document}

\setcopyright{acmcopyright}
\acmJournal{CSUR}
\acmYear{2019} \acmVolume{53} \acmNumber{1} \acmArticle{15} \acmMonth{12} \acmPrice{15.00}\acmDOI{10.1145/xxxxxxx}

\title{Change Detection and Notification of Web Pages: A Survey}

\author{Vijini Mallawaarachchi}
\affiliation{%
  \institution{The Australian National University}
  \city{Canberra}
  \country{Australia}
}

\author{Lakmal Meegahapola}
\affiliation{%
  \institution{Idiap Research Institute \& École Polytechnique Fédérale de Lausanne (EPFL)}
  \city{Lausanne}
  \country{Switzerland}
}

\author{Roshan Madhushanka}
\affiliation{%
  \institution{University of Moratuwa}
  \city{Moratuwa}
  \country{Sri Lanka}
}

\author{Eranga Heshan}
\affiliation{%
  \institution{University of Moratuwa}
  \city{Moratuwa}
  \country{Sri Lanka}
}

\author{Dulani Meedeniya}
\affiliation{%
  \institution{University of Moratuwa}
  \city{Moratuwa}
  \country{Sri Lanka}
}

\author{Sampath Jayarathna}
\affiliation{%
  \institution{Old Dominion University}
  \city{Norfolk}
  \country{USA}
}

\renewcommand{\shortauthors}{Mallawaarachchi et al.}

\begin{abstract}
The majority of currently available webpages are dynamic in nature and are changing frequently. New content gets added to webpages, and existing content gets updated or deleted. Hence, people find it useful to be alert for changes in webpages that contain information that is of value to them. In the current context, keeping track of these webpages and getting alerts about different changes have become significantly challenging. Change Detection and Notification (CDN) systems were introduced to automate this monitoring process, and to notify users when changes occur in webpages. This survey classifies and analyzes different aspects of CDN systems and different techniques used for each aspect. Furthermore, the survey highlights the current challenges and areas of improvement present within the field of research.
\end{abstract}



\maketitle

\section{Introduction}
\label{section1}
The World Wide Web (WWW or Web in simpler terms) is being evolved at a rapid pace, and keeping track of changes is becoming more challenging. Many websites are being created and updated daily with the advancement of tools and web technologies. Hence, websites at present have become more dynamic, and their content keeps changing continuously. Many users are interested in keeping track of the changes occurring on websites such as news websites, booking websites, wiki pages and blogs. Back in the 1990s, people used to register to Really Simple Syndication (RSS) feeds, originated from the Resource Description Framework (RDF) specification \cite{RDFSiteSummary2000} to keep track of frequently updated content. Later in 2005, RSS was replaced by Atom \cite{RFC4287Atom2005}. Currently, the majority of the users keep track of websites and get the latest updates using bookmarks in web browsers.

Web syndication technologies (e.g., RSS and Atom) emerged as a popular means of delivering frequently updated web content on time \cite{Hmedeh2011}. Users can subscribe to RSS or Atom feeds, and get the latest updates. However, when considered from a perspective of webpage change detection, RSS feeds have many potential issues. A study carried out to characterize web syndication behavior and content \cite{Hmedeh2011} shows that the utilization of fields specified in the XML specification of RSS is less, which can result in missing information, errors and uncategorized feeds. Furthermore, services such as Google Reader have been discontinued due to the declining popularity of RSS feeds \cite{OfficialGoogleBlog2013} caused by the rising popularity of \emph{microblogging} (also known as \emph{social media}), shifting of formats from XML to JSON, and market forces promoting proprietary interfaces and de-emphasizing interoperability.

In the current context, managing and using bookmarked websites and RSS feeds have become a significant challenge, and people are continuously seeking better and convenient methods. Change Detection and Notification (CDN) systems \cite{Meegahapola2017a} make it easy for users to get notified about changes that have occurred on webpages, without having to refresh the webpage continuously. 
Google Alerts \cite{GoogleAlerts2003}, Follow That Page \cite{FollowThatPage2008} and Visualping \cite{VisualPing2017} are some of the most popular CDN services which are used by many users to get updates about content changes that occur on webpages. 

\subsection{Change Detection \& Notification Systems}

CDN systems \cite{Meegahapola2017b} automatically detect changes made to pages in the web, and notifies about the changes to interested parties. The significant difference between search engines and CDN systems is that search engines are developed for searching webpages, whereas CDN systems are developed for monitoring changes that occur on webpages. In theory, most of the search engines also have an underlying change detection mechanism to determine which sites to crawl and keep their search results up-to-date \cite{GoogleBot2013}. The use of CDN systems allows users to reduce the browsing time, and facilitates users with the ability to check for changes on webpages of their interest \cite{Yadav2007}.

CDN systems emerged in the WWW with the introduction of Mind-it (currently discontinued) \cite{Mindit1996}, the first CDN tool which was developed by NetMind in 1996. Since then, several services were introduced such as ChangeDetection (introduced in 1999, now known as Visualping \cite{VisualPing2017}), ChangeDetect \cite{ChangeDetect2002} (introduced in 2002), Google Alerts \cite{GoogleAlerts2003} (introduced in 2003), Follow That Page \cite{FollowThatPage2008} and Wachete \cite{Wachete2014}. CDN systems have evolved throughout the past few decades, with improvements in detection rates, efficient crawling mechanisms and user-friendly notification techniques.

CDN systems available at present have become easier to use, and are more flexible to incorporate user requirements. The majority of the currently available CDN systems such as Wachete \cite{Wachete2014} and VisualPing \cite{VisualPing2017} provide various monitoring options such as;

\begin{enumerate}
    \item Multiple webpage monitoring: multiple parts of a webpage, an entire webpage, multiple webpages or an entire website.
    \item Content to monitor: text, images, links, documents (PDF, DOC, DOCX).
    \item The threshold of changes: the percentage of changes occurring on a given webpage.
    \item Frequency of monitoring: hourly, daily, weekly, monthly or on-demand monitoring.
    \item Frequency of notification: twice a day, once a day, once a week or as changes occur.
\end{enumerate}

\subsection{Categories of Change Detection and Notification}

Based on the architecture involved, change detection can be segregated into two main subdomains. The first branch is server-side change detection, and the other is client-side change detection \cite{Meegahapola2017a}. Server-side change detection systems use servers that poll webpages, track changes, and notify them to users. The client-side change detection systems make the client-side infrastructure poll the webpages, and track changes on their own.

CDN systems obtain versions of webpages by crawling them, and saving the data to version repositories. These data are saved in an unstructured manner, mostly in the format of documents with tags, to allow easy storage and retrieval. Then, changes are detected by comparing a previously saved version with the latest version of a particular webpage using similarity computations. The majority of the change detection mechanisms convert the data of a saved version into an XML-like format where an element represents opening and closing HTML tags (e.g., \emph{<h>} and \emph{</h>}). Certain CDN systems allow the user to define a threshold of change, and the user gets notified about a change, only if the change exceeds this threshold.

The majority of the CDN systems operate on predefined monitoring schedules, that are specified by the user or determined by the system itself. Detected changes are visualized using many methods. A common means of visualizing text changes is by showing the newly added text in green color, and the deleted text in red color (often with strikethrough formatting) \cite{Wachete2014}.

Another prominent factor discussed in CDN systems is their crawling schedules. Most of the currently available CDN systems crawl webpages under predefine schedules \cite{Meegahapola2017c}. However, webpages can be updated at different time schedules (certain webpages may be frequently updated, whereas some webpages may get updated rarely), thus how often they change can vary. Hence, CDN systems require mechanisms for estimating the change frequency to create efficient checking schedules for webpages. This will reduce the number of page checks where no changes were detected, and maximize the probability of achieving optimum resource utilization.

\subsection{Survey Motivation}

According to our study, only a limited number of surveys have been carried out regarding webpage CDN techniques. Additionally, it is challenging to find comprehensive evaluations of existing CDN systems which discuss different aspects of such systems. Oita et al. \cite{Oita2011} have reviewed major approaches used for the change detection process in webpages. They have reviewed temporal aspects of webpages, types of webpage changes, models used to represent webpages to facilitate change detection and various similarity metrics that are used for detecting changes. Shobhna and Chaudhary~\cite{Shobhna2013} discuss about a selected set of CDN systems with different types of change detection algorithms in a summarized manner. However, there is a requirement to explore and improve CDN systems by comprehensively considering the various aspects of CDN such as the architecture, monitoring frequency, estimation of change frequency, change notification methods and change visualization mechanisms.

Several CDN systems have been developed, and are available for public use \cite{Meegahapola2017b}. However, we discovered that still there are issues related to these systems, and limited evaluations have been carried out. Hence, the first objective of this survey would be to deliver a comprehensive overview of the different characteristics of CDN systems. The second objective is to study existing CDN systems, and evaluate their features and various performance aspects. Our final objective is to evaluate the different aspects of CDN, study new trends and highlight the areas which require improvement. We believe that this survey can shed light on the relevant research areas, and possibly, pave the way for new scientific research fields.

\begin{figure}[!ht]
    \centering
    \includegraphics[width=\textwidth]{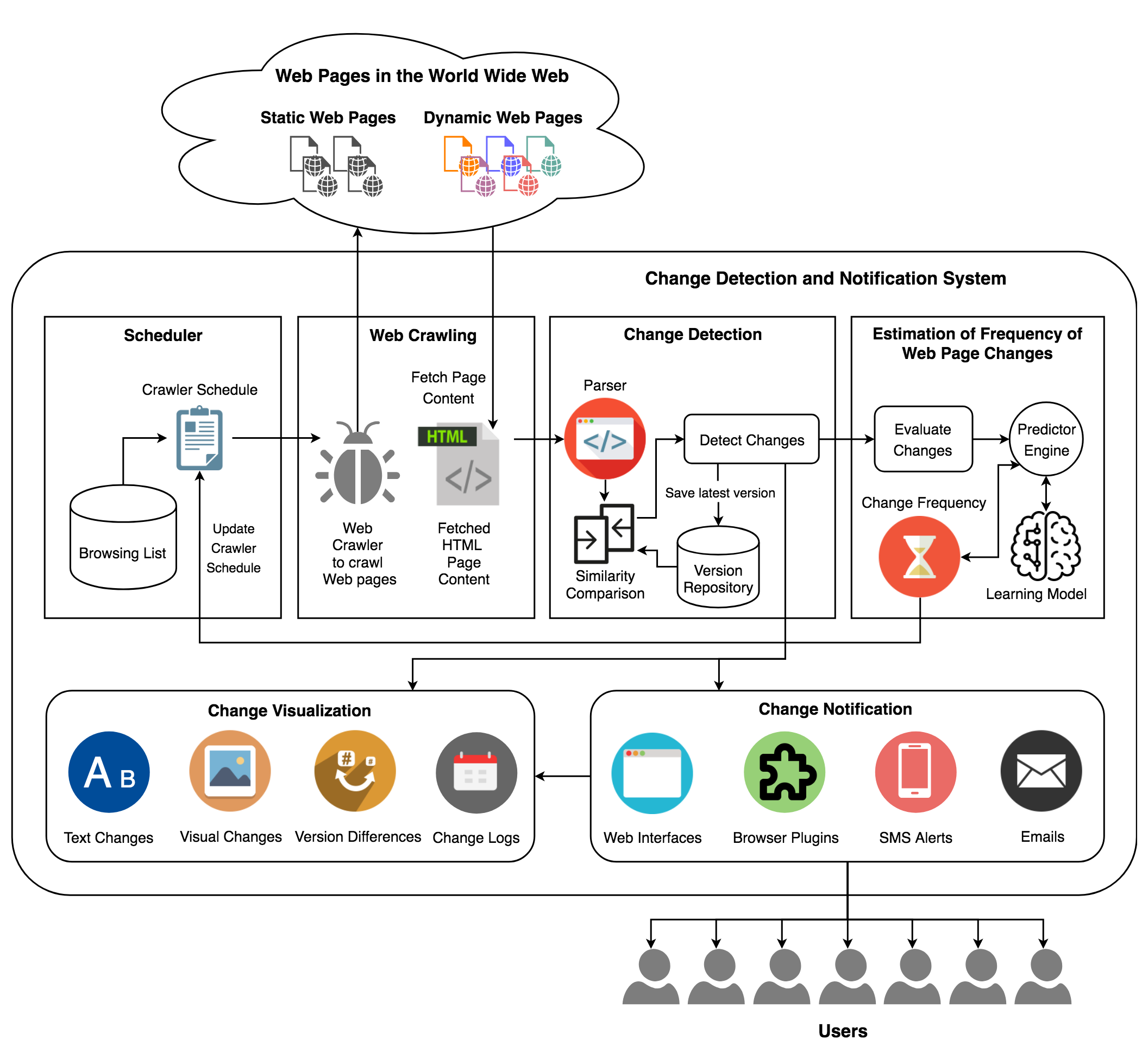}
    \caption{Organization of the survey and the aspects discussed.}
    \label{fig1:survey-organization}
\end{figure}

The organization and the aspects discussed in this survey are summarized in Figure~\ref{fig1:survey-organization}. Section~\ref{section2} discusses the dynamic nature of webpages, and the experiments that have been conducted to understand the changing behavior of webpages. Section~\ref{section3} presents different architectural approaches, which have been introduced to develop CDN systems. Further, various traditional architectures and several architectures that have been customized to improve the efficiency of the CDN mechanisms are presented.  Section~\ref{section4} confers about the techniques used for detecting changes on webpages. It includes different crawling techniques, change detection techniques, scheduling algorithms and methods to detect the frequency of webpage changes. Section~\ref{section5} presents different notification techniques to notify users when changes have been detected on webpages of their interest, whereas Section~\ref{section6} describes how these changes are visualized to the user. Section~\ref{section7} compares and evaluates the different features of publicly available CDN systems and Section~\ref{section8} discusses current trends and issues in the field of CDN. Finally, Section~\ref{section9} concludes the survey paper with further improvements which can be incorporated into existing CDN systems, and presents the identified future research directions for CDN systems.

\section{Dynamics of Web-based Content}
\label{section2}
The World Wide Web (WWW) keeps on growing larger and more diverse every day as new content is being added with the advancement of web content creation tools. The most common units of information on the web are pages, documents and resources \cite{Sebesta2001}. These units can include textual as well as non-textual information such as audio files, video files and images. They can undergo various changes since the time they were added to the WWW. Hence, it is crucial to understand the changing frequency and the dynamic nature of webpages to provide efficient solutions to detect such changes.

\subsection{What are Webpages and their Models of Change?}

Webpages are individual files that consist of text, graphics and other features such as links to scripts and style sheets \cite{Smith2008}. Webpages are implemented using HyperText Markup Language (HTML) or a comparable markup language. The WWW is considered as a collection of such webpages. Webpages are linked together to form websites. Once a webpage is requested on a web browser, the web browser obtains the relevant content, coordinates the resources and presents the webpage. The web browser uses Hypertext Transfer Protocol (HTTP) to send such requests to retrieve webpages. Webpages fall into two broad categories namely, (1) static and (2) dynamic.

\subsubsection{Static Webpages}

\begin{figure}[!ht]
    \centering
    \includegraphics[width=0.7\textwidth]{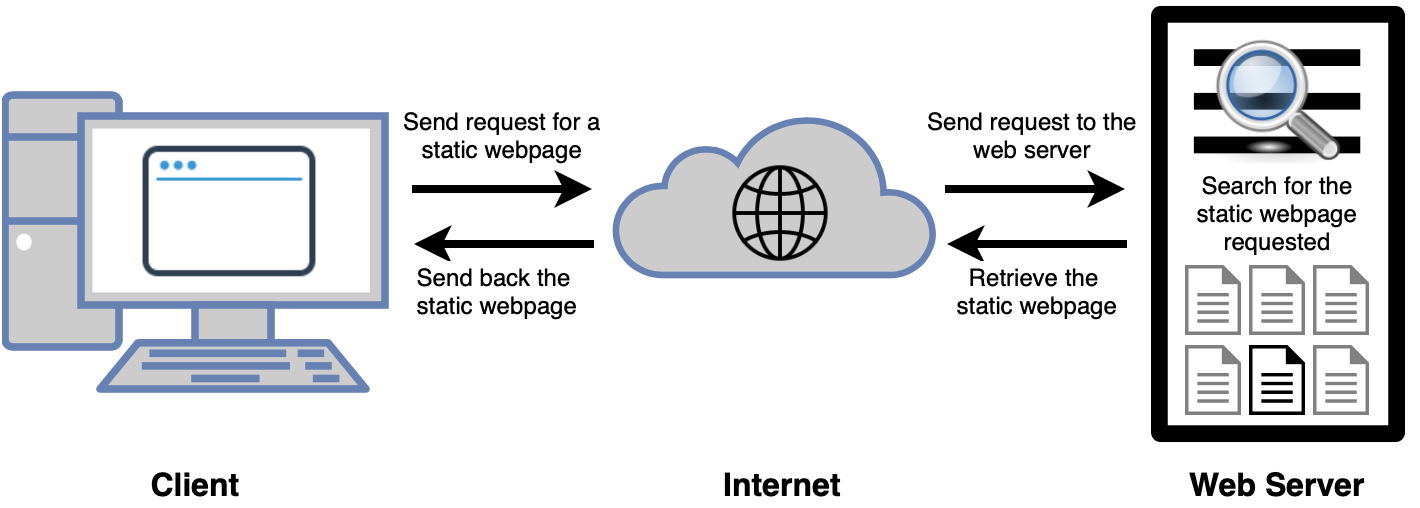}
    \caption{Process of retrieving a static webpage.}
    \label{fig2:static-webpages}
\end{figure}

Static webpages have content that does not change after the developer has saved the webpage to the web server \cite{Keogh2005}. The webpage remains the same until the developer replaces it with an updated static webpage in the server. Static webpages are not tailored to each visitor. They appear the same to all the users who view it. Figure~\ref{fig2:static-webpages} depicts how a static webpage is displayed once the client requests it. 

Static webpages can be created easily as existing web development tools allow us to drag and drop HTML elements as required. Similarly, it is easy to host because only a web server is required to host, without requiring any additional software. Furthermore, static webpages have the advantages of fast loading and improved performance for end-users. However, if dynamic functionalities such as personalized features are required, they have to be added separately.

\subsubsection{Dynamic Webpages}

\begin{figure}[!ht]
    \centering
    \includegraphics[width=0.95\textwidth]{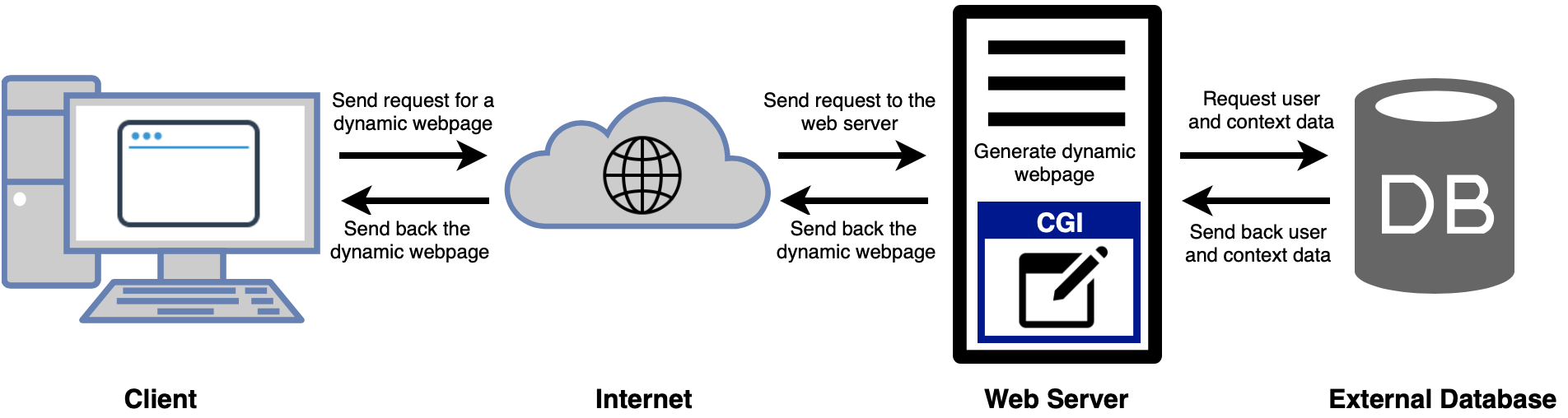}
    \caption{Process of retrieving a dynamic webpage.}
    \label{fig3:dynamic-webpages}
\end{figure}

Dynamic webpages are pages that do not exist until it is generated by a program in response to a request from a client while guaranteeing that the content is up-to-date \cite{Keogh2005}. Their content changes in response to different contexts or conditions. As the information obtained from the client is used to generate the webpage to be shown, it can be tailored according to the client. Figure~\ref{fig3:dynamic-webpages} illustrates the process of generating and displaying a dynamic webpage when a request is made by a client. 

Dynamic webpages pave the way for more functionality and better usability, which is not available via static webpages. They allow the creation of web applications which can be stored on a central server, and can be authorized to access from any place via a web browser. Details related to the user and context can be retrieved from external databases to tailor the webpage. Moreover, it reduces the cost and burden for updating the entire webpage whenever changes occur frequently \cite{Sproul2008}. However, dynamic webpages may face security risks as corporate applications and databases can be exposed to unintended parties. Furthermore, additional software should be installed and maintained, which is required to generate tailored websites to clients.

The Common Gateway Interface (CGI) \cite{Keogh2005} is the software, which runs in the server that is invoked when a client requests to retrieve a webpage. It is responsible for generating webpages dynamically. PHP, ASP.NET and JSP are some of the common languages that are often used to create webpages, and they use the CGI to generate dynamic webpages. The majority of the webpages available at present are dynamic. Dynamic webpages have become popular as many services (e.g., Content Management Systems (CMS) such as WordPress~\cite{WordPress2003} and Drupal~\cite{Drupal2000}) are available at present, where anyone, even a person with limited programming knowledge, can create a website with a few webpages via a control panel, update pages as required and deploy them instantly.

The content found in webpages may get outdated or maybe no longer required. Hence, timely maintenance should be done to ensure that the webpages are up-to-date. Three key events: creations, updates and deletions are considered to cause webpages to change \cite{Levene2004,BaezaYates2004}.

\begin{enumerate}
    \item \textit{Creations}: Once a webpage is created, it cannot be seen publicly on the Web until it is linked by an existing webpage. Hence, adding a new webpage causes at least an update of an existing one; adding a new link to the newly created webpage. However, at present, search engines such as Google provide the facility to add the Uniform Resource Locator (URL) of a newly created webpage, so that it can be indexed, and made available to the public \cite{SubmitURL2018}. 
    \item \textit{Updates}: Updates can be made to the existing content of webpages. Updates can be of two types. The first type is a minor change, where the content of the webpage is almost the same as its previous version, but slight modifications may have been done, such as at the sentence or paragraph level. The second type is a major change, where all the content of the webpage is drastically different from its previous version.
    \item \textit{Deletions}: A page is considered to be deleted if there are no links available to traverse to that particular page or if the page itself is removed from the Web. However, there may be instances where the webpage has been removed but still the link to that webpage exists (known as a \emph{broken link}). Furthermore, content can be deleted from a webpage as well.
\end{enumerate}

\subsection{Detecting the Dynamic Nature of Webpages}

Identifying the amount of change and changing patterns of webpages has been of great interest to researchers. Many studies have been carried out to understand the changing nature of webpages. The content of webpages and their change frequencies have been highly focused areas in this research scope. The available literature demonstrates the ever-changing nature of webpages and various reasons for those changes. Different factors have been considered regarding the dynamic behavior of web content.

Cho and Garcia-Molina \cite{Cho2000a} have conducted an experiment with 720,000 webpages from 270 websites to study how webpages evolve over time. They have crawled these webpages every day for 4 months. Based on the results, the researchers have found that more than 20\% of the webpages had shown changes whenever they visited them, and 40\% of the webpages had changed in a week. Over 50\% of the .com and .edu webpages had not changed at all during the time frame of the experiment. A massive-scale experiment which extended the study done by Cho and Garcia-Molina was performed by Fetterly et al. \cite{Fetterly2003}. They have studied how frequently webpages change, and the quantity of change occurred, using approximately 150 million webpages over eleven weeks. According to their findings, approximately 65\% of the pages had zero change over the considered time. Further, it shows the relationships in-between certain top-level domain types (e.g., .org, .gov, .com), and their respective frequencies of changing. It was revealed that .org and .gov domains are less likely to change than .edu domains. Furthermore, it was shown that pages including spams would change more frequently. 

Olston and Pandey~\cite{Olston2008} have crawled 10,000 web URLs from the Yahoo crawled collection and 10,000 web URLs from the Open Directory listing. According to the results, from the dynamic content on webpages of the Open Directory listing, about a third has shown a scroll behavior. Adar et al. \cite{Adar2009} have crawled and analyzed approximately 55,000 webpages, which are revisited by users, to characterize the various changes occurring in them. The authors have tracked the frequency and amount of change that has occurred in the webpages individually. 34\% of the webpages had not shown any changes whereas the remaining pages had displayed at least one change every 123 hours. This study has shown that popular webpages change more frequently when compared to less popular webpages. Webpages falling under categories such as sports, news and personal pages change most frequently, and webpages with government and educational domain addresses have no frequent changes.

\begin{table}[!ht]
    \centering
    \caption{A summary of the work done to detect the dynamic nature of webpages.}
    \begin{tabular}{|l|l|l|l|}
    \hline
    \textbf{\makecell{Work}} & \textbf{\makecell{Websites crawled}} & \textbf{\makecell{Pages that have \\ changed significantly}} & \textbf{\makecell{Pages that have \\ not changed}} \\
    \hline
    \makecell[l]{Cho and Garcia-\\Molina 2000~\cite{Cho2000a}} & \makecell[l]{720,000 webpages \\from 270 websites} & \makecell[l]{40\% of the crawled \\webpages} & \makecell[l]{Over 50\% of .com and \\.edu webpages}  \\
    \hline
    \makecell[l]{Fetterly et al. 2003\\ \cite{Fetterly2003}} & \makecell[l]{150 million \\webpages} & \makecell[l]{Webpages of .edu \\domain and spam} & \makecell[l]{Webpages of .org and \\.gov domains} \\
    \hline
    \makecell[l]{Adar et al. 2009~\cite{Adar2009}} & \makecell[l]{55,000 webpages} & \makecell[l]{Popular webpages \\(e.g. sports, news, etc.)} & \makecell[l]{Webpages of .gov \\and .edu domains} \\
    \hline
    \makecell[l]{Elsas and Dumais \\ 2010~\cite{Elsas2010}} & \makecell[l]{2,000,000 HTML \\pages} & \makecell[l]{Highly relevant \\documents} & \makecell[l]{62\% of the crawled \\webpages} \\
    \hline
    \makecell[l]{Saad and Gançarski \\2012~\cite{Saad2012}} & \makecell[l]{100 webpages from \\FranceTV archive} & \makecell[l]{Webpages at the root \\level of the archive} & \makecell[l]{Webpages in deepest \\levels of the archive} \\
    \hline
    \end{tabular}
    \label{tab1:dynamic-web}
\end{table}

Furthermore, the work carried out by Elsas and Dumais~\cite{Elsas2010} describes the temporal aspects of changing web documents. The authors describe news websites to consist of highly changing webpages. Whenever new stories are available, or the old stories are updated, the news websites would change. These changes occur in different amounts and at different frequencies. To observe how documents are changed, the authors have created a set of approximately 2,000,000 HTML pages, and crawled them every week for 10 weeks. Over the sampled period, over 62\% of pages had no virtual difference. They have also pointed out that highly relevant documents are both more likely to change, and contain a higher amount of change than general documents. As the final outcome, the authors have proposed a ranked retrieval model for dynamic documents based on the temporal aspect of content which lets differential weighting of content. Work done by Saad and Gançarski~\cite{Saad2012} has monitored over 100 webpages from the France Televisions (FranceTV) archive, which depicts the evolution of changes within the French Web. Each page was crawled every hour, and over 3000 versions were obtained daily. The results have shown that the pages at the root level of the archive, such as homepages changed significantly, whereas pages in the deepest levels did not change.

Table~\ref{tab1:dynamic-web} presents a summary of the various research work carried out to detect the dynamic nature of web content. From Table~\ref{tab1:dynamic-web}, it can be seen that webpages belonging to popular websites such as sports and news websites tend to change frequently whereas webpages belonging to government and educational domains change less frequently. Hence, it can be concluded that webpages belonging to popular websites tend to change more frequently than those of less popular websites that target specific functions and niche audiences.

\section{Architectural Aspects}
\label{section3}
Several studies have proposed different architectures for change detection systems. The two main architectures that are being used widely within current CDN systems are server-based architecture and client-based architecture \cite{Meegahapola2017a} and they have their advantages and disadvantages.

\subsection{Server-based Architecture}

\begin{figure}[!ht]
    \centering
    \includegraphics[width=0.6\textwidth]{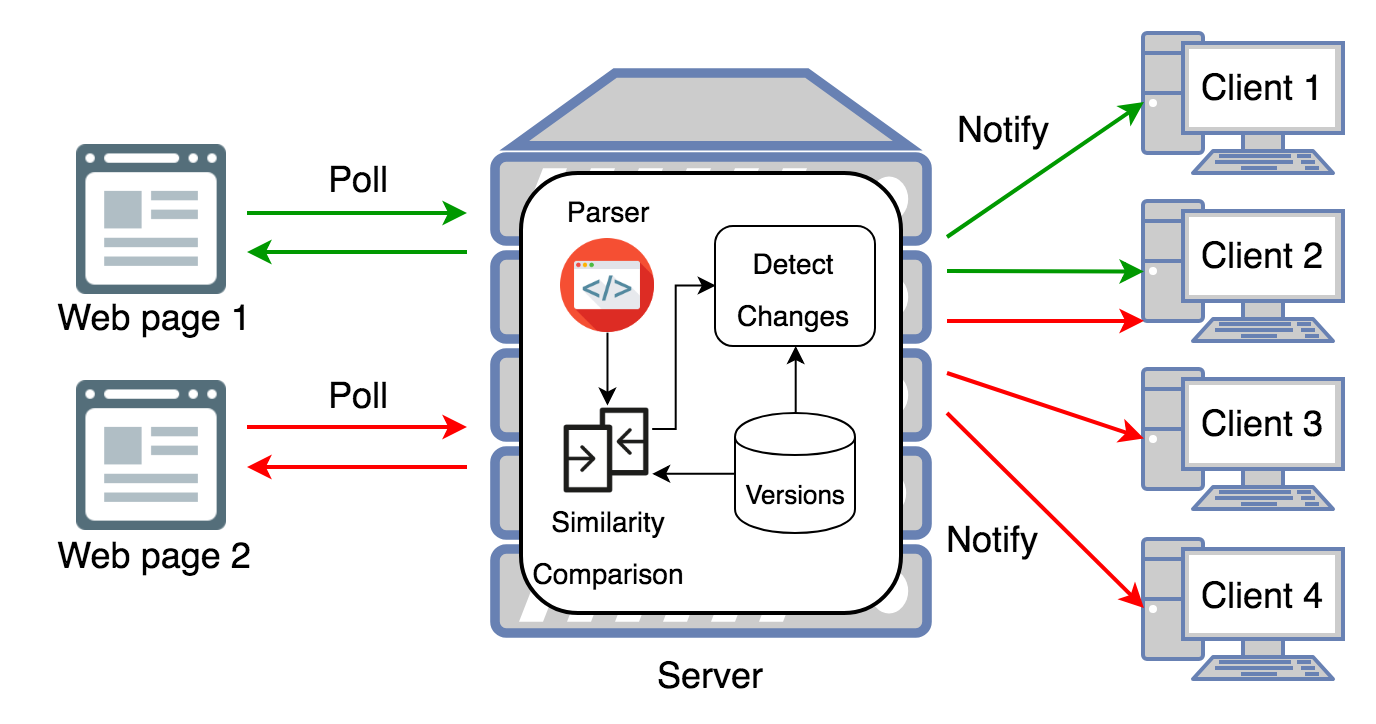}
    \caption{Server-based architecture for CDN systems.}
    \label{fig4:server-archi}
\end{figure}

The server-based architecture, as depicted in Figure~\ref{fig4:server-archi} consists of the main server, which polls webpages periodically to track changes, and sends alerts about these changes to the subscribed users (clients) by email notifications. If a large number of webpages exist, the computational load for the server will increase as the server must identify changes in each of the webpages added by users. This can also lead to reduced detection frequencies. The process of scaling such tools with the server-based architecture becomes expensive and makes the system less efficient. Due to these issues, the frequency in which changes are detected on webpages will not be sufficient and the server might fail to observe some changes that have occurred in frequently-changing webpages.

Sitemaps~\cite{sitemaps2008} is used by servers to inform search engines about the changes that it thinks are important, and are made available for crawling. This allows crawlers to identify webpages that have changed recently without having to download and process HTML pages~\cite{Brandman2000}. Support for Sitemaps protocol by the search engine industry was announced by Google, Yahoo and Microsoft in the year 2006~\cite{Schonfeld2009}. Since then, many websites such as Amazon, CNN and Wikipedia have begun supporting Sitemaps. Studies have shown that the use of sitemaps has improved crawling results~\cite{Schonfeld2009}. Data from Sitemaps can be used by CDN systems to get updated content. The use of Sitemaps helps to eliminate difficulties faced by crawlers and expose data regarding changes only.

\subsection{Client-based Architecture}

\begin{figure}[!ht]
    \centering
    \includegraphics[width=0.6\textwidth]{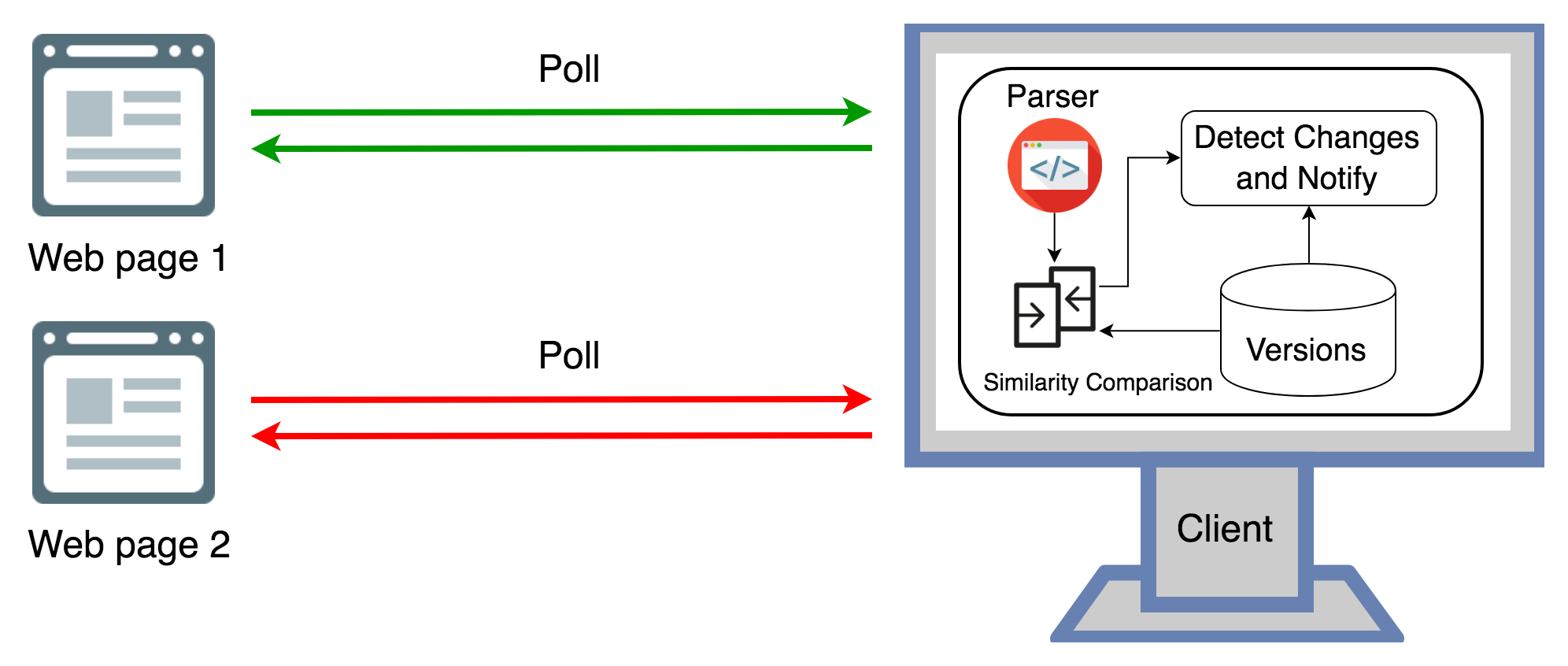}
    \caption{Client-based architecture for CDN systems.}
    \label{fig5:client-archi}
\end{figure}

The client-based architecture involves clients who wish to track changes occurring on webpages, and these machines poll webpages in iterations to identify changes. Users having extra computational resources can detect frequent changes occurring on webpages, and the other users might be unable to do so. The client-based architecture has been implemented in the form of browser plugins, and hence, may get bottlenecks due to network connectivity and machine performance. Figure~\ref{fig5:client-archi} illustrates the client-based architecture for a CDN system. 

Out of the currently available CDN systems, a limited number of them are built using the client-based architecture. They come in the form of browser plugins/extensions such as Distill~\cite{Distill2013} and Check4Change~\cite{Check4Change2006}. Once the extension is installed and enabled successfully, the user can add webpages to monitor. It will regularly check the webpages the user has added to be monitored, within the user’s browser. If any changes are detected, the user will get a browser notification.

\subsection{Customized Architectures}

Several customized architectures of CDN can be found in which researchers have tried to improve the efficiency from an architectural perspective. A design to crawl webpages in parallel using several machines, and integrate the problems with crawling has been proposed by Yadav et al.~\cite{Yadav2007}. They have designed a new architecture to build a parallel crawler. The proposed architecture consists of two main components. They are the Multi-threaded Server (MT-Server) and the client crawlers. MT-Server is the principal component that manages a collection of client machine connections, which downloads and crawls webpages. It has a special methodology to distribute URLs among the clients when their "priority index" is determined. The client crawlers are considered as the different machines which interact with the server. The client count that can be supported may vary according to the available resources and their implementation. Furthermore, there exists no communication between clients, and the clients interact only with the server. However, scaling the system can result in high costs as the count of server nodes has to grow.

Work carried out by Prieto et al.~\cite{Prieto2015} presents a system with a collaborative architecture to detect changes in distributed systems. The authors have proposed the Web Change Detection system (WCD). The four major components of this system’s architecture are the web client, web server, WCD agent and WCD server. Web client is a general browser of the web, which loads the webpages. Web server is a general server that caches the webpages that were monitored. WCD agent is an application operating in the browser that sends information about modifications that have occurred on webpages to the WCD server. WCD server stores and sends the WCD agents to clients, and stores the data about monitored webpages. To detect the near-duplicates, PageRank~\cite{Page1999} values have been considered along with the shash tool~\cite{ShashTool2015}. High change detection rates and a low cost of maintenance have been produced by this tool. However, in times of excessive usage, if the system gets many requests, it may fail to process them in real-time.

An "approach for distributed aggregation, faster scalable content classification and change detection in web content" has been proposed by Nadaraj~\cite{Nadaraj2016}. The author has presented an algorithm to determine webpage changes, and a method to identify what field has been altered with the help of a data structure named Bloom filters~\cite{Bloom1970}. Web crawlers are being used to collect details about pages and URLs. It uses Bloom filters to identify the duplicate URLs in a webpage. Hash keys for every visited URL are saved in the bloom filter. Bloom filter entries will be validated when new URLs are discovered. This prevents the crawling mechanism from repeating in loops. The system creates a hash key for the content that has been crawled, and checks the presence of the hash within the Bloom lookup. If present, the content is the same as the existing content; otherwise, the content has been updated. If the hash key for the URL is not found, then the URL is added to the Bloom filter lookup file, and a hash for the crawled content is created and inserted. This method increases the efficiency of the crawling process as it distributes the workload. Furthermore, the strong association of crawlers to working machines will be minimized, and the crawlers will be able to function without any restrictions in distributed networks.

A hybrid architecture for CDN is proposed by Meegahapola et al.~\cite{Meegahapola2017b,Meegahapola2017a}. This architecture is a hybrid of server-based and client-based architectures. It has two independent crawling schedules for the server and the clients. The server will crawl all the webpages it has registered, and the clients will crawl the webpages which they want to track. The change detection process occurs independently in the server and clients. If the server detects a change, it will be notified to the interested clients. If a change is detected by a client, then it will directly report back to the server, and the server will notify the interested clients. According to this architecture, the time elapsed in between two consecutive server poll actions is divided among the available clients which in turn speeds up the detection process. Table~\ref{tab2:cdn-archi} presents a summary of the various architectures that are being used by commercially available CDN systems and that have been proposed by researchers.

\begin{table}[!ht]
    \centering
    \caption{A summary of the architectures used/proposed for CDN systems.}
    \resizebox{\textwidth}{!}{%
    \begin{tabular}{|l|l|l|l|l|l|}
    \hline
    \textbf{\makecell{Architecture}} & \textbf{\makecell{Advantage}} & \textbf{\makecell{Disadvantage}} & \textbf{\makecell{Inter-Client \\Communi-\\cation}} & \textbf{\makecell{Parallel \\Processing}} & \textbf{\makecell{Cost}} \\
     \hline
    \makecell[l]{Server-based~\cite{Meegahapola2017a}} & \makecell[l]{Centralized \\monitoring} & \makecell[l]{High load on \\the server} & \makecell[l]{No} & \makecell[l]{No} & \makecell[l]{High} \\
    \hline
    \makecell[l]{Client-based~\cite{Meegahapola2017a}} & \makecell[l]{Clients have \\control} & \makecell[l]{The network \\bottleneck} & \makecell[l]{No} & \makecell[l]{No} & \makecell[l]{Medium} \\
    \hline
    \makecell[l]{Parallel \\crawling~\cite{Yadav2007}} & \makecell[l]{Prioritize URLs \\to crawl} & \makecell[l]{High overheads \\in communication} & \makecell[l]{No} & \makecell[l]{Yes} & \makecell[l]{High} \\
    \hline
    \makecell[l]{Collaborative \\WCD~\cite{Prieto2015}} & \makecell[l]{Operate in \\distributed \\networks} & \makecell[l]{Can fail to \\process requests \\in real-time} & \makecell[l]{No} & \makecell[l]{No} & \makecell[l]{Medium} \\
    \hline
    \makecell[l]{Distributed \\aggregation~\cite{Nadaraj2016}} & \makecell[l]{Distributes the \\workload} & \makecell[l]{Overhead of \\hashing} & \makecell[l]{Yes} & \makecell[l]{Yes} & \makecell[l]{Low} \\
    \hline
    \makecell[l]{Hybrid~\cite{Meegahapola2017b,Meegahapola2017a}} & \makecell[l]{Optimal use of \\client resources} & \makecell[l]{Need to co-\\ordinate clients} & \makecell[l]{Yes} & \makecell[l]{Yes} & \makecell[l]{Low} \\
    \hline
    \end{tabular}}
    \label{tab2:cdn-archi}
\end{table}

\section{Detecting Changes of Webpages}
\label{section4}
\subsection{Web Crawlers and Crawling Techniques}

A web crawler (also known as a spider) is "a software or a programmed script that browses the WWW in a systematic, automated manner"~\cite{Kausar2013}, and systematically downloads numerous webpages starting from a seed URL~\cite{Boldi2018}. Web crawlers date back to the 1990s, where they were introduced when the WWW was invented. The World Wide Web Worm~\cite{McBryan1994} and MOMspider~\cite{Fielding1994} were among the early web crawlers. Moreover, commercial search engines developed their own web crawlers as well such as Google crawler~\cite{Brin1998} and  AltaVista~\cite{Silverstein1999}. Later on, web crawlers that could efficiently download millions of webpages were built. 

We can consider the Internet to be a "directed graph" where each node represents a webpage, and the edges represent hyperlinks connecting these webpages~\cite{Kausar2013}. Web crawlers traverse over this graph-like structure of the Internet, go to webpages, and download their content for indexing purposes. It is crucial to identify which crawling technique should be used according to the purpose of the application.

\begin{figure}[!ht]
    \centering
    \includegraphics[width=0.9\textwidth]{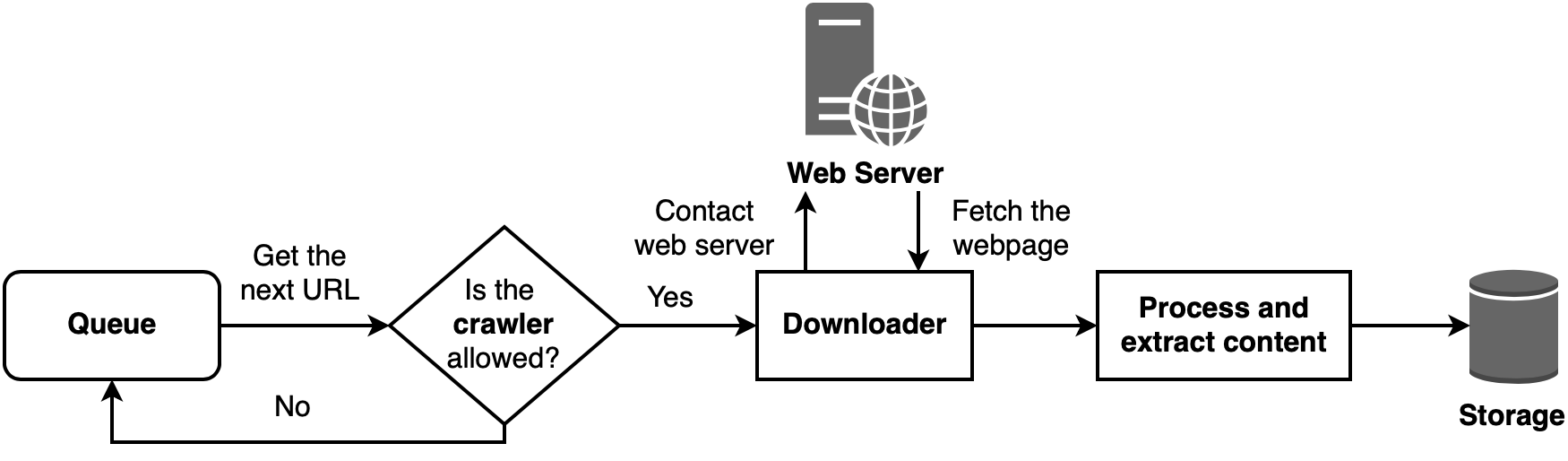}
    \caption{Overview of the web crawling process.}
    \label{fig7:web-crawling}
\end{figure}

Web crawling can be considered as the main process behind web caching, search engines and web information retrieval. A web crawler begins crawling from a seed URL and visits pages. Then it downloads the page, and retrieves the URLs in it. The discovered URLs will be kept in a queue, and this process repeats as the crawler travels from page to page. Figure~\ref{fig7:web-crawling} shows an overview of the web crawling process. Many crawling techniques are being used by web crawlers at present~\cite{Kausar2013} such as (1) general-purpose crawling, (2) focused crawling and (3) distributed crawling.

\subsubsection{General-Purpose Crawling}

In the general-purpose crawling technique, the web crawlers collect all the possible webpages from a set of URLs and their links to a central location. It can fetch numerous pages from many different locations. However, this technique can slow down the crawling process and reduce the network bandwidth as all the pages are being fetched.

\subsubsection{Focused Crawling}

Focused crawling is used to collect pages and documents that belong to a specific topic. This can reduce the workload and network traffic during download as the required pages are only downloaded. It crawls only the relevant regions of the Web. This method saves hardware and network resources significantly.

Initially, focused crawling was introduced by Chakrabarti et al.~\cite{Chakrabarti1999} as "a new approach to topic-specific Web resource discovery". The focused crawler has three major components. They are as follows.

\begin{enumerate}
    \item "Classifier decides whether to expand links on the webpages crawled".
    \item "Distiller determines visit priorities of crawled pages".
    \item "Crawler consists of dynamically reconfigurable priority controls which are controlled by the classifier and distiller"~\cite{Chakrabarti1999}.
\end{enumerate}

Diligenti et al.~\cite{Diligenti2000} have highlighted the importance of assigning credits to different documents across crawling paths that produce larger sets of topically relevant pages. The authors have proposed a focused crawling algorithm with a model that captures link hierarchies containing relevant pages. Later on, Menczer et al.~\cite{Menczer2001} have discussed different ways to compare topic-driven crawling techniques, whereas "a general framework to evaluate topical crawlers" was presented by Srinivasan et al.~\cite{Srinivasan2005}. Pant and Menczer~\cite{Pant2003} have introduced a topical crawling technique to gather documents related to business intelligence, which can support organizations to identify competitors, partners or acquisitions. The authors have tested four crawlers; Breadth-First, Naive Best-First, DOM and Hub-Seeking. The results of precision and recall on a given topic show that the Hub-Seeking crawler outperforms other crawlers. A popular technique to design-focused crawlers is the utilization of the "link structure" of the documents. Li et al.~\cite{Li2005} have proposed a focused crawling technique using a decision tree created by anchor texts found in hyperlinks. Jamali et al. ~\cite{Jamali2006} have presented a novel "hybrid focused crawler", which utilizes the "link structure" and similarity of the content to a particular topic. 

Mali and Meshram~\cite{Mali2011} have proposed another approach for a web crawler with focused crawling features. Three layers are present in this architecture; "page relevance computation", "determination of page change" and "updating the URL repository". During the crawling mechanism, all the pages are not downloaded. Instead, it extracts the URLs and the words of interest. Frequency of related words and the number of forward links and backward links to and from the webpage collaboratively decide the importance of that webpage being parsed. Certain parameters such as topic vectors and relevance scores are used to check the importance of the page. If the relevance level exceeds a predefined threshold, it is downloaded for extraction. Otherwise, the page will be discarded. 

Work done by Bhatt et al.~\cite{Bhatt2015} has studied focused web crawlers with their advantages and disadvantages. Additionally, the authors have suggested methods to further improve the efficiency of web crawlers. With the advancement in optimization algorithms, researchers have turned their focus on optimizing web crawlers. Optimizations allow web crawlers to select more suitable webpages to be fetched. Focused crawlers have been optimized to increase the efficiency and performance of the crawler using search algorithms such as Breadth-First Search~\cite{Bhatt2015}, Fish-Search~\cite{DeBra1997}, Shark-Search~\cite{Hersovici1998} and evolutionary algorithms such as Genetic algorithms~\cite{Yohanes2011,Zheng2011} and Ant algorithms~\cite{Zheng2011}.

\begin{figure}[!ht]
    \centering
    \includegraphics[width=0.7\textwidth]{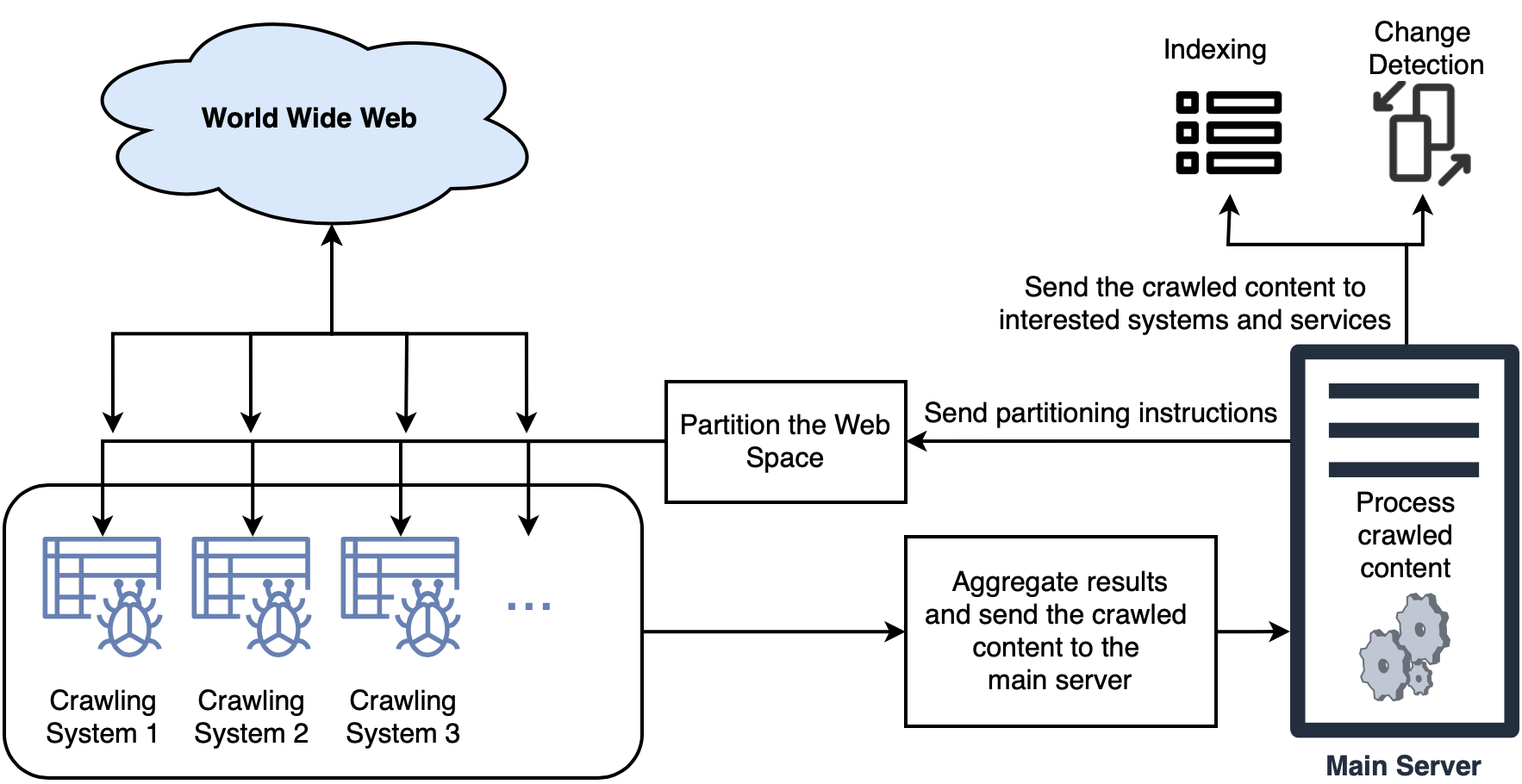}
    \caption{Arrangement of components in a distributed crawling system.}
    \label{fig9:distributed-crawl}
\end{figure}

\subsubsection{Distributed Crawling}

This method uses multiple processes to crawl webpages and download their content. Web crawlers are distributed over different systems, allowing them to operate independently. Distributed crawling techniques were introduced due to the inherent problems faced by centralized crawling solutions, such as reduced throughput of the crawl and link congestion~\cite{Kc2008}. 

Figure~\ref{fig9:distributed-crawl} denotes the general arrangement of components in a distributed crawling system. The crawlers are distributed across different systems, and they crawl webpages in different parts of the Web. The Web can be partitioned using different graph partitioning algorithms, and each crawler is provided with a seed URL to start the crawling process~\cite{AhmadiAbkenari2012}. All the crawled content is then aggregated and sent to the main server. This content can be processed within the main server or they can be sent to other services for different purposes such as indexing and version comparison within CDN systems. One of the most popular high-performance distributed crawlers found at present is the Googlebot~\cite{GoogleBot2013}. Googlebot is the web crawler used by Google to fetch webpages for indexing. It is designed to be distributed on several machines so that the crawler can scale as the Web grows. Table~\ref{tab3:crawling-techniques} summarizes the three main crawling techniques with their advantages, disadvantages and comparison of features.

\begin{table}[!ht]
    \centering
    \caption{A summary of the main crawling techniques.}
    \resizebox{\textwidth}{!}{%
    \begin{tabular}{|c|c|c|c|c|c|}
    \hline
    \textbf{\makecell{Crawling \\Technique}} & \textbf{\makecell{Crawling \\Pattern}} & \textbf{\makecell{Prioritize \\links}} & \textbf{\makecell{Partition \\link \\space}} & \textbf{\makecell{Advantage}} & \textbf{\makecell[l]{Disadvantage}} \\
    \hline
    \makecell[l]{General \\purpose\\ \cite{Kausar2013}} & \makecell[l]{Crawl all the \\available webpages \\by going through \\link URLs} & \makecell[l]{No} & \makecell[l]{No} & \makecell[l]{Can fetch pages \\from many \\different \\locations} & \makecell[l]{Affects the \\network \\bandwidth} \\
    \hline
    \makecell[l]{Focused \\crawling\\ \cite{Chakrabarti1999}} & \makecell[l]{Crawl pages \\belonging to a \\given topic} & \makecell[l]{Yes} & \makecell[l]{No} & \makecell[l]{Can retrieve \\content related \\to a given topic} & \makecell[l]{Becomes more \\time consuming \\as the topics \\expand} \\
    \hline
    \makecell[l]{Distributed \\crawling\\ \cite{Kc2008}} & \makecell[l]{Crawlers are \\distributed across \\systems \\independently} & \makecell[l]{Varies} & \makecell[l]{Yes} & \makecell[l]{Provide more \\throughput} & \makecell[l]{Difficult to \\manage and \\coordinate \\crawlers} \\
    \hline
    \end{tabular}}
    \label{tab3:crawling-techniques}
\end{table}

Detection of crashed agents is one of the important concerns of distributed web crawlers. Several distributed crawler solutions have been presented during the past~\cite{Shkapenyuk2002,Boldi2004}. To address this concern, crawlers can be designed as reliable failure detectors~\cite{Chandra1996}, in which timeouts can be used to detect crashed agents. UbiCrawler~\cite{Boldi2004} is an example of a web crawler with a reliable failure detector. It is a distributed web crawler, which consists of several agents that are responsible to crawl their own share of the Web. However, it does not guarantee that the same webpage is not visited more than once. Based on the experience with UbiCrawler, the authors have built BUbiNG~\cite{Boldi2018}, a fully distributed web crawler which can detect (near-)duplicate webpages based on the content. However, it does not support URL prioritization. 

Another aspect that has drawn the attention of researchers is the efficient partitioning mechanisms of the Web space. Work done by Exposto et al.~\cite{Exposto2008} has presented a multi-objective approach for partitioning the Web space by modeling the Web hosts and IP hosts as graphs. These graphs are partitioned, and a new graph is created with the weights calculated using the original weights and the edge-cuts. Results show that efficient partitioning techniques have improved download times, exchange times and relocation times.

Kc et al.~\cite{Kc2008} have introduced LiDi Crawl (which stands for Lightweight Distributed Crawler), which is a centralized crawling application with limited resources. It consists of a central node and several individual crawling components. The distributed crawling nature results in the reduced dependence on expensive resources. Kumar and Neelima~\cite{Kumar2011} have proposed a scalable, fully-distributed web crawler, without a central node. It consists of multiple agents, where each agent is coordinated so that they crawl their own region on the Web. An agent runs several threads, where each thread is made responsible to visit a single host. The main objective of having multiple agents is to break down the task of crawling into separate parts, and allow specialized modules to carry them out efficiently. Anjum et al.~\cite{Anjum2012} state that web crawlers should be aware of webpage modifications, and have discussed techniques to retrieve information on such modifications. However, the authors have found that the presence of multiple JavaScript and CSS files can reduce the efficiency of certain retrieval techniques. 

Efficient crawling of Rich Internet Applications (RIA) is an open problem as RIAs consist of many characteristics such as, the use of JavaScript and browser plugins, which makes the crawling process complex. Model-Based Crawling~\cite{Benjamin2011} was introduced to determine an optimal crawling strategy for RIA. An extension of this model is presented by Dincturk et al.~\cite{Dincturk2012}, which uses a statistical approach in determining a crawling strategy. The recent work carried out by Mirtaheri et al.~\cite{Mirtaheri2013}, intends to lower the time taken to crawl RIAs by introducing Dist-RIA crawler, which is a distributed crawler to crawl RIAs in parallel. However, it assumes that all the nodes have equal processing power, and assigns an equal number of tasks to each node. This can be problematic when there is heterogeneity.

\begin{table}[!ht]
    \centering
    \caption{A summary of various crawling solutions presented in the literature.}
    \begin{tabular}{|l|l|l|l|}
    \hline
    \textbf{\makecell{Crawler}} & \textbf{\makecell{Distributed}} & \textbf{\makecell{Advantage}} & \textbf{\makecell{Disadvantage}} \\
    \hline
    \makecell[l]{UbiCrawler~\cite{Boldi2004}} & \makecell[l]{Yes} & \makecell[l]{Failure detection} & \makecell[l]{Does not guarantee that \\duplications are not present} \\
    \hline
    \makecell[l]{BUbiNG~\cite{Boldi2018}} & \makecell[l]{Yes} & \makecell[l]{No coupling to machines} & \makecell[l]{Does not allow prioritization \\of URLs} \\
    \hline
    \makecell[l]{LiDi Crawl~\cite{Kc2008}} & \makecell[l]{No} & \makecell[l]{Can work with limited \\resources} & \makecell[l]{Affected by a single point of \\failure}\\
    \hline
    \makecell[l]{Dist-RIA crawler\\ \cite{Mirtaheri2013}} & \makecell[l]{Yes} & \makecell[l]{Can crawl Rich Internet \\Applications in parallel} & \makecell[l]{Lack of adaptive load-\\balancing for diverse nodes}\\
    \hline
    \makecell[l]{MTCCDW~\cite{Meegahapola2017d}} & \makecell[l]{No} & \makecell[l]{Optimized for the process of \\change detection} & \makecell[l]{Affected by a single point of \\failure} \\
    \hline
    \end{tabular}
    \label{tab4:crawling-solutions}
\end{table}

Multi-Threaded Crawler for Change Detection of Web (MTCCDW) has been introduced by Meegahapola et al.~\cite{Meegahapola2017d} to optimize the change detection process of webpages. Many threads were used to carry out the tasks of (1) "retrieving current versions of webpages", (2) "retrieving old versions from a version repository" and (3) "comparison of the two versions to detect the changes"~\cite{Meegahapola2017d}. Two thread-safe queues were used in between these three tasks to optimize them. The authors have determined the optimum number of threads to be used in each of these tasks, to ensure that the CDN system works optimally without idling. This multi-threading-based algorithm differs from standard process optimization tasks because of the I/O operations which are involved in fetching a webpage from the Web. Hence this algorithm provides a more significant improvement in processing times over doing I/O operations and processing in a central processing unit. Table~\ref{tab4:crawling-solutions} summarizes the various crawling solutions presented in this subsection.

\subsection{Scheduling Algorithms}

Among the important research problems observed in the web information retrieval domain, the scheduling process of visiting webpages along hierarchies of links has become significant. The frequency of change may differ in different webpages as they get updated at different time schedules. Certain webpages such as wiki pages may get updated rarely whereas news websites and blogs may get updated more frequently. To crawl these webpages more efficiently, dynamic mechanisms are required to detect the change frequency, and create checking schedules accordingly. Users will be notified immediately after changes have occurred on webpages that they are interested in. This will ensure computational resources are used optimally. However, most of these algorithms are kept proprietary, and a limited amount of details are published about them to prevent being reproduced~\cite{Castillo2004}. 

Various solutions have been proposed to determine optimal webpage crawling schedules based on different assumptions and different statistical frameworks. Work done by Coffman et al.~\cite{Coffman1998} has described crawler scheduling policies by assuming independent Poisson page-change processes. Studies carried out by Cho and Garcia-Molina~\cite{Cho2000b} have addressed the problem of determining the optimal number of crawls per page by using a staleness metric and Poisson process, where they assume uniform synchronization over time. Work done by Wolf et al.~\cite{Wolf2002} has proposed a technique to detect optimal crawl frequencies and theoretical optimal times to crawl webpages based on probability, resource allocation theory and network flow theory. Pandey and Olston~\cite{Pandey2005} have proposed a scheduling policy based on the "user-centric search repository quality".

\begin{table}[!ht]
    \centering
    \caption{A summary of various scheduling solutions.}
    \begin{tabular}{|l|l|l|}
    \hline
    \textbf{\makecell{Scheduling \\Solution}} & \textbf{\makecell{Advantage}} & \textbf{\makecell{Limitations}}\\
    \hline
    \makecell[l]{Coffman et al. 1998\\ \cite{Coffman1998}} & \makecell[l]{Schedule crawls when webpages \\change and reduce unnecessary \\crawls} & \makecell[l]{Can be sensitive to parameter \\changes in the Poisson process} \\
    \hline
    \makecell[l]{Cho and Garcia-\\Molina 2000~\cite{Cho2000b}} & \makecell[l]{Significantly improves the \\freshness of webpage indexes} & \makecell[l]{Affected when resource \\synchronization is not uniform} \\
    \hline
    \makecell[l]{Wolf et al. 2002~\cite{Wolf2002}} & \makecell[l]{Minimizes the number of \\non-relevant webpages for a \\user query} & \makecell[l]{Not suitable to be used \\online with large input \\dimensions}\\
    \hline
    \makecell[l]{Pandey and Olston \\2005~\cite{Pandey2005}} & \makecell[l]{Improves the quality of user \\experience with fewer resources} & \makecell[l]{Does not revisit webpages \\based on change frequency}\\
    \hline
    \makecell[l]{Santos et al. 2013\\ \cite{Santos2013}} & \makecell[l]{Prioritizes crawls according \\to how frequently webpages \\change} & \makecell[l]{Has not considered other \\features such as PageRank \\and crawling cost}\\
    \hline
    \end{tabular}
    \label{tab5:scheduling-solutions}
\end{table}

Various scheduling strategies for web crawling have been studied and compared in previous studies~\cite{Castillo2004,Filipowski2014}. Among these strategies are optimal, depth, length, batch, partial~\cite{Castillo2004}, depth-first search, breadth-first search, best-first search and best n-first search~\cite{Filipowski2014}. However, some of the strategies which have been considered cannot determine the temporal aspects such as how frequently webpages are undergoing changes, and prioritize the crawls accordingly. Evolutionary programming techniques such as genetic programming~\cite{Santos2013} can be used to solve this issue by facilitating schedulers to rank webpages according to how likely they are being modified. Moreover, certain algorithms may get affected as webpages are crawled even though nothing has changed, where computational resources are used inefficiently. Table~\ref{tab5:scheduling-solutions} compares the various scheduling solutions presented previously.

\subsection{Change Detection Algorithms}

Changes occurring on webpages can be divided into five categories as discussed by Oita and Senellart~\cite{Oita2011}. They are (1) content changes, (2) structural (or layout) changes, (3) attribute (or presentation) changes, (4) behavioural changes and (5) type changes. Content changes include changes occurring in the text, images, etc., whereas structural (or layout) changes consist of changes occurring to the placement of HTML tags. Attribute (or presentation) changes include changes done to the design and presentation of a webpage, such as changes in the fonts, colors or captions. Behavioral changes occur when active components such as embedded applications and scripts of webpages are changed. Type changes occur when modifications are done to the existing HTML tags, such as when a \emph{p} tag becomes a \emph{div} tag. Studies have been focused on all of these types of changes~\cite{Oita2011} and different methods and algorithms have been proposed~\cite{Borgolte2014,Jain2014}. 

A major concern of change detection in webpages is the ability to identify relevant and irrelevant changes. The research carried out by Borgolte et al.~\cite{Borgolte2014} has focused to ignore irrelevant changes such as change of advertisements, and try to detect important changes occurring on webpages using different methods. The  Delta framework introduced in this research, consists of four precise steps. In the initial step, it retrieves the currently available version of a website, and normalizes the DOM tree. Then in the next step, similarities are measured in comparison to the base version of the website. Similar changes are clustered to lower the analysis engine submissions. The final step is the "analysis of identified and novel modifications with a potential computationally costly analysis engine". Compared to many other pieces of research that focus on detecting changes that have been done to a website, the Delta framework focuses more on detecting significant changes occurring on a website.

Changes occurring within information collections can be difficult to track due to the vast amount of data being available. Unexpected changes within the collection can make the content unorganized and outdated, which can cause burdens such as the removal of outdated resources and replacement for lost resources. Jayarathna and Poursardar~\cite{Jayarathna2016} have presented a study that focuses on a categorization and classification framework to manage and curate distributed collections of web-based resources. The selected pages were categorized with their relationships between the anchor text to identify the relevant information of the target page. They have proposed a digital collection manager that addresses webpages with textual content only. It has been shown that due to unexpected changes, documents in these collections can get problematic. Further research is necessary to detect changes occurring to other resources in digital collections, such as audio files, video files and documents.

Various change detection approaches have been proposed across the available literature to detect the different changes occurring on webpages. Elaborated in Table~\ref{tab6:change-diff-algo} are a few popular algorithms that were considered in this survey. One of the common approaches that have been used for change detection in hierarchically structured documents such as webpages is the use of tree-structured data with \emph{diff} algorithms~\cite{Cobena2002}. Many variations of these diff algorithms have been presented in the available literature~\cite{Wang2003,Borgolte2014}. The basic idea on top of which all of these algorithms are built on is the process of modeling the hierarchically structured documents in the form of a tree with nodes, and compare the trees to identify changes that have occurred. Similarly, two trees can be developed; one with nodes from the parsed content of a previous version and the other with nodes from the parsed content of the currently available version. The two trees could be compared to identify which nodes have changed so that the changes can be determined.

\subsubsection{Fuzzy Tree Difference Algorithm}

The Delta framework proposed in~\cite{Borgolte2014} involves tree difference algorithms. Firstly, the modifications that have occurred in a website are extracted using fuzzy tree difference algorithms. Secondly, a machine learning algorithm is used to cluster similar modifications. The authors have employed a tree difference algorithm in which they have generalized to remove irrelevant modifications during the early phases of the analysis.

\subsubsection{BULD Diff Algorithm}

The BULD Diff algorithm~\cite{Cobena2002} has been used in computing the differences among the given two XML documents. It matches nodes in the two XML documents, and constructs a delta that represents the changes between the two compared documents. BULD stands for "Bottom-Up, Lazy-Down" propagation~\cite{Cobena2002}, which has been derived from its matchings that are propagated "bottom-up" and "lazily down". This algorithm can run in linear time. First, the algorithm matches the largest identical parts of both the XML documents. Then the algorithm tries to match subtrees and the parents of the matched subtrees. This process is repeated until no more matches are made. Then the remaining unmatched nodes can be identified as insertions or deletions.

\subsubsection{X-Diff Algorithm}

The research carried out by Wang et al.~\cite{Wang2003} has investigated how XML documents change and methods to perform change detection of XML documents. The authors have pointed out the fact that XML is becoming important as it is at present the de-facto standard in publishing web applications and data transfer/transportation. Internet query systems, search engines and continuous query systems heavily rely on efficiently detecting webpage changes in XML-documents because of the frequent rate at which webpages change in present days. The authors have described an algorithm named X-Diff~\cite{Wang2003}, which is an algorithmic model used to compute the difference in-between two versions of an XML-document. Unordered trees, structural information of XML-documents and high performance are among the main features of X-diff algorithm. The authors have tested the X-Diff algorithm for its accuracy and efficiency of change detection by considering three main nodes in the DOM tree namely element nodes, text nodes and attribute nodes. Going beyond what has currently being done, the authors have compared the ordered tree model of change detection and unordered tree model which they have used. They have also discussed the characteristics of XML domain, and established few key concepts such as "node signature" and "XHash".

\subsubsection{Tree Difference Algorithms}

The research carried out by Jain and Khandagale~\cite{Jain2014}, has focused on detecting changes in a specific location on a website or any document. This method involves tree comparison techniques. The majority of the existing techniques/systems check for changes in the whole webpage without allowing the user to select specific areas to monitor. When considering frequent changes, the cost of communication (or the information exchange) will cause inefficiencies by disturbing the focus to a given context. Hence, the authors have proposed a tree comparison mechanism to overcome these difficulties. They have considered the management of zone changes (changes in a particular place on a webpage), and it is quite achievable using a tree mechanism because once a webpage is converted into XML format, it can be converted into a tree according to the HTML tags and attributes. When it has to localize the detection mechanism, the focus is given only to a particular node, and the detection process will continue to their child nodes. However, the limit for the depth of the tree is not specified, which can be inefficient with large trees.

Yadav et al.~\cite{Yadav2007} describe a methodology to build a model to efficiently monitor webpage changes. The authors have suggested an algorithm that efficiently identifies and extracts the changes on various versions of a particular webpage, and evaluates the significance/importance of the change. This algorithmic model has three main sections. Initially, a document-tree is created for the webpage, and then it encodes the tree. Finally, it matches the current version with the previous version of the encoded tree. The algorithm searches for two change types occurring in web content. They are namely, structural changes and content changes.

\begin{figure}[!ht]
    \centering
    \includegraphics[width=\textwidth]{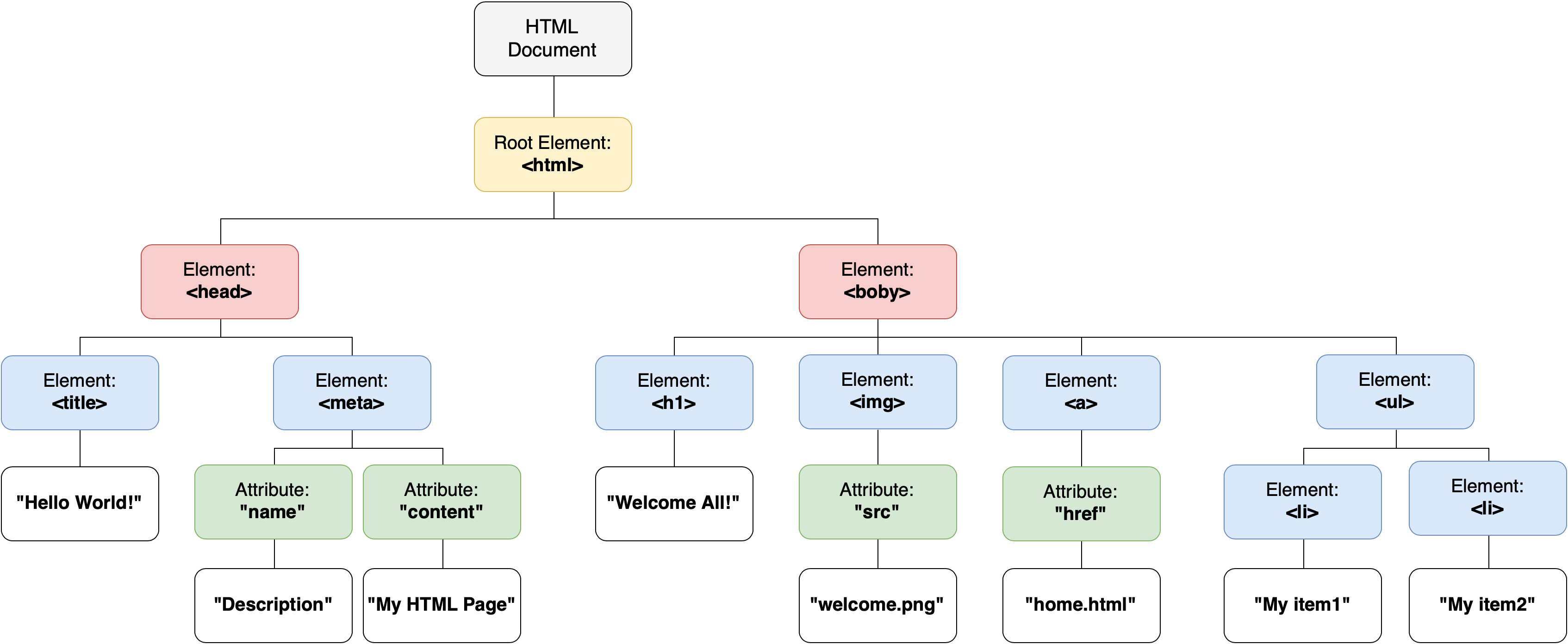}
    \caption{HTML document tree hierarchy of a sample webpage.}
    \label{fig10:html-doc-tree}
\end{figure}

Tree models can be of two types: the first type is the ordered tree model where the left-to-right order between nodes is crucial, and this order directly contributes to the result of change detection; the second type is unordered tree model, where only the ancestor relationships are significant. Different studies have been carried out using these two tree models. Level Order Traversal~\cite{Yadav2007} is a form of breadth-first search that uses an ordered tree model. First, it constructs a document tree by taking in an HTML file and parses the elements. Then, the opening tags are identified as tree nodes, and the tree is constructed for the webpage as illustrated in Figure~\ref{fig10:html-doc-tree}, while maintaining parent-child relationships and left-to-right order in-between siblings. Then the algorithm traverses through the tree to detect changes, and identify at which level of the tree the changes have occurred. However, some researches~\cite{Wang2003} have argued that the unordered tree model can generate more accurate results.

\subsubsection{Johnson's Algorithm}

The Johnson's algorithm~\cite{Johnson1977} was originally proposed as a method to "find the shortest paths between all the pairs of vertices in a sparse weighted directed graph". The same algorithm has been introduced by Johnson and Tanimoto~\cite{Johnson1999} to detect changes to content. The authors have tested a prototype to manage and deliver the latest tutorial content on the web to students. The system was designed to anticipate changes to documents. The difference between documents stored is evaluated and updated accordingly. This algorithm has been used in a tool named as Walden’s Paths Path Manager~\cite{FranciscoRevilla2001}. It computes the distance between two documents based on paragraphs, headings and keywords. Each signature difference is calculated, and the sum is taken as the total distance. It categorizes the change type, and it is easy to identify which type of content is mostly changed on a webpage. However, it does not consider the changes that occur to links. Furthermore, the results produced by this algorithm can be hard for a normal user to understand without the assistance of a computing device.

\begin{table}[!ht]
    \centering
    \caption{A summary of existing change detection algorithms.}
    \resizebox{\textwidth}{!}{%
    \begin{tabular}{|l|l|l|l|l|}
    \hline
    \makecell{Algorithm} & \makecell{Methodology/Function} & \makecell{Advantages} & \makecell{Disadvantages} & \makecell{Associated \\Work} \\
    \hline
    \makecell[l]{Shingling \\algorithm} & \makecell[l]{Group a sequence of terms in a \\given document and encode by \\a 64-bit Rabin fingerprint, referred \\as a "shingle". Use the Jaccard \\coefficient between the shingle \\vectors to compute the similarity \\ between the two documents.} & \makecell[l]{Fast detection due \\to the speed of the \\algorithm.} & \makecell[l]{If the content considered \\in the document is small, \\it will not be able to \\generate enough shingles.} & \makecell[l]{Detecting \\duplicate web \\documents \\using click-\\through data \\ \cite{Radlinski2011} } \\
    \hline
    \makecell[l]{Johnson’s \\algorithm} & \makecell[l]{Computes the distance between \\two documents based on \\paragraphs, headings and \\keywords. Each signature \\difference is calculated, and the \\sum is taken as the total distance.} & \makecell[l]{Categorizes the \\change type. Easy \\identification of the \\mostly changed \\content type.} & \makecell[l]{Does not include links \\when determining \\changes.} & \makecell[l]{Walden’s Paths \\Path Manager\\ \cite{FranciscoRevilla2001}}\\
    \hline
    \makecell[l]{Proportional \\algorithm} & \makecell[l]{Computes the distance that is \\normalized and symmetric using \\each individual signature. \\Proportional change of each \\signature is used with regards to \\the total number of changes.} & \makecell[l]{Can evaluate changes \\efficiently in the page \\without analyzing \\all the related pages in \\a given path.} & \makecell[l]{There is a slight \\performance trade-off \\when compared to \\Johnson’s Algorithm.} & \makecell[l]{Walden’s Paths \\Path Manager\\ \cite{FranciscoRevilla2001}}\\
    \hline
    \makecell[l]{Cosine \\algorithm} & \makecell[l]{Compute a cosine value between \\two vectors of the page \\considered and all the pages for \\the similar topic except the page \\considered. Then measure the \\change in cosine value.} & \makecell[l]{Provides a more \\meaningful cut-off to \\identify the level of \\change.} & \makecell[l]{The context-based \\algorithm gives a mid-\\level outcome as it tends \\to generate false positives \\and false negatives.} & \makecell[l]{Managing \\distributed \\collections~\cite{Dalai2004}} \\
    \hline
    \makecell[l]{Fuzzy tree \\difference \\algorithm} & \makecell[l]{A tree difference algorithm \\which is generalized into a \\fuzzy-notion.} & \makecell[l]{Can eliminate trivial \\changes momentarily.} & \makecell[l]{Have to define a suitable \\(fuzzy) hash function.} & \makecell[l]{Delta \\framework~\cite{Borgolte2014}} \\
    \hline
    \makecell[l]{BULD Diff \\algorithm} & \makecell[l]{Match subtrees of two XML \\documents till no more matches \\can be made. Unmatched subtrees \\are considered to be insertions or \\deletions.} & \makecell[l]{The difference can be \\computed in linear \\time.} & \makecell[l]{Reduce performance \\proportionally to the \\tree-depth.} & \makecell[l]{Detect changes \\in XML \\documents~\cite{Cobena2002}} \\
    \hline
    \makecell[l]{CX-DIFF} & \makecell[l]{Identifies user specific changes \\on XML documents. Objects of \\interest are extracted, filtered \\for unique insertions and \\deletions, and the common order \\subsequence is computed.} & \makecell[l]{Gives notifications to \\users on time. \\Optimized space \\usage. Efficient \\monitoring of types.} & \makecell[l]{Can lead to overloading \\of the servers due to the \\highly expensive \\computational cost.} & \makecell[l]{WebVigil~\cite{Jacob2004}} \\
    \hline
    \makecell[l]{X-Diff~} & \makecell[l]{Compares two XML documents \\depending on their equivalent \\trees. It generates a "minimum-\\cost edit script" including a \\series of fundamental edit \\operations that transform a \\given tree to the other at a \\minimum cost.} & \makecell[l]{Provides accurate \\results by maintaining \\an unordered view of \\the document where \\ancestor links add \\value to the result.} & \makecell[l]{May require a large \\amount of statistical \\information for the \\detection process. } & \makecell[l]{X-Diff~\cite{Wang2003} } \\
    \hline
    \makecell[l]{Vi-DIFF} & \makecell[l]{Detects content and structural \\changes including the visual \\representation of webpages.} & \makecell[l]{Can detect semantic \\differences between \\two versions.} & \makecell[l]{May cost more resources \\and time when compared \\to other methods.} & \makecell[l]{Vi-DIFF~\cite{Pehlivan2010}}\\
    \hline
    \makecell[l]{Level order \\traversal} & \makecell[l]{This is a breadth first traversal \\algorithm and considers the \\changes in the document-tree to \\detect the changes.} & \makecell[l]{Simple. Low cost of \\computation. Helps to \\reduce network traffic \\due to the usage of \\HTTP metadata.} & \makecell[l]{May not give sufficient \\information to the user.} & \makecell[l]{Change Detection \\in Webpages\\ \cite{Yadav2007}} \\
    \hline
    \end{tabular}}
    \label{tab6:change-diff-algo}
\end{table}

\subsubsection{Proportional Algorithm}

The Proportional algorithm~\cite{FranciscoRevilla2001} is based on signatures, and gives a much simple calculation of the distance. It computes a distance that is normalized and symmetric using each signature (paragraphs, headings, keywords and links). The proportional change of each signature is used with regards to the total number of changes. The measured distance has properties such as normalized properties and symmetric properties. These properties help the user of the algorithm significantly by providing a listing or visualization of various webpage changes. Having such listings, and visualizations makes it possible for the user to easily analyze changes of the webpage without necessarily reading and reviewing all the pages in the path. However, there is a slight performance trade-off when compared to the Johnson's Algorithm as changes to each signature are computed individually.

Table~\ref{tab6:change-diff-algo} summarizes the change detection algorithms discussed in this subsection, and compares the methodology/functions, advantages, disadvantages and associated research work. According to Table~\ref{tab6:change-diff-algo}, it can be observed that there are many change detection algorithms based on difference calculation and traversal techniques. Different algorithms detect different changes such as changes occurring to text, visual representation and XML structure. The majority of the algorithms perform at fairly good speeds, and provide faster detection rates along with efficient resource utilization and low computational costs. Algorithms such as level order traversal have a low overhead, which will in return reduce the network traffic in communication. Algorithms such as the Cosine algorithm and the Fuzzy Tree Difference algorithm provide different levels of changes so that the threshold can be decided upon the specific usage of the algorithms. Algorithms such as the Johnson’s algorithm can be used to categorize the change types so that it is easy to identify which type of content is mostly changed in a webpage. 

However, there are certain shortcomings in each of these algorithms. If the content being compared is very small the Shingling algorithm will not be able to generate sufficient shingles. The Johnson’s algorithm does not identify changes occurring in links. Algorithms such as CX-DIFF and Vi-DIFF consist of highly expensive computations, and can even lead to overloading of the servers. Level order traversal algorithms do not provide sufficient information to the user about changes occurring.

\subsection{Frequency of Webpage Changes}

Since the 1970s, several studies have been carried out based on statistics, to estimate the change frequency~\cite{Misra1975,Canavos1972}. Nevertheless, the majority of these studies have been done under the assumption that the history of changes is available for the webpage, which might not be true always in the area of change detection of webpages. When analyzing CDN systems, it is visible that the complete change histories will not be available for a webpage being tracked. In several related studies~\cite{Fetterly2003,Grimes2008,Olston2008}, the Poisson model has been introduced as a model that can be used to estimate the rate of change of a webpage (also known as change frequency). Most of the work which has been carried out is with an assumption: "changes arrive as a Poisson process, and that the average rate of change can be estimated under this model". Brewington and Cybenko~\cite{Brewington2000} have used an  exponential probabilistic model to infer the times between changes occurring in individual webpages. The ages of webpages that have gone through many changes over a time period can be closely modeled using an exponential distribution. However, it models all the webpages as dynamic, even if the webpages change rarely and their only changes are their removal from the Web.

According to Olston and Najork~\cite{Olston2010}, the binary freshness model can be used to measure the freshness in a webpage. This model which is also known as obsolescence is a function that is of the form
\begin{equation}
    f(p,t) \in \{0,1\}
\end{equation}

where $f(p, t)$ denotes the freshness of a particular webpage $p$ over a $t$ time period. It compares the live copy of a specific webpage $p$ with a cached version of the webpage across a time period $t$ to check if they are identical (or near-identical). Under this model, if $f(p, t)$ equals to one, then the webpage $p$ can be called fresh over the time $t$, whereas otherwise it is considered as a stale webpage which has gone through some change. This model is simple, but effective, and provides readers with a very good understanding of webpage changes. The first most study regarding the freshness maximization problem was done by Coffman et al.~\cite{Coffman1998}, where the authors have proposed a Poisson model for webpage changes. A set of random and independent events that occur in a fixed-rate can be modeled using the Poisson Process. A webpage can undergo changes that cause the cached copy of the web crawler to go stale. If $\lambda(p)$ is the rate parameter of a Poisson distribution where $p$ is the webpage, that specific distribution can be used to denote the occurrence of changes in that specific webpage. This also suggests that the changes happen independently and randomly with a mean rate of $\lambda(p)$ changes per unit of time. 

However, the binary freshness model lacks the ability to determine whether one page is fresher than the other since the model outputs a binary value; fresh or stale. Hence Cho and Garcia-Molina~\cite{Cho2003a} have introduced a non-binary freshness model known as the temporal freshness metric which is,
\begin{equation}
    f(p,t) \propto age(p,t)
\end{equation}

in which $age(p, t)$ represents the age of a page $p$ up to time $t$ and $age(p, t) \in \{0, a\}$ where $a$ is the time duration the copies differed. If a cached-copy of webpage $p$ is indistinguishable from its live-copy, then $age(p, t) = 0$. The intuition of this methodology is that "the more time a cached page stays unsynchronized with its live-copy, the more their content tends to drift away".

Cho and Garcia-Molina~\cite{Cho2003b} have proposed several frequency estimators for different online applications that require frequency estimation with different accuracy levels. The authors have proposed frequency estimators for scenarios, where the existence of a change is known, and the last date of change is known. Furthermore, the authors have proposed a model to categorize elements into different classes based on their change frequencies. These aspects can be modeled in CDN systems to categorize webpages once their change frequencies have been estimated.

\begin{table}[!ht]
    \centering
    \caption{A summary of existing change frequency detection techniques.}
    \resizebox{\textwidth}{!}{%
    \begin{tabular}{|l|l|l|l|}
    \hline
    \textbf{\makecell{Method}} & \textbf{\makecell{Citation}} & \textbf{\makecell{Characteristics}} & \textbf{\makecell{Limitation}}\\
    \hline
    \makecell[l]{Poisson \\distribution} & \makecell[l]{Coffman et al. \\1998~\cite{Coffman1998}} & \makecell[l]{Webpage changes occur as a \\Poisson process and determine \\the mean rate at which \\changes occur.} & \makecell[l]{Sensitive to parameter \\changes}  \\
    \hline
    \makecell[l]{Exponential \\probabilistic model} & \makecell[l]{Brewington \\and Cybenko \\2000~\cite{Brewington2000}} & \makecell[l]{The ages of webpages that have \\gone through many changes over \\a time period are modeled using \\an exponential PDF.} & \makecell[l]{Sensitive to parameter \\changes. \\Assumes all webpages \\are dynamic.}  \\
    \hline
    \makecell[l]{Temporal freshness \\metric} & \makecell[l]{Cho and \\Garcia-Molina \\2003~\cite{Cho2003a}} & \makecell[l]{Determine the level of freshness \\of a page} & \makecell[l]{Cannot incorporate \\information about the \\type of changes} \\
    \hline
    \makecell[l]{Binary freshness \\model} & \makecell[l]{Olston and \\Najork \\2010~\cite{Olston2010}} & \makecell[l]{Determine whether a page is \\fresh or stale (provides a binary \\value)} & \makecell[l]{Cannot determine the \\level of freshness}  \\
    \hline
    \makecell[l]{Change Frequency \\Estimator Module \\(CFEM)} & \makecell[l]{Meegahapola \\et al. 2017~\cite{Meegahapola2017c}} & \makecell[l]{Use a machine learning model to \\determine the time interval \\between two crawls} & \makecell[l]{Can result in \\overfitting}  \\
    \hline
    \end{tabular}}
    \label{tab7:freq-detect-tech}
\end{table}

Grimes and O’Brien~\cite{Grimes2008} state that for every webpage, the hourly update rate can be represented by a Poisson distribution together with $\lambda = 1 / \Delta$ where $\lambda$ is a parameter and $\Delta$ is the mean time between changes. The authors have also described a mechanism to identify webpage changes and a model for the computation of the rate of change of a given webpage. 

Work carried out by Meegahapola et al.~\cite{Meegahapola2017c} has proposed a methodology to pre-predict the update frequency of webpages (the rate at which changes happen), and reschedule the crawling schedule to make the process of crawling webpages more efficient in search engines and web crawlers used for change detection. The authors have introduced a change frequency detection system named Change Frequency Estimator Module (CFEM). It includes two processes. Whenever a fresh webpage is added, that webpage will be crawled to detect a suitable change frequency, and it will be recorded in the system. Then these values are sent to a machine learning model~\cite{Meegahapola2018} which will predict the time interval between two crawls for a particular webpage. The change values together with change frequencies for a webpage is sent to the machine learning model, it would output a time interval called a \emph{loop time} corresponding to that particular webpage. This value is an estimation of the average time taken by the webpage between two changes or in other terms, the \emph{refresh rate}. It has also been observed by the authors that frequently changing websites obtained lower loop times in comparison to webpages which do not change often. Table~\ref{tab7:freq-detect-tech} summarizes the different change frequency detection techniques with their characteristics and limitations.

\section{Change Notification}
\label{section5}
CDN systems provide facilities to notify users about information changes or occurrence of events in a particular webpage of interest. From our studies, we have determined three main characteristics that should be considered when designing a notification system. They are (1) when to notify, (2) how to notify and (3) what to notify.

When to notify changes is an important aspect to decide on when developing a change notification system. Users may want to get notifications as soon as they occur, or some may want to get periodic notifications as they may not want to have a clutter of notifications. The way notifications are sent decides how useful the notification system will be. The system should be able to send the notifications to the user in a way that the user will not be overwhelmed with the notifications. The content to be notified may depend on the user as they may have different information needs. Hence, it is wise to allow the user to customize the content.

The change notification process of CDN systems available at present consists of a wide range of notification methods. A few of the popular methods are web interfaces, browser plugin notifications, emails, SMS alerts and popup alerts~\cite{Distill2013}.

\subsection{Web Interfaces}

WebCQ~\cite{Liu2000} is a CDN system that is among the earliest to become popular back at the beginning of the 2000s. The notification system of WebCQ runs on the server-side, and hence, it uses periodic notifications so that it can run efficiently with a large user base and webpage entries. The interval to send notifications is defined by the user when registering a webpage to track. WebCQ allows users to view changes on a web interface, where they can query to find reports detailing the changes or reports with a summary of changes. 

\begin{figure}[!ht]
    \centering
    \begin{minipage}{1\textwidth}
    \begin{minipage}{0.43\textwidth}
    \centering
    \includegraphics[width=\textwidth]{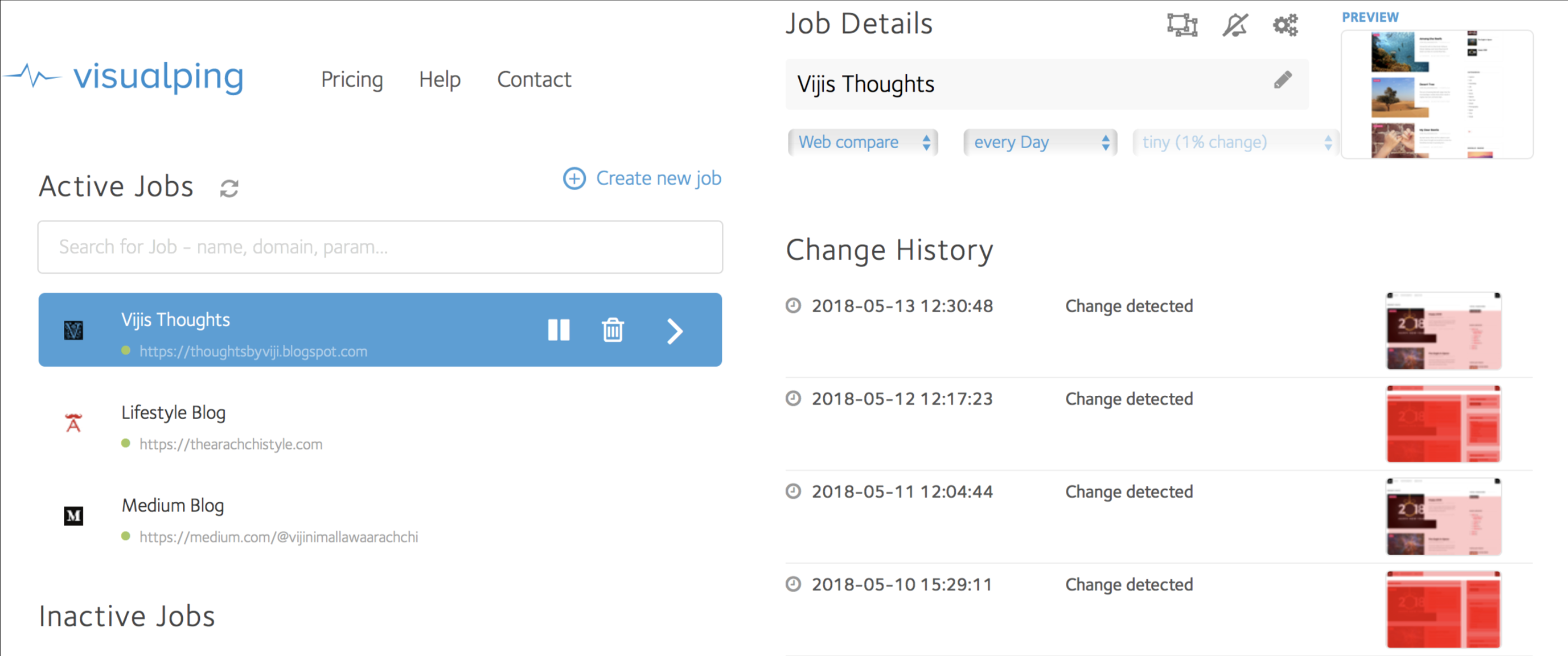}\\
    (a)
    \end{minipage}
    \begin{minipage}{0.57\textwidth}
    \centering
    \includegraphics[width=\textwidth]{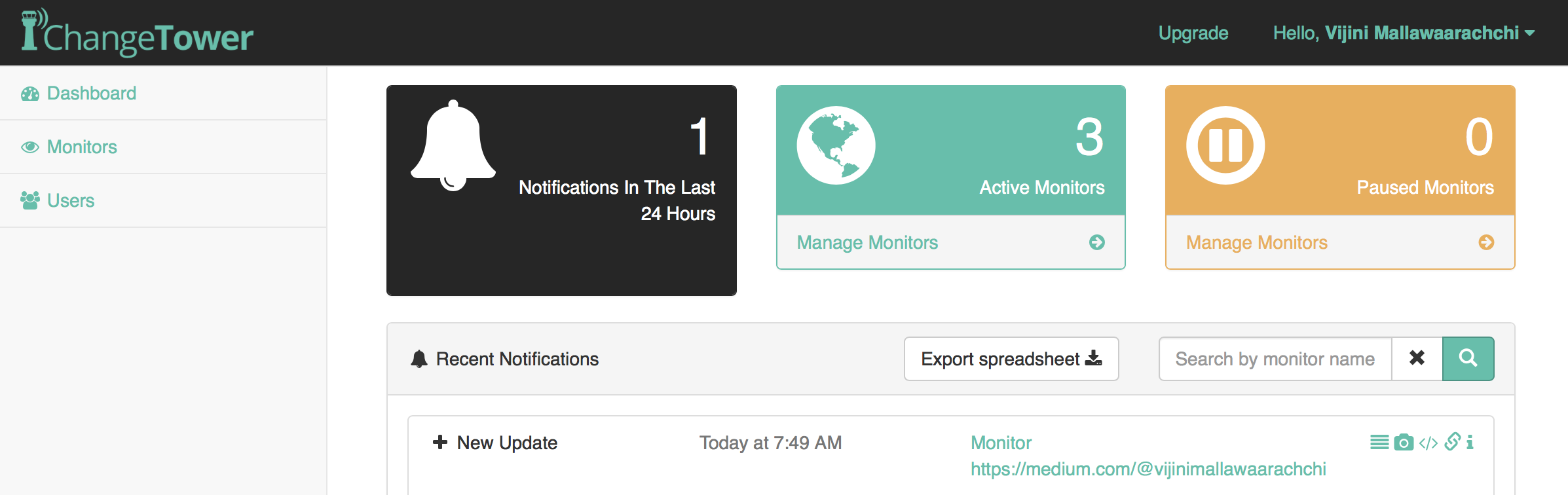}\\
    (b)
    \end{minipage}
    \end{minipage}
    \caption{Dashboard of (a) Visualping~\cite{VisualPing2017} and (b) ChangeTower~\cite{ChangeTower2017}}
    \label{fig11:VisualPingChangeTowe}
\end{figure}

Throughout the past two decades, CDN systems have evolved, by utilizing modern front-end development frameworks, to provide more appealing and user-friendly interfaces with improved user experience. These interfaces have been able to convey useful information to the user in a more efficient and readable manner. Recently introduced change detection systems such as Visualping~\cite{VisualPing2017} and ChangeTower~\cite{ChangeTower2017} provide more advanced features for notifying changes within the web interfaces (as shown in Figure~\ref{fig11:VisualPingChangeTowe}). The majority of the systems provide a dashboard for the user with summaries of recent changes that have occurred to the webpages they are tracking.

\subsection{Browser Plugins}

\begin{figure}[!ht]
    \centering
    \includegraphics[width=0.85\textwidth]{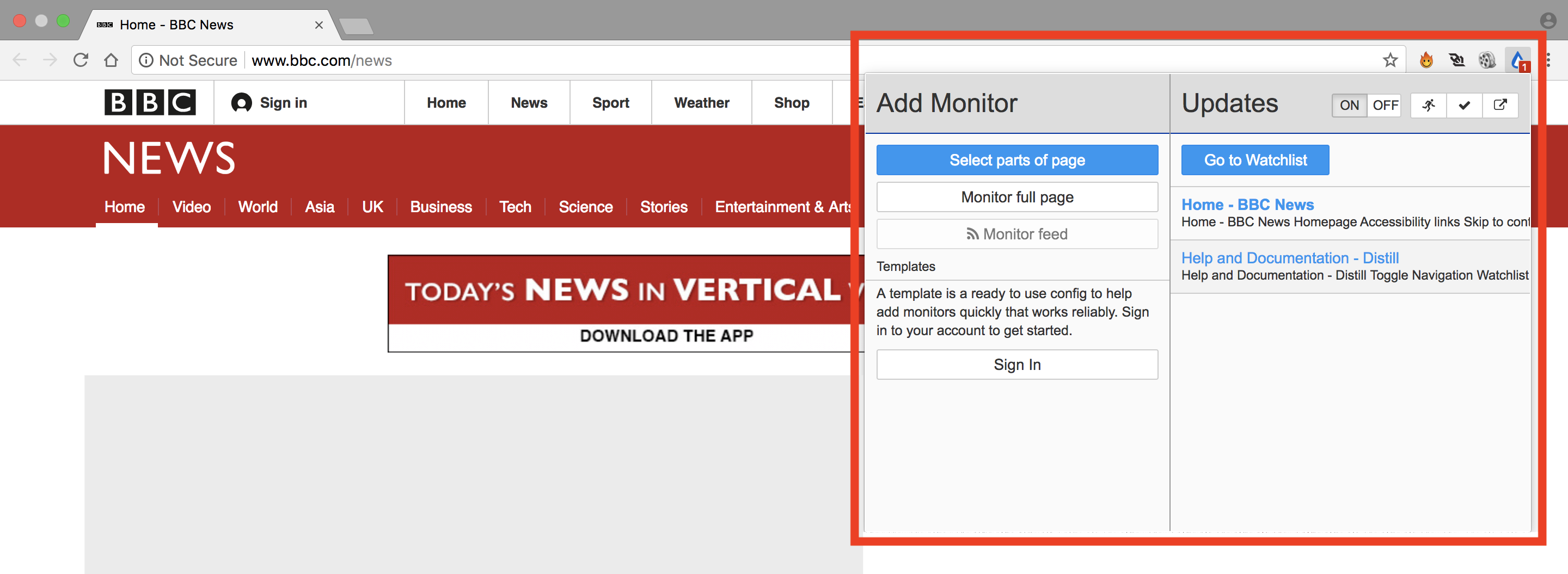}
    \caption{Browser plugin of Distill Web Monitor~\cite{Distill2013}}
    \label{fig12:distill}
\end{figure}

Distill Web Monitor~\cite{Distill2013} is a CDN system, which is available as a browser plugin. Figure~\ref{fig12:distill} illustrates the browser plugin of Distill Web Monitor. It allows users to view changes highlighted on the webpage itself. Furthermore, this system provides various notification options for users including, emails, browser notifications, SMS, sounds and popup alerts.

\subsection{Emails}

Email notifications have become popular in all the online services as a means to convey new updates and news to the users. Similarly, most of the CDN systems provide the facility to get notified via emails, once changes occur in monitored webpages. Emails generally contain links which when clicked by the user, will be redirected to a page with further information about the relevant change.

\begin{table}[!ht]
    \centering
    \caption{A summary of the change notification techniques.}
    \begin{tabular}{|l|l|l|l|l|}
    \hline
    \textbf{\makecell{Change Notification \\Method}} & \textbf{\makecell{Amount of \\information \\delivered}} & \textbf{\makecell{Payload \\size}} & \textbf{\makecell{Level of \\accessibility}} & \textbf{\makecell{Cost to \\implement}}\\
    \hline
    \makecell[l]{Web interfaces} & \makecell[l]{High} & \makecell[l]{High} & \makecell[l]{Medium} & \makecell[l]{Low} \\
    \hline
    \makecell[l]{Browser plugins} & \makecell[l]{High} & \makecell[l]{Low} & \makecell[l]{Medium} & \makecell[l]{Low} \\
    \hline
    \makecell[l]{Emails} & \makecell[l]{Medium} & \makecell[l]{Medium} & \makecell[l]{Medium} & \makecell[l]{Medium} \\
    \hline
    \makecell[l]{SMS Alerts} & \makecell[l]{Low} & \makecell[l]{Low} & \makecell[l]{High} & \makecell[l]{High} \\
    \hline
    \end{tabular}
    \label{tab8:change-notify}
\end{table}

\subsection{SMS Alerts}

Certain CDN systems provide the facility to get notifications via Short Message Service (SMS). The system requests the user to enter a mobile phone number to which the user wants to have the notifications delivered. Google Alerts~\cite{GoogleAlerts2003} and Distill Web Monitor~\cite{Distill2013} are two popular services that provide this facility for its users. SMS alerts are very useful for users who wish to get notifications about webpage updates while traveling and when they do not have access to the Internet to log into the online system. However, the SMS alert feature may require the user to upgrade to their paid versions. Table~\ref{tab8:change-notify} compares the different features of the various change notification techniques.

\section{Change Visualization}
\label{section6}
Visualization of changes is an important aspect of CDN systems. Proper visualization techniques will allow the users to easily identify the changes of the webpages being tracked. Publicly available CDN systems visualize changes occurring on webpages in different ways~\cite{VisualPing2017,Wachete2014}. Most of the changes are depicted in the original interface itself that is loaded by the CDN system.

\begin{figure}[!ht]
    \centering
    \begin{minipage}{1\textwidth}
    \begin{minipage}{0.46\textwidth}
    \centering
    \includegraphics[width=\textwidth]{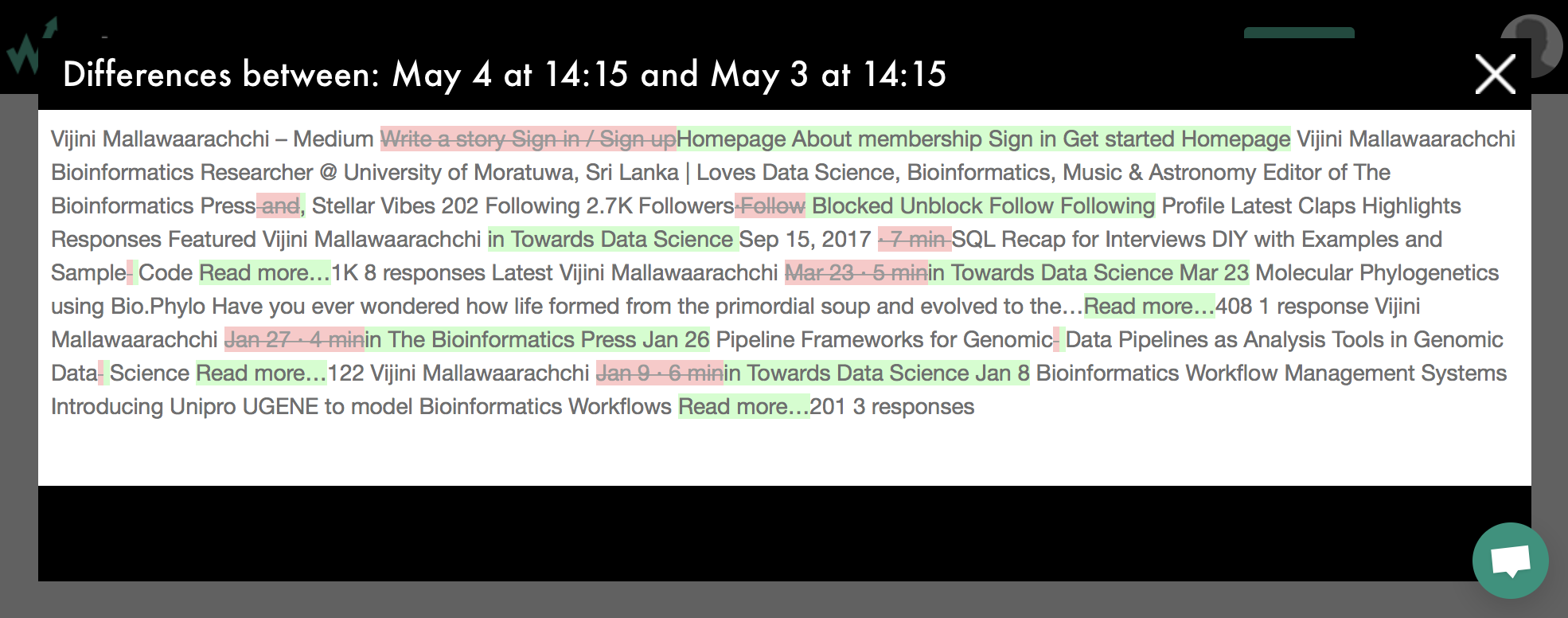}\\
    (a)
    \end{minipage}
    \begin{minipage}{0.54\textwidth}
    \centering
    \includegraphics[width=\textwidth]{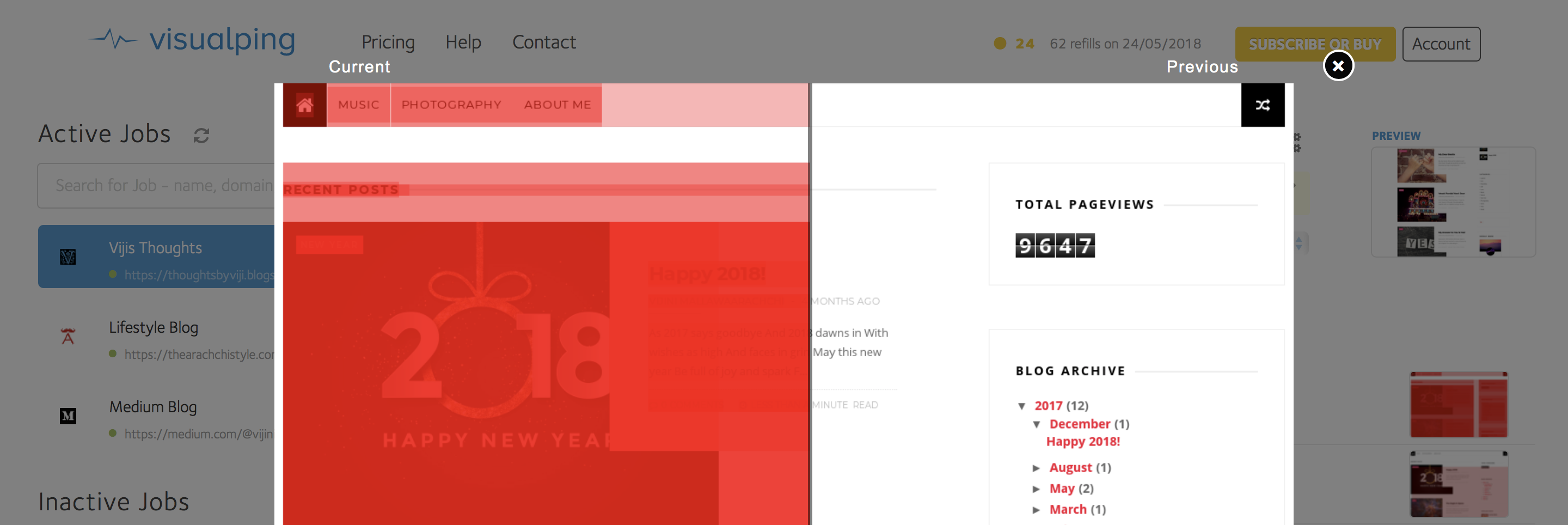}\\
    (b)
    \end{minipage}
    \end{minipage}
    \caption{(a) Text change visualization in Wachete~\cite{Wachete2014} and (b) visual comparison of changes in Visualping~\cite{VisualPing2017}}
    \label{fig11:wachete_visualping}
\end{figure}

\subsection{HTML Differencing}

One popular means of visualizing textual content is by graphically annotating the differences in the HTML content. This was first introduced as the HtmlDiff tool~\cite{Douglis1996}, which marks up the HTML text to indicate how it has changed from a previous version. This tool was introduced in the AT\&T Internet Difference Engine (AIDE)~\cite{Douglis1998} which was developed to detect and visualize changes to webpages. Most of the currently available CDN tools use the HtmlDiff technique or its variants that highlight the changes in different colors. The most commonly used color convention is that deleted text is highlighted in red color with strike-through formatting, whereas newly added text is highlighted in green color. The text which has not changed is not highlighted. This method of change visualization is more straightforward as the changes have already been highlighted and are shown to the user. Figure~\ref{fig11:wachete_visualping} (a) shows the visualization used in Wachete~\cite{Wachete2014}.

\subsection{Visual Comparison}

Another interactive method of visualizing changes is by showing the current version and previous version of a webpage side by side on an interface allowing a user to observe the changes for himself. Visualping~\cite{VisualPing2017} provides this facility for its users, as shown in Figure~\ref{fig11:wachete_visualping} (b). The two versions are shown side by side, and the cursor can be moved horizontally to slide the visible window to view and compare a particular area on the webpage in the two versions. This method is more interactive as the user is involved in identifying the changes which cannot be seen at once. However, certain users may find this method too tedious and time-consuming as the user himself has to find the changes by making comparisons between the two versions of a webpage.


\subsection{Change Logs}

\begin{figure}[!ht]
    \centering
    \includegraphics[width=0.9\textwidth]{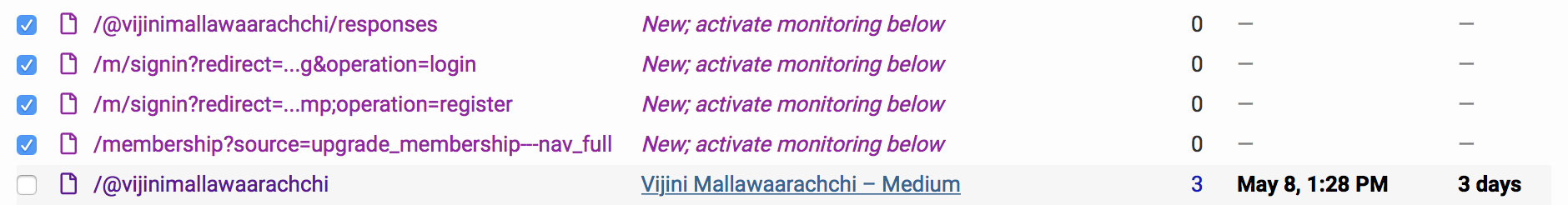}
    \caption{Sample change log of Versionista~\cite{Versionista2007}}
    \label{fig15:versionista}
\end{figure}

Versionista~\cite{Versionista2007} is a CDN tool, where the user can view the modifications in the form of a log. Figure~\ref{fig15:versionista} illustrates a sample change log from Versionista. However, some users may find this format hard to read and understand as the changes are not visible immediately.

Table~\ref{tab9:change-visualize} denotes a feature comparison of the various change visualization methods. It can be seen that different techniques have different levels of information representation, understandability, ease of identifying, and cost.

\begin{table}[!ht]
    \centering
    \caption{A summary of the change visualizations techniques.}
    \begin{tabular}{|l|l|l|l|l|}
    \hline
    \textbf{\makecell{Change Visualization \\Technique}} & \textbf{\makecell{Amount of \\information \\shown}} & \textbf{\makecell{Ease of \\understanding}} & \textbf{\makecell{Ease of \\identifying \\changes}} & \textbf{\makecell{Cost}}\\
    \hline
    \makecell[l]{HTML differencing} & \makecell[l]{Medium} & \makecell[l]{High} & \makecell[l]{High} & \makecell[l]{Medium}\\
    \hline
    \makecell[l]{Visual comparison} & \makecell[l]{Medium} & \makecell[l]{Medium} & \makecell[l]{Medium} & \makecell[l]{High}\\
    \hline
    \makecell[l]{Change logs} & \makecell[l]{High} & \makecell[l]{Low} & \makecell[l]{Low} & \makecell[l]{Medium}\\
    \hline
    \end{tabular}
    \label{tab9:change-visualize}
\end{table}

\section{Publicly Available CDN Systems}
\label{section7}
Different CDN systems are available at present, and each of them has its own supported set of features. Table~\ref{tab10:public-cdn} denotes twelve popular CDN systems, and compares their features such as pages monitored, detection architecture and notification methods.

\begin{table}[!ht]
    \centering
    \caption{Comparison of features of publicly available CDN Systems.}
    \resizebox{\textwidth}{!}{%
    \begin{tabular}{|l|c|c|c|c|c|c|c|c|}
    \hline
    \textbf{\makecell{CDN System}} & \textbf{\makecell{Monitor \\a single \\page}} & \textbf{\makecell{Monitor \\multiple \\pages}} & \textbf{\makecell{Server-\\side \\detection}} & \textbf{\makecell{Client-\\side \\detection}} & \textbf{\makecell{Fixed \\interval \\checks}} & \textbf{\makecell{Email}} & \textbf{\makecell{Browser \\plugin}} & \textbf{\makecell{SMS \\alerts}}\\
    \hline
    Google Alerts~\cite{GoogleAlerts2003} & X & X & X & - & X & X & - & X$^c$ \\
    \hline
    Distill~\cite{Distill2013} & X & X & - & X & X & X & X & X$^a$ \\
    \hline
    Visualping~\cite{VisualPing2017} & X & X & X & X & X & X & X & - \\
    \hline
    FollowThatPage~\cite{FollowThatPage2008} & X & X & X & X & X & X & X & - \\
    \hline
    Trackly~\cite{Trackly2016} & X & X & X & - & X & X & - & - \\
    \hline
    Versionista~\cite{Versionista2007} & X & X$^b$ & X & - & X & X & - & - \\
    \hline
    ChangeDetect~\cite{ChangeDetect2002} & X & X & X & - & X & X & - & - \\
    \hline
    Wachete~\cite{Wachete2014} & X & X$^b$ & X & X & X & X & X & - \\
    \hline
    ChangeTower~\cite{ChangeTower2017} & X & X$^b$ & X & - & X & X & - & - \\
    \hline
    OnWebChange~\cite{OnWebChange2009} & X & X$^a$ & X & - & X & X & - & - \\
    \hline
    ChangeMon~\cite{ChangeMon2015} & X & X & X & - & X & X & - & - \\
    \hline
    Pagescreen~\cite{Pagescreen2018} & X & X$^b$ & X & - & X & X & - & - \\
    \hline
    \end{tabular}}
    \label{tab10:public-cdn}
    \makecell[l]{
    \footnotesize{$^a$ This feature is available in the paid version only.} \\
    \footnotesize{$^b$ Only a number of limited webpages can be tracked with the free version} \\
    \footnotesize{$^c$ SMS alerts provided by third party applications.}}
\end{table}

Most of the systems support change detection of a single page and multiple pages for a single user freely. However, systems such as Versionista~\cite{Versionista2007}, Wachete~\cite{Wachete2014} and PageScreen~\cite{Pagescreen2018} offer a limited number of webpages which can be checked under the trial version. If a user wants to track more webpages than the given limit, then the user has to upgrade to the paid version where unlimited tracking is provided.

The majority of the systems use server-side change detection approaches, whereas a few systems use client-side change detection approaches. According to studies, most of the commercial systems available at present use server-side change detection due to the easy implementation in a central location, where users can call the functionality as a service via a web interface. There are a few tools such as Visualping~\cite{VisualPing2017} and Follow that Page~\cite{FollowThatPage2008}, where they use both server-side and client-side change detection approaches. However, tools such as Distill~\cite{Distill2013} use client-side change detection via browser plugins.

It is evident that all the systems support fixed interval checks. This can cause issues because the user has no knowledge of how often a particular webpage will get changed, and the user may fail to observe important updates by selecting arbitrary checking intervals. Hence, dynamic scheduling mechanisms should be addressed to enhance the efficiency of CDN systems.

Most systems support a wide range of fixed interval checks such as twice a day, daily, weekly and monthly. More frequent checks such as hourly checks, 3 hourly checks and 6 hourly checks are provided in the paid versions of many CDN systems. However, the browser plugin of Distill supports checks varying from every 5 seconds up to 29 days. Such high checking frequencies are possible as the Distill system runs on the client, where the client may have ample resources. However, server-based systems may not support such high checking frequencies as the server can get overloaded with the growing user base and the number of pages to be tracked.

When considering the notification methods, it can be seen that all the tools provide email alerts. Emails have become popular due to its simplicity, easy implementation and extensive use among the clients. A few systems provide browser plugin alerts and SMS alerts. However, with the development of mobile devices, certain services such as Watchete and Visualping have provided mobile applications that can be installed on smartphones. This has allowed the user to get updates and manage monitors via his/her smartphone. Some systems such as ChangeDetect~\cite{ChangeDetect2002} and PageScreen provide free trials for users to try out their features for a limited time. After the trial period has passed the users must upgrade to continue to get the services. Trackly~\cite{Trackly2016} provides a free plan where it allows a user to track three webpages. Trackly also provides 30-day free trials for all its plans and consists of the same features as the paid version. ChangeMon~\cite{ChangeMon2015} provides a 7-day free trial where a user is not required to sign up for an account and can create up to 1000 monitors.

\section{Discussion}
\label{section8}
In the modern world, webpage change detection has become very complicated due to many reasons such as (1) evolution of technologies used in webpage creation, (2) addition of rich and dynamic content to webpages and (3) privacy concerns and regulations. In this section, we will discuss some of the trends, concerns and insights as to what modern researchers/developers should pay attention to when building solutions related to webpage CDN.

\subsection{Ethical Aspects and Security Concerns of Web Crawling}

Change detection of webpages primarily relies on web crawling. Ethical aspects and security concerns of web crawling are important considerations when building web crawler-based applications although they are mostly neglected. Research and guidelines in this area are limited, and work done by Thelwall and Stuart~\cite{Thelwall2006} is one of the modern works regarding this topic. After conducting an extensive analysis regarding applications of web crawlers, they have come up with four main types of issues that web crawlers may cause to the society or individuals. They are namely (1) denial of service – due to repetitive requests to web servers, failures may occur causing disturbances to normal users of the website; (2) cost – increasing number of requests may incur additional costs to the website owner depending on the cost of the web hosting plan which is used; (3) privacy – even though data in webpages are public, gathering of information in large scale might lead to privacy violations; (4) copyright – crawlers create copies of webpages/datasets from websites without prior consent, which may directly involve copyright violations. The authors have further explained how Robots Exclusion Protocol (robots.txt)~\cite{Koster1993} can be used to overcome the above issues from the perspectives of website owners and web crawler owners. In doing so, they have emphasized the limitations of the set of guidelines, and elaborated on alternative techniques to overcome the limitations.

Even though many commercial search engines and CDN systems have adopted the Robots Exclusion Protocol to various extents, the degree to which each web crawler abides by the guidelines set by the protocol differs. A study carried out by Sun et al.~\cite{Sun2010} has come up with a mechanism to quantify the ethicality of web crawlers using a vector space model. While acknowledging the fact that the unethicality of crawling may differ from webpage to webpage, they have used a common set of ethical guidelines set by Eichmann~\cite{Eichmann1995} and Thelwall et al.~\cite{Thelwall2006} in creating this mechanism. More research on this avenue would be interesting as a lot of governments, and policy regulating agencies have shown an increasing interest regarding ethical aspects and data privacy of online content during the last 5 years. To come up with widely adopted guidelines and regulations regarding web crawling, having such ethicality quantification measurements of web crawlers would be crucial. Further, regardless of the guidelines set by the Robots Exclusion Protocol that has been adopted by over 30\% of webpages by 2007~\cite{Sun2007}, many web crawlers do not abide by the regulations. Because of this, many websites with crucial and private data deploy anti-crawling techniques such as (1) IP address ban; (2) captcha; (3) obfuscated JavaScript code; (4) frequent structural changes; (5) limit the frequency of requests and downloadable data allowances and (6) mapping important data like images, which is considered one of the most effective ways. 

\subsection{Change Detection in Dynamic Webpages and Webpages with Rich Content}

Modern webpages developed using technologies such as HTML5 with rich content such as images and videos, are constantly changing their elements to provide rich and interactive functionality to users~\cite{Ryou2018}. Cascading Style Sheets (CSS) is used to manage their layouts, and JavaScript is used to manage their user interactions. The content and layouts of such webpages have become more dynamic to adapt to different user actions and devices. Moreover, dynamic webpages can have temporary information such as help pop-ups, stored in a set of stacked layers apart from the basic two-dimensional layout, hence giving the name three-dimensional webpages~\cite{ThreeDWeb2017}. These layers can undergo many changes. Hence, detecting changes in such dynamic webpages has become challenging.

Currently, a limited amount of research work can be found for change detection in dynamic webpages within the available literature. However, web tripwires~\cite{Reis2008}, CRAWLJAX~\cite{Mesbah2012}, ReDeCheck~\cite{Walsh2017} and detection of visibility faults caused by dynamic layout changes~\cite{Moyeen2016,Ryou2018} can be considered as significant work for crawling and detecting changes in dynamic webpages. These make use of the client-side code to determine state changes occurred within the browser’s dynamically built DOM tree when user events are triggered. The DOM structure of webpages created using JavaScript frameworks such as angular, ember and react can change as events are triggered, and relevant DOM elements are rendered. Crawler scripts can determine such changes by accessing the rendered content via calling the innerHTML property of the elements, as the HTML content is not readily visible~\cite{Reis2008}. Since accessing the client-side code is a critical aspect in detecting changes of dynamic webpages, it is worth to explore more efficient methods to perform this task.

When considering from an industry perspective, not many commercially available CDN systems support the monitoring of dynamic webpages. A handful of CDN systems such as Wachete~\cite{Wachete2014}, Visualping~\cite{VisualPing2017} and Versionista~\cite{Versionista2007} allow users to monitor dynamic and JavaScript rendered pages. Moreover, the algorithms used are kept as trade secrets by these companies, and are not available publicly. However, the monitoring process of highly dynamic JavaScript webpages can timeout due to large amounts of data that have to be processed. Hence, there is an opportunity for researchers to contribute for the development of efficient algorithms to determine changes occurring in highly dynamic webpages.

Modern webpages have rich content such as images, audio and video to enhance the user experience. Webpages may be changed by changing these rich content. Most of the currently available CDN systems can detect changes within the HTML content. In the case of images, a change can be detected if the contents of the \emph{<img>} tag change. Sometimes the actual image may be changed but the image has the same file name as before, and hence, such changes are not detected. However, to the best of our knowledge, currently available systems do not digitally analyze images, and detect whether an image is actually changed or not. However, several image change detection algorithms such as image differencing, image ratioing and change vector analysis can be found in the research domain of image processing. These algorithms are used for applications such as aerial analysis of images of forest areas and urban environments~\cite{Minu2015}. Similar ideas can be utilized in CDN systems for webpages to detect changes occurring in images. However, if such sophisticated methods are implemented within CDN systems, they will require more computational resources to operate efficiently.

\subsection{Resource Synchronization}

The resource synchronization problem has been identified as an issue with frequently changing web sources~\cite{Umbrich2010}. It is defined as the "need for one or more destination servers to remain synchronized with (some of the) resources made available by a source server". From a CDN perspective, we can state this problem as clients wanting to stay up to date with the new content of frequently changing webpages. Previous work includes the use of Open Archives Initiative Protocol for Metadata Harvesting (OAI-PMH)~\cite{OAIPMH2002} to harvest digital resources~\cite{VandeSompel2004}, where new and updated resources are collected incrementally, but it was not adopted broadly. 

ResourceSync is an emerging framework~\cite{VandeSompel2012,Klein2013} that describes means for both client-pull and server-push of change notifications. It consists of several modules that allow intended parties to stay in sync with changing resources. The framework allows source servers to publish up-to-date lists of resources and changes to the resources. Destination servers can access this published data and synchronize the resources. This framework can be adapted for large repositories. The framework represents changed resources or differences between the previous and new representations as a bitstream, allowing to obtain changes to different media types. The use of the ResourceSync framework in CDN systems can make the synchronizing process of clients more efficient, and deliver change notifications in a timely manner.

\subsection{Linked Data Notifications}

With the emergence of User-Generated Content (UGC) proprietorship and anti-scraping tools, commercial crawlers are not allowed to crawl content, and bots are blocked~\cite{Muscovitch2012}. Hence, the details about webpages are unknown before crawling. Despite these challenges, if the changes of webpages can be made available to web crawlers in a standard manner, web crawlers can easily access these data to detect changes that have occurred. An emerging idea that can be used by CDN systems is the Linked Data Notifications (LDN) protocol.

The LDN protocol describes how servers (termed as receivers) can make applications (termed as senders) push messages to them, and how other applications (termed as consumers) can retrieve those messages~\cite{Capadisli2017}. This protocol allows us to share and reuse notifications across different applications while paying less attention to how they were created or what their contents are. Notifications are implemented in such a manner that they can run on different technology stacks, and continuously work together while supporting decentralized communication in the web. The idea was first brought forward by Capadisli~\cite{Capadisli2017}, and has been considered as a W3C recommendation. LDN identifies a notification as "an individual entity with its own Uniform Resource Identifier (URI)", and hence, they can be retrieved and reused. The protocol stores notifications in a manner so that they are compatible with the Linked Data Platform (LDP) standard. 

\begin{figure}[!ht]
    \centering
    \includegraphics[width=0.6\textwidth]{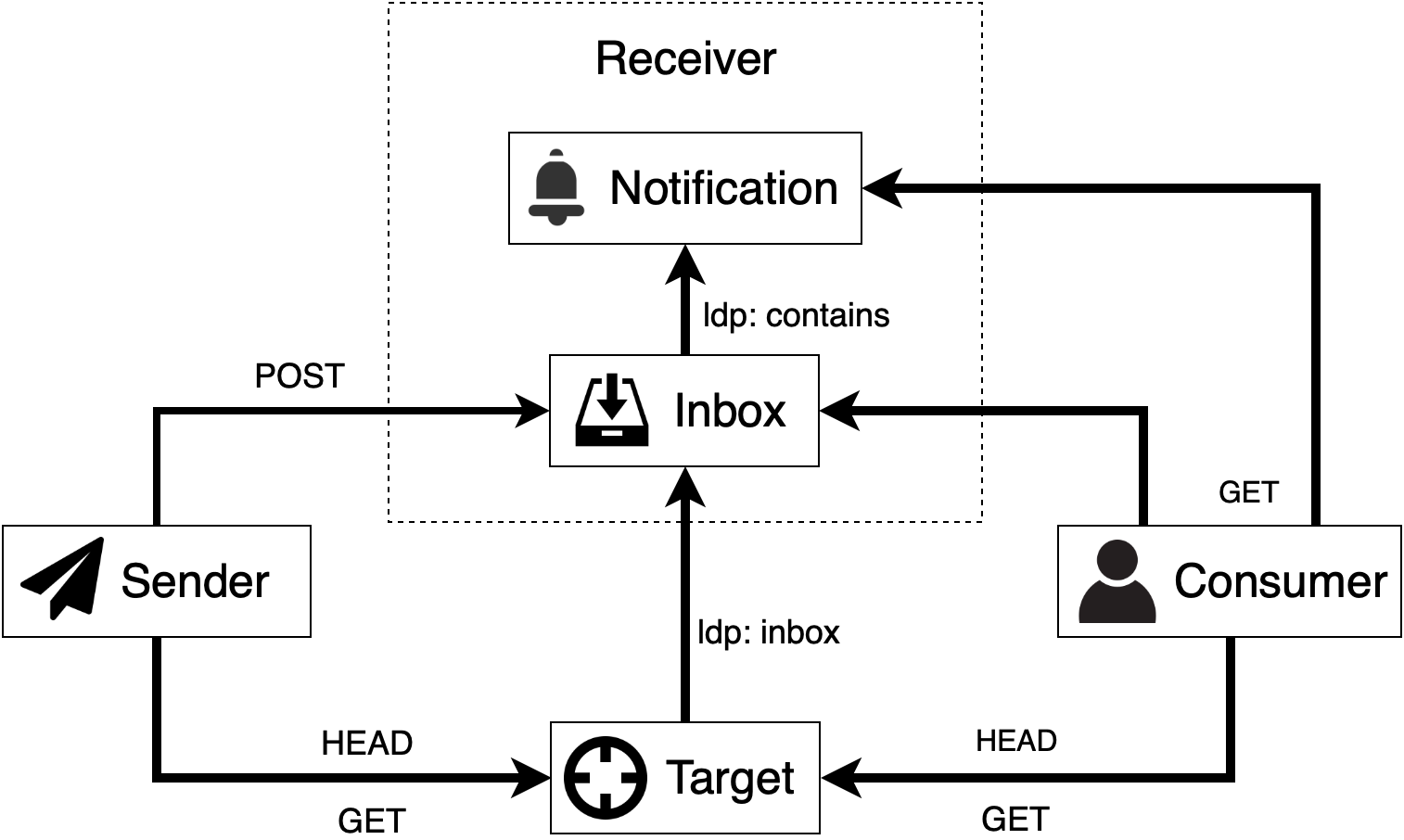}
    \caption{Overview of Linked Data Notifications~\cite{Capadisli2017}.}
    \label{fig16:LDN}
\end{figure}

The overview of LDN is illustrated in Figure~\ref{fig16:LDN}. A sender wants to deliver a notification to the server which is intended to a receiver. The sender selects a target source, finds the location of the target’s inbox, and sends the notification to that particular inbox. Then the receiver allows consumers to access the notification in its inbox. Similarly, when considering the perspective of CDN, CDN systems can make use of the LDN protocol to implement notifications sent to subscribed users. Since the notifications are reused, the system becomes more efficient, and improves the system productivity. LDN will become the next trend in CDN systems.

\section{Conclusion}
\label{section9}
The history of CDN systems dates to the 1990s when they were introduced to automate the detection process of webpage changes and notify interested users. Since then, various CDN systems and techniques have been introduced to improve the efficiency of the CDN process. This paper presented a survey on CDN systems for webpages and various related techniques. We have reviewed and compared different techniques in the literature involving various aspects of CDN systems. Among them, techniques used in web crawler scheduling, change detection and change frequency were identified as significant research areas, where extensive research has been carried out. The most common change detection algorithms are based on difference calculation between tree structures of documents. The process of identifying the frequency of changes occurring on webpages plays a significant role in optimizing crawler schedules to retrieve updated content.

We have also compared different change notification and change visualization techniques that are being used by currently available CDN systems. Most of such systems show notifications on a web interface, and use email notifications as their main method of notifying the user, whereas some systems provide the facility to get notifications via SMS alerts or browser plugin notifications. Additionally, a majority of applications use HTML differencing techniques for change visualization between two variants of a webpage, whereas some systems provide a visual separation among versions, which allows the user to identify the changes by observing the two versions. Moreover, we have compared different features of twelve popular CDN systems that are publicly available at present. According to the comparison results, it is evident that most of the systems support checks at fixed intervals, but not checks at random intervals. These systems can be improved by introducing intelligent crawling schedules to optimize the crawling process by crawling webpages at their estimated change frequency. 

Finally, we have discussed new trends such as LDN, and issues such as determining changes in dynamic webpages, privacy concerns and ethical aspects in CDNs. Throughout this survey, we have identified four important directions of research. The first research direction focuses on improving the architecture of CDN systems, where computing resources and temporal resources can be utilized efficiently while overcoming the limitations of traditional server-based and client-based architectures. The second research direction focuses on improving change detection algorithms to track webpage changes quickly with high accuracy. The third research direction focuses on identifying the change frequency of webpages, and designing optimized crawler schedules so that computing resources can be used efficiently by deploying crawlers when required. The final research direction is improving and developing methods and algorithms to detect changes in dynamic and JavaScript rendered webpages, that can efficiently handle large amounts of data.

\bibliographystyle{ACM-Reference-Format}
\bibliography{10-references}


\begin{thebibliography}{128}


\ifx \showCODEN    \undefined \def \showCODEN     #1{\unskip}     \fi
\ifx \showDOI      \undefined \def \showDOI       #1{#1}\fi
\ifx \showISBNx    \undefined \def \showISBNx     #1{\unskip}     \fi
\ifx \showISBNxiii \undefined \def \showISBNxiii  #1{\unskip}     \fi
\ifx \showISSN     \undefined \def \showISSN      #1{\unskip}     \fi
\ifx \showLCCN     \undefined \def \showLCCN      #1{\unskip}     \fi
\ifx \shownote     \undefined \def \shownote      #1{#1}          \fi
\ifx \showarticletitle \undefined \def \showarticletitle #1{#1}   \fi
\ifx \showURL      \undefined \def \showURL       {\relax}        \fi
\providecommand\bibfield[2]{#2}
\providecommand\bibinfo[2]{#2}
\providecommand\natexlab[1]{#1}
\providecommand\showeprint[2][]{arXiv:#2}

\bibitem[\protect\citeauthoryear{Adar, Teevan, Dumais, and Elsas}{Adar
  et~al\mbox{.}}{2009}]%
        {Adar2009}
\bibfield{author}{\bibinfo{person}{E. Adar}, \bibinfo{person}{J. Teevan},
  \bibinfo{person}{S.~T. Dumais}, {and} \bibinfo{person}{J.~L. Elsas}.}
  \bibinfo{year}{2009}\natexlab{}.
\newblock \showarticletitle{The Web Changes Everything: Understanding the
  Dynamics of Web Content}. In \bibinfo{booktitle}{\emph{Proceedings of the
  Second ACM International Conference on Web Search and Data Mining}}
  \emph{(\bibinfo{series}{WSDM '09})}. \bibinfo{publisher}{ACM},
  \bibinfo{address}{New York, NY, USA}, \bibinfo{pages}{282--291}.
\newblock
\showISBNx{978-1-60558-390-7}
\urldef\tempurl%
\url{https://doi.org/10.1145/1498759.1498837}
\showDOI{\tempurl}


\bibitem[\protect\citeauthoryear{Ahmadi-Abkenari and Selamat}{Ahmadi-Abkenari
  and Selamat}{2012}]%
        {AhmadiAbkenari2012}
\bibfield{author}{\bibinfo{person}{F. Ahmadi-Abkenari} {and}
  \bibinfo{person}{A. Selamat}.} \bibinfo{year}{2012}\natexlab{}.
\newblock \showarticletitle{An architecture for a focused trend parallel Web
  crawler with the application of clickstream analysis}.
\newblock \bibinfo{journal}{\emph{Information Sciences}} \bibinfo{volume}{184},
  \bibinfo{number}{1} (\bibinfo{year}{2012}), \bibinfo{pages}{266--281}.
\newblock
\showISSN{0020-0255}
\urldef\tempurl%
\url{https://doi.org/10.1016/j.ins.2011.08.022}
\showDOI{\tempurl}


\bibitem[\protect\citeauthoryear{{Anjum} and {Anjum}}{{Anjum} and
  {Anjum}}{2012}]%
        {Anjum2012}
\bibfield{author}{\bibinfo{person}{A. {Anjum}} {and} \bibinfo{person}{A.
  {Anjum}}.} \bibinfo{year}{2012}\natexlab{}.
\newblock \showarticletitle{Aiding web crawlers; projecting web page last
  modification}. In \bibinfo{booktitle}{\emph{2012 15th International
  Multitopic Conference (INMIC)}}. \bibinfo{pages}{245--252}.
\newblock
\urldef\tempurl%
\url{https://doi.org/10.1109/INMIC.2012.6511443}
\showDOI{\tempurl}


\bibitem[\protect\citeauthoryear{Baeza-Yates, Castillo, and
  Saint-Jean}{Baeza-Yates et~al\mbox{.}}{2004}]%
        {BaezaYates2004}
\bibfield{author}{\bibinfo{person}{R. Baeza-Yates}, \bibinfo{person}{C.
  Castillo}, {and} \bibinfo{person}{F. Saint-Jean}.}
  \bibinfo{year}{2004}\natexlab{}.
\newblock \bibinfo{booktitle}{\emph{Web Dynamics, Structure, and Page
  Quality}}.
\newblock \bibinfo{publisher}{Springer Berlin Heidelberg},
  \bibinfo{address}{Berlin, Heidelberg}, \bibinfo{pages}{93--109}.
\newblock
\showISBNx{978-3-662-10874-1}
\urldef\tempurl%
\url{https://doi.org/10.1007/978-3-662-10874-1_5}
\showDOI{\tempurl}


\bibitem[\protect\citeauthoryear{Benjamin, von Bochmann, Dincturk, Jourdan, and
  Onut}{Benjamin et~al\mbox{.}}{2011}]%
        {Benjamin2011}
\bibfield{author}{\bibinfo{person}{K. Benjamin}, \bibinfo{person}{G. von
  Bochmann}, \bibinfo{person}{M.~E. Dincturk}, \bibinfo{person}{G.-V. Jourdan},
  {and} \bibinfo{person}{I.~V. Onut}.} \bibinfo{year}{2011}\natexlab{}.
\newblock \showarticletitle{A Strategy for Efficient Crawling of Rich Internet
  Applications}. In \bibinfo{booktitle}{\emph{Web Engineering}},
  \bibfield{editor}{\bibinfo{person}{S.~Auer}, \bibinfo{person}{O.~D{\'i}az},
  {and} \bibinfo{person}{G.~A. Papadopoulos}} (Eds.).
  \bibinfo{publisher}{Springer Berlin Heidelberg}, \bibinfo{address}{Berlin,
  Heidelberg}, \bibinfo{pages}{74--89}.
\newblock
\showISBNx{978-3-642-22233-7}


\bibitem[\protect\citeauthoryear{Bhatt, Vyas, and Pandya}{Bhatt
  et~al\mbox{.}}{2015}]%
        {Bhatt2015}
\bibfield{author}{\bibinfo{person}{D. Bhatt}, \bibinfo{person}{D.~A. Vyas},
  {and} \bibinfo{person}{S. Pandya}.} \bibinfo{year}{2015}\natexlab{}.
\newblock \showarticletitle{Focused Web Crawler}.
\newblock \bibinfo{journal}{\emph{Advances in Computer Science and Information
  Technology (ACSIT)}} \bibinfo{volume}{2}, \bibinfo{number}{11}
  (\bibinfo{date}{April} \bibinfo{year}{2015}), \bibinfo{pages}{1--6}.
\newblock


\bibitem[\protect\citeauthoryear{Bloom}{Bloom}{1970}]%
        {Bloom1970}
\bibfield{author}{\bibinfo{person}{B.~H. Bloom}.}
  \bibinfo{year}{1970}\natexlab{}.
\newblock \showarticletitle{{Space/Time Trade-offs in Hash Coding with
  Allowable Errors}}.
\newblock \bibinfo{journal}{\emph{Commun. ACM}} \bibinfo{volume}{13},
  \bibinfo{number}{7} (\bibinfo{date}{July} \bibinfo{year}{1970}),
  \bibinfo{pages}{422--426}.
\newblock
\showISSN{0001-0782}
\urldef\tempurl%
\url{https://doi.org/10.1145/362686.362692}
\showDOI{\tempurl}


\bibitem[\protect\citeauthoryear{Boldi, Codenotti, Santini, and Vigna}{Boldi
  et~al\mbox{.}}{2004}]%
        {Boldi2004}
\bibfield{author}{\bibinfo{person}{P. Boldi}, \bibinfo{person}{B. Codenotti},
  \bibinfo{person}{M. Santini}, {and} \bibinfo{person}{S. Vigna}.}
  \bibinfo{year}{2004}\natexlab{}.
\newblock \showarticletitle{UbiCrawler: a scalable fully distributed Web
  crawler}.
\newblock \bibinfo{journal}{\emph{Software: Practice and Experience}}
  \bibinfo{volume}{34}, \bibinfo{number}{8} (\bibinfo{year}{2004}),
  \bibinfo{pages}{711--726}.
\newblock
\urldef\tempurl%
\url{https://doi.org/10.1002/spe.587}
\showDOI{\tempurl}
\showeprint{https://onlinelibrary.wiley.com/doi/pdf/10.1002/spe.587}


\bibitem[\protect\citeauthoryear{Boldi, Marino, Santini, and Vigna}{Boldi
  et~al\mbox{.}}{2018}]%
        {Boldi2018}
\bibfield{author}{\bibinfo{person}{P. Boldi}, \bibinfo{person}{A. Marino},
  \bibinfo{person}{M. Santini}, {and} \bibinfo{person}{S. Vigna}.}
  \bibinfo{year}{2018}\natexlab{}.
\newblock \showarticletitle{BUbiNG: Massive Crawling for the Masses}.
\newblock \bibinfo{journal}{\emph{ACM Trans. Web}} \bibinfo{volume}{12},
  \bibinfo{number}{2}, Article \bibinfo{articleno}{12} (\bibinfo{date}{June}
  \bibinfo{year}{2018}), \bibinfo{numpages}{26}~pages.
\newblock
\showISSN{1559-1131}
\urldef\tempurl%
\url{https://doi.org/10.1145/3160017}
\showDOI{\tempurl}


\bibitem[\protect\citeauthoryear{Borgolte, Kruegel, and Vigna}{Borgolte
  et~al\mbox{.}}{2014}]%
        {Borgolte2014}
\bibfield{author}{\bibinfo{person}{K. Borgolte}, \bibinfo{person}{C. Kruegel},
  {and} \bibinfo{person}{G. Vigna}.} \bibinfo{year}{2014}\natexlab{}.
\newblock \showarticletitle{Relevant Change Detection: A Framework for the
  Precise Extraction of Modified and Novel Web-based Content As a Filtering
  Technique for Analysis Engines}. In \bibinfo{booktitle}{\emph{Proceedings of
  the 23rd International Conference on World Wide Web}}
  \emph{(\bibinfo{series}{WWW '14 Companion})}. \bibinfo{publisher}{ACM},
  \bibinfo{address}{New York, NY, USA}, \bibinfo{pages}{595--598}.
\newblock
\showISBNx{978-1-4503-2745-9}
\urldef\tempurl%
\url{https://doi.org/10.1145/2567948.2578039}
\showDOI{\tempurl}


\bibitem[\protect\citeauthoryear{Brandman, Cho, Garcia-Molina, and
  Shivakumar}{Brandman et~al\mbox{.}}{2000}]%
        {Brandman2000}
\bibfield{author}{\bibinfo{person}{O. Brandman}, \bibinfo{person}{J. Cho},
  \bibinfo{person}{H. Garcia-Molina}, {and} \bibinfo{person}{N. Shivakumar}.}
  \bibinfo{year}{2000}\natexlab{}.
\newblock \showarticletitle{{Crawler-Friendly Web Servers}}.
\newblock \bibinfo{journal}{\emph{SIGMETRICS Perform. Eval. Rev.}}
  \bibinfo{volume}{28}, \bibinfo{number}{2} (\bibinfo{date}{Sept.}
  \bibinfo{year}{2000}), \bibinfo{pages}{9--14}.
\newblock
\showISSN{0163-5999}
\urldef\tempurl%
\url{https://doi.org/10.1145/362883.362894}
\showDOI{\tempurl}


\bibitem[\protect\citeauthoryear{Brewington and Cybenko}{Brewington and
  Cybenko}{2000}]%
        {Brewington2000}
\bibfield{author}{\bibinfo{person}{B.~E. Brewington} {and} \bibinfo{person}{G.
  Cybenko}.} \bibinfo{year}{2000}\natexlab{}.
\newblock \showarticletitle{How dynamic is the Web?1This research was partially
  supported by AFOSR grant F49620-97-1-0382, DARPA grant F30602-98-2-0107 and
  NSF grant CCR-9813744. Any opinions, findings, and conclusions are those of
  the authors and do not necessarily reflect the views of the above
  agencies.1}.
\newblock \bibinfo{journal}{\emph{Computer Networks}} \bibinfo{volume}{33},
  \bibinfo{number}{1} (\bibinfo{year}{2000}), \bibinfo{pages}{257--276}.
\newblock
\showISSN{1389-1286}
\urldef\tempurl%
\url{https://doi.org/10.1016/S1389-1286(00)00045-1}
\showDOI{\tempurl}


\bibitem[\protect\citeauthoryear{Brin and Page}{Brin and Page}{1998}]%
        {Brin1998}
\bibfield{author}{\bibinfo{person}{S. Brin} {and} \bibinfo{person}{L. Page}.}
  \bibinfo{year}{1998}\natexlab{}.
\newblock \showarticletitle{{The anatomy of a large-scale hypertextual Web
  search engine}}.
\newblock \bibinfo{journal}{\emph{Computer Networks and ISDN Systems}}
  \bibinfo{volume}{30}, \bibinfo{number}{1} (\bibinfo{year}{1998}),
  \bibinfo{pages}{107--117}.
\newblock
\showISSN{0169-7552}
\urldef\tempurl%
\url{https://doi.org/10.1016/S0169-7552(98)00110-X}
\showDOI{\tempurl}
\newblock
\shownote{Proceedings of the Seventh International World Wide Web Conference.}


\bibitem[\protect\citeauthoryear{Buytaert}{Buytaert}{2000}]%
        {Drupal2000}
\bibfield{author}{\bibinfo{person}{D. Buytaert}.}
  \bibinfo{year}{2000}\natexlab{}.
\newblock \bibinfo{booktitle}{\emph{{Drupal - Open Source CMS | Drupal.org}}}.
\newblock {Drupal community}.
\newblock
\urldef\tempurl%
\url{https://www.drupal.org/}
\showURL{%
Retrieved November 8, 2019 from \tempurl}


\bibitem[\protect\citeauthoryear{{Canavos}}{{Canavos}}{1972}]%
        {Canavos1972}
\bibfield{author}{\bibinfo{person}{G.~C. {Canavos}}.}
  \bibinfo{year}{1972}\natexlab{}.
\newblock \showarticletitle{A Bayesian Approach to Parameter and Reliability
  Estimation in the Poisson Distribution}.
\newblock \bibinfo{journal}{\emph{IEEE Transactions on Reliability}}
  \bibinfo{volume}{R-21}, \bibinfo{number}{1} (\bibinfo{date}{Feb}
  \bibinfo{year}{1972}), \bibinfo{pages}{52--56}.
\newblock
\urldef\tempurl%
\url{https://doi.org/10.1109/TR.1972.5216172}
\showDOI{\tempurl}


\bibitem[\protect\citeauthoryear{Capadisli, Guy, Lange, Auer, Sambra, and
  Berners-Lee}{Capadisli et~al\mbox{.}}{2017}]%
        {Capadisli2017}
\bibfield{author}{\bibinfo{person}{S. Capadisli}, \bibinfo{person}{A. Guy},
  \bibinfo{person}{C. Lange}, \bibinfo{person}{S. Auer}, \bibinfo{person}{A.
  Sambra}, {and} \bibinfo{person}{T. Berners-Lee}.}
  \bibinfo{year}{2017}\natexlab{}.
\newblock \showarticletitle{Linked Data Notifications: A Resource-Centric
  Communication Protocol}. In \bibinfo{booktitle}{\emph{The Semantic Web}},
  \bibfield{editor}{\bibinfo{person}{E.~Blomqvist},
  \bibinfo{person}{D.~Maynard}, \bibinfo{person}{A.~Gangemi},
  \bibinfo{person}{R.~Hoekstra}, \bibinfo{person}{P.~Hitzler}, {and}
  \bibinfo{person}{O.~Hartig}} (Eds.). \bibinfo{publisher}{Springer
  International Publishing}, \bibinfo{address}{Cham},
  \bibinfo{pages}{537--553}.
\newblock
\showISBNx{978-3-319-58068-5}


\bibitem[\protect\citeauthoryear{{Castillo}, {Marin}, {Rodriguez}, and
  {Baeza-Yates}}{{Castillo} et~al\mbox{.}}{2004}]%
        {Castillo2004}
\bibfield{author}{\bibinfo{person}{C. {Castillo}}, \bibinfo{person}{M.
  {Marin}}, \bibinfo{person}{A. {Rodriguez}}, {and} \bibinfo{person}{R.
  {Baeza-Yates}}.} \bibinfo{year}{2004}\natexlab{}.
\newblock \showarticletitle{Scheduling algorithms for Web crawling}. In
  \bibinfo{booktitle}{\emph{WebMedia and LA-Web, 2004. Proceedings}}.
  \bibinfo{pages}{10--17}.
\newblock
\urldef\tempurl%
\url{https://doi.org/10.1109/WEBMED.2004.1348139}
\showDOI{\tempurl}


\bibitem[\protect\citeauthoryear{Chakrabarti, van~den Berg, and
  Dom}{Chakrabarti et~al\mbox{.}}{1999}]%
        {Chakrabarti1999}
\bibfield{author}{\bibinfo{person}{S. Chakrabarti}, \bibinfo{person}{M. van~den
  Berg}, {and} \bibinfo{person}{B. Dom}.} \bibinfo{year}{1999}\natexlab{}.
\newblock \showarticletitle{Focused crawling: a new approach to topic-specific
  Web resource discovery}.
\newblock \bibinfo{journal}{\emph{Computer Networks}} \bibinfo{volume}{31},
  \bibinfo{number}{11} (\bibinfo{year}{1999}), \bibinfo{pages}{1623--1640}.
\newblock
\showISSN{1389-1286}
\urldef\tempurl%
\url{https://doi.org/10.1016/S1389-1286(99)00052-3}
\showDOI{\tempurl}


\bibitem[\protect\citeauthoryear{Chandra and Toueg}{Chandra and Toueg}{1996}]%
        {Chandra1996}
\bibfield{author}{\bibinfo{person}{T.~D. Chandra} {and} \bibinfo{person}{S.
  Toueg}.} \bibinfo{year}{1996}\natexlab{}.
\newblock \showarticletitle{Unreliable Failure Detectors for Reliable
  Distributed Systems}.
\newblock \bibinfo{journal}{\emph{J. ACM}} \bibinfo{volume}{43},
  \bibinfo{number}{2} (\bibinfo{date}{March} \bibinfo{year}{1996}),
  \bibinfo{pages}{225--267}.
\newblock
\showISSN{0004-5411}
\urldef\tempurl%
\url{https://doi.org/10.1145/226643.226647}
\showDOI{\tempurl}


\bibitem[\protect\citeauthoryear{{ChangeDetect}}{{ChangeDetect}}{2002}]%
        {ChangeDetect2002}
\bibfield{author}{\bibinfo{person}{{ChangeDetect}}.}
  \bibinfo{year}{2002}\natexlab{}.
\newblock \bibinfo{booktitle}{\emph{ChangeDetect web page monitoring}}.
\newblock ChangeDetect.
\newblock
\urldef\tempurl%
\url{https://www.changedetect.com/}
\showURL{%
Retrieved March 24, 2017 from \tempurl}


\bibitem[\protect\citeauthoryear{{ChangeMon}}{{ChangeMon}}{2015}]%
        {ChangeMon2015}
\bibfield{author}{\bibinfo{person}{{ChangeMon}}.}
  \bibinfo{year}{2015}\natexlab{}.
\newblock \bibinfo{booktitle}{\emph{{ChangeMon - Monitor Any Web Page For
  Changes }}}.
\newblock {ChangeMon.com}.
\newblock
\urldef\tempurl%
\url{https://changemon.com/}
\showURL{%
Retrieved Nonember 8, 2019 from \tempurl}


\bibitem[\protect\citeauthoryear{{ChangeTower}}{{ChangeTower}}{2017}]%
        {ChangeTower2017}
\bibfield{author}{\bibinfo{person}{{ChangeTower}}.}
  \bibinfo{year}{2017}\natexlab{}.
\newblock \bibinfo{booktitle}{\emph{{ChangeTower - Monitor Website Changes, Get
  Alerts, Archive Website History}}}.
\newblock {ChangeTower LLC}.
\newblock
\urldef\tempurl%
\url{https://changetower.com/}
\showURL{%
Retrieved April 3, 2018 from \tempurl}


\bibitem[\protect\citeauthoryear{Cho and Garcia-Molina}{Cho and
  Garcia-Molina}{2000a}]%
        {Cho2000b}
\bibfield{author}{\bibinfo{person}{J. Cho} {and} \bibinfo{person}{H.
  Garcia-Molina}.} \bibinfo{year}{2000}\natexlab{a}.
\newblock \showarticletitle{The Evolution of the Web and Implications for an
  Incremental Crawler}. In \bibinfo{booktitle}{\emph{Proceedings of the 26th
  International Conference on Very Large Data Bases}}
  \emph{(\bibinfo{series}{VLDB '00})}. \bibinfo{publisher}{Morgan Kaufmann
  Publishers Inc.}, \bibinfo{address}{San Francisco, CA, USA},
  \bibinfo{pages}{200--209}.
\newblock
\showISBNx{1-55860-715-3}
\urldef\tempurl%
\url{http://dl.acm.org/citation.cfm?id=645926.671679}
\showURL{%
\tempurl}


\bibitem[\protect\citeauthoryear{Cho and Garcia-Molina}{Cho and
  Garcia-Molina}{2000b}]%
        {Cho2000a}
\bibfield{author}{\bibinfo{person}{J. Cho} {and} \bibinfo{person}{H.
  Garcia-Molina}.} \bibinfo{year}{2000}\natexlab{b}.
\newblock \showarticletitle{Synchronizing a Database to Improve Freshness}. In
  \bibinfo{booktitle}{\emph{Proceedings of the 2000 ACM SIGMOD International
  Conference on Management of Data}} \emph{(\bibinfo{series}{SIGMOD '00})}.
  \bibinfo{publisher}{ACM}, \bibinfo{address}{New York, NY, USA},
  \bibinfo{pages}{117--128}.
\newblock
\showISBNx{1-58113-217-4}
\urldef\tempurl%
\url{https://doi.org/10.1145/342009.335391}
\showDOI{\tempurl}


\bibitem[\protect\citeauthoryear{Cho and Garcia-Molina}{Cho and
  Garcia-Molina}{2003a}]%
        {Cho2003a}
\bibfield{author}{\bibinfo{person}{J. Cho} {and} \bibinfo{person}{H.
  Garcia-Molina}.} \bibinfo{year}{2003}\natexlab{a}.
\newblock \showarticletitle{Effective Page Refresh Policies for Web Crawlers}.
\newblock \bibinfo{journal}{\emph{ACM Trans. Database Syst.}}
  \bibinfo{volume}{28}, \bibinfo{number}{4} (\bibinfo{date}{Dec.}
  \bibinfo{year}{2003}), \bibinfo{pages}{390--426}.
\newblock
\showISSN{0362-5915}
\urldef\tempurl%
\url{https://doi.org/10.1145/958942.958945}
\showDOI{\tempurl}


\bibitem[\protect\citeauthoryear{Cho and Garcia-Molina}{Cho and
  Garcia-Molina}{2003b}]%
        {Cho2003b}
\bibfield{author}{\bibinfo{person}{J. Cho} {and} \bibinfo{person}{H.
  Garcia-Molina}.} \bibinfo{year}{2003}\natexlab{b}.
\newblock \showarticletitle{Estimating Frequency of Change}.
\newblock \bibinfo{journal}{\emph{ACM Trans. Internet Technol.}}
  \bibinfo{volume}{3}, \bibinfo{number}{3} (\bibinfo{date}{Aug.}
  \bibinfo{year}{2003}), \bibinfo{pages}{256--290}.
\newblock
\showISSN{1533-5399}
\urldef\tempurl%
\url{https://doi.org/10.1145/857166.857170}
\showDOI{\tempurl}


\bibitem[\protect\citeauthoryear{{Cobena}, {Abiteboul}, and {Marian}}{{Cobena}
  et~al\mbox{.}}{2002}]%
        {Cobena2002}
\bibfield{author}{\bibinfo{person}{G. {Cobena}}, \bibinfo{person}{S.
  {Abiteboul}}, {and} \bibinfo{person}{A. {Marian}}.}
  \bibinfo{year}{2002}\natexlab{}.
\newblock \showarticletitle{Detecting changes in XML documents}. In
  \bibinfo{booktitle}{\emph{Proceedings 18th International Conference on Data
  Engineering}}. \bibinfo{pages}{41--52}.
\newblock
\urldef\tempurl%
\url{https://doi.org/10.1109/ICDE.2002.994696}
\showDOI{\tempurl}


\bibitem[\protect\citeauthoryear{Coffman~Jr., Liu, and Weber}{Coffman~Jr.
  et~al\mbox{.}}{1998}]%
        {Coffman1998}
\bibfield{author}{\bibinfo{person}{E.~G. Coffman~Jr.}, \bibinfo{person}{Z.
  Liu}, {and} \bibinfo{person}{R.~R. Weber}.} \bibinfo{year}{1998}\natexlab{}.
\newblock \showarticletitle{Optimal robot scheduling for Web search engines}.
\newblock \bibinfo{journal}{\emph{Journal of Scheduling}} \bibinfo{volume}{1},
  \bibinfo{number}{1} (\bibinfo{year}{1998}), \bibinfo{pages}{15--29}.
\newblock
\urldef\tempurl%
\url{https://doi.org/10.1002/(SICI)1099-1425(199806)1:1<15::AID-JOS3>3.0.CO;2-K}
\showDOI{\tempurl}


\bibitem[\protect\citeauthoryear{{D3S}}{{D3S}}{2015}]%
        {ShashTool2015}
\bibfield{author}{\bibinfo{person}{{D3S}}.} \bibinfo{year}{2015}\natexlab{}.
\newblock \bibinfo{booktitle}{\emph{{Shash Tool}}}.
\newblock D3S.
\newblock
\urldef\tempurl%
\url{http://d3s.mff.cuni.cz/ ~holub/sw/shash/}
\showURL{%
Retrieved March 4, 2017 from \tempurl}


\bibitem[\protect\citeauthoryear{{Dalai}, {Dash}, {Dave}, {Francisco-Revilla},
  {Furuta}, {Karadkar}, and {Shipma}}{{Dalai} et~al\mbox{.}}{2004}]%
        {Dalai2004}
\bibfield{author}{\bibinfo{person}{Z. {Dalai}}, \bibinfo{person}{S. {Dash}},
  \bibinfo{person}{P. {Dave}}, \bibinfo{person}{L. {Francisco-Revilla}},
  \bibinfo{person}{R. {Furuta}}, \bibinfo{person}{U. {Karadkar}}, {and}
  \bibinfo{person}{F. {Shipma}}.} \bibinfo{year}{2004}\natexlab{}.
\newblock \showarticletitle{Managing distributed collections: evaluating Web
  page changes, movement, and replacement}. In
  \bibinfo{booktitle}{\emph{Proceedings of the 2004 Joint ACM/IEEE Conference
  on Digital Libraries, 2004.}} \bibinfo{pages}{160--168}.
\newblock
\urldef\tempurl%
\url{https://doi.org/10.1109/JCDL.2004.240012}
\showDOI{\tempurl}


\bibitem[\protect\citeauthoryear{De~Bra and Calvi}{De~Bra and Calvi}{1997}]%
        {DeBra1997}
\bibfield{author}{\bibinfo{person}{P. De~Bra} {and} \bibinfo{person}{L.
  Calvi}.} \bibinfo{year}{1997}\natexlab{}.
\newblock \showarticletitle{{Creating Adaptive Hyperdocuments for and on the
  Web}}. In \bibinfo{booktitle}{\emph{Proceedings of the WebNet '97
  Conference}} \emph{(\bibinfo{series}{WebNet '97})}.
  \bibinfo{pages}{149--165}.
\newblock


\bibitem[\protect\citeauthoryear{Diligenti, Coetzee, Lawrence, Giles, and
  Gori}{Diligenti et~al\mbox{.}}{2000}]%
        {Diligenti2000}
\bibfield{author}{\bibinfo{person}{M. Diligenti}, \bibinfo{person}{F. Coetzee},
  \bibinfo{person}{S. Lawrence}, \bibinfo{person}{C. Giles}, {and}
  \bibinfo{person}{M. Gori}.} \bibinfo{year}{2000}\natexlab{}.
\newblock \showarticletitle{{Focused Crawling Using Context Graphs}}. In
  \bibinfo{booktitle}{\emph{Proceedings of the 26th VLDB Conference}}
  \emph{(\bibinfo{series}{VLDB 2000})}. \bibinfo{pages}{527--534}.
\newblock


\bibitem[\protect\citeauthoryear{Dincturk, Choudhary, von Bochmann, Jourdan,
  and Onut}{Dincturk et~al\mbox{.}}{2012}]%
        {Dincturk2012}
\bibfield{author}{\bibinfo{person}{M.~E. Dincturk}, \bibinfo{person}{S.
  Choudhary}, \bibinfo{person}{G. von Bochmann}, \bibinfo{person}{G.-V.
  Jourdan}, {and} \bibinfo{person}{I.~V. Onut}.}
  \bibinfo{year}{2012}\natexlab{}.
\newblock \showarticletitle{A Statistical Approach for Efficient Crawling of
  Rich Internet Applications}. In \bibinfo{booktitle}{\emph{Web Engineering}},
  \bibfield{editor}{\bibinfo{person}{M.~Brambilla},
  \bibinfo{person}{T.~Tokuda}, {and} \bibinfo{person}{R.~Tolksdorf}} (Eds.).
  \bibinfo{publisher}{Springer Berlin Heidelberg}, \bibinfo{address}{Berlin,
  Heidelberg}, \bibinfo{pages}{362--369}.
\newblock
\showISBNx{978-3-642-31753-8}


\bibitem[\protect\citeauthoryear{{Distill.io}}{{Distill.io}}{2013}]%
        {Distill2013}
\bibfield{author}{\bibinfo{person}{{Distill.io}}.}
  \bibinfo{year}{2013}\natexlab{}.
\newblock \bibinfo{booktitle}{\emph{Monitor websites for changes, get SMS
  alerts and email alerts | Distill.io}}.
\newblock {Neemb LLC}.
\newblock
\urldef\tempurl%
\url{https://distill.io}
\showURL{%
Retrieved April 3, 2018 from \tempurl}


\bibitem[\protect\citeauthoryear{Douglis and Ball}{Douglis and Ball}{1996}]%
        {Douglis1996}
\bibfield{author}{\bibinfo{person}{F. Douglis} {and} \bibinfo{person}{T.
  Ball}.} \bibinfo{year}{1996}\natexlab{}.
\newblock \showarticletitle{{Tracking and Viewing Changes on the Web}}. In
  \bibinfo{booktitle}{\emph{Proceedings of the USENIX Annual Technical
  Conference}} \emph{(\bibinfo{series}{USENIX 1996})}.
  \bibinfo{publisher}{USENIX Association}, \bibinfo{address}{Berkley, CA, USA}.
\newblock


\bibitem[\protect\citeauthoryear{Douglis, Ball, Chen, and Koutsofios}{Douglis
  et~al\mbox{.}}{1998}]%
        {Douglis1998}
\bibfield{author}{\bibinfo{person}{F. Douglis}, \bibinfo{person}{T. Ball},
  \bibinfo{person}{Y.‐F. Chen}, {and} \bibinfo{person}{E. Koutsofios}.}
  \bibinfo{year}{1998}\natexlab{}.
\newblock \showarticletitle{The AT{\&}T Internet Difference Engine: Tracking
  and viewing changes on the web}.
\newblock \bibinfo{journal}{\emph{World Wide Web}} \bibinfo{volume}{1},
  \bibinfo{number}{1} (\bibinfo{date}{01 Mar} \bibinfo{year}{1998}),
  \bibinfo{pages}{27--44}.
\newblock
\showISSN{1573-1413}
\urldef\tempurl%
\url{https://doi.org/10.1023/A:1019243126596}
\showDOI{\tempurl}


\bibitem[\protect\citeauthoryear{Eichmann}{Eichmann}{1995}]%
        {Eichmann1995}
\bibfield{author}{\bibinfo{person}{D. Eichmann}.}
  \bibinfo{year}{1995}\natexlab{}.
\newblock \showarticletitle{Ethical Web agents}.
\newblock \bibinfo{journal}{\emph{Computer Networks and ISDN Systems}}
  \bibinfo{volume}{28}, \bibinfo{number}{1} (\bibinfo{year}{1995}),
  \bibinfo{pages}{127--136}.
\newblock
\showISSN{0169-7552}
\urldef\tempurl%
\url{https://doi.org/10.1016/0169-7552(95)00107-3}
\showDOI{\tempurl}
\newblock
\shownote{Selected Papers from the Second World-Wide Web Conference.}


\bibitem[\protect\citeauthoryear{Elsas and Dumais}{Elsas and Dumais}{2010}]%
        {Elsas2010}
\bibfield{author}{\bibinfo{person}{J.~L. Elsas} {and} \bibinfo{person}{S.~T.
  Dumais}.} \bibinfo{year}{2010}\natexlab{}.
\newblock \showarticletitle{Leveraging Temporal Dynamics of Document Content in
  Relevance Ranking}. In \bibinfo{booktitle}{\emph{Proceedings of the Third ACM
  International Conference on Web Search and Data Mining}}
  \emph{(\bibinfo{series}{WSDM '10})}. \bibinfo{publisher}{ACM},
  \bibinfo{address}{New York, NY, USA}, \bibinfo{pages}{1--10}.
\newblock
\showISBNx{978-1-60558-889-6}
\urldef\tempurl%
\url{https://doi.org/10.1145/1718487.1718489}
\showDOI{\tempurl}


\bibitem[\protect\citeauthoryear{Exposto, Macedo, Pina, Alves, and
  Rufino}{Exposto et~al\mbox{.}}{2008}]%
        {Exposto2008}
\bibfield{author}{\bibinfo{person}{J. Exposto}, \bibinfo{person}{J. Macedo},
  \bibinfo{person}{A. Pina}, \bibinfo{person}{A. Alves}, {and}
  \bibinfo{person}{J. Rufino}.} \bibinfo{year}{2008}\natexlab{}.
\newblock \showarticletitle{Efficient Partitioning Strategies for Distributed
  Web Crawling}. In \bibinfo{booktitle}{\emph{Information Networking. Towards
  Ubiquitous Networking and Services}},
  \bibfield{editor}{\bibinfo{person}{T.~Vaz{\~a}o}, \bibinfo{person}{M.~M.
  Freire}, {and} \bibinfo{person}{I.~Chong}} (Eds.).
  \bibinfo{publisher}{Springer Berlin Heidelberg}, \bibinfo{address}{Berlin,
  Heidelberg}, \bibinfo{pages}{544--553}.
\newblock
\showISBNx{978-3-540-89524-4}


\bibitem[\protect\citeauthoryear{Fetterly, Manasse, Najork, and
  Wiener}{Fetterly et~al\mbox{.}}{2003}]%
        {Fetterly2003}
\bibfield{author}{\bibinfo{person}{D. Fetterly}, \bibinfo{person}{M. Manasse},
  \bibinfo{person}{M. Najork}, {and} \bibinfo{person}{J. Wiener}.}
  \bibinfo{year}{2003}\natexlab{}.
\newblock \showarticletitle{A Large-scale Study of the Evolution of Web Pages}.
  In \bibinfo{booktitle}{\emph{Proceedings of the 12th International Conference
  on World Wide Web}} \emph{(\bibinfo{series}{WWW '03})}.
  \bibinfo{publisher}{ACM}, \bibinfo{address}{New York, NY, USA},
  \bibinfo{pages}{669--678}.
\newblock
\showISBNx{1-58113-680-3}
\urldef\tempurl%
\url{https://doi.org/10.1145/775152.775246}
\showDOI{\tempurl}


\bibitem[\protect\citeauthoryear{Fielding}{Fielding}{1994}]%
        {Fielding1994}
\bibfield{author}{\bibinfo{person}{R.~T. Fielding}.}
  \bibinfo{year}{1994}\natexlab{}.
\newblock \showarticletitle{{Maintaining distributed hypertext infostructures:
  Welcome to MOMspider's Web}}.
\newblock \bibinfo{journal}{\emph{Computer Networks and ISDN Systems}}
  \bibinfo{volume}{27}, \bibinfo{number}{2} (\bibinfo{year}{1994}),
  \bibinfo{pages}{193--204}.
\newblock
\showISSN{0169-7552}
\urldef\tempurl%
\url{https://doi.org/10.1016/0169-7552(94)90133-3}
\showDOI{\tempurl}
\newblock
\shownote{Selected Papers of the First World-Wide Web Conference.}


\bibitem[\protect\citeauthoryear{{Filipowski}}{{Filipowski}}{2014}]%
        {Filipowski2014}
\bibfield{author}{\bibinfo{person}{K. {Filipowski}}.}
  \bibinfo{year}{2014}\natexlab{}.
\newblock \showarticletitle{Comparison of Scheduling Algorithms for Domain
  Specific Web Crawler}. In \bibinfo{booktitle}{\emph{2014 European Network
  Intelligence Conference}}. \bibinfo{pages}{69--74}.
\newblock
\urldef\tempurl%
\url{https://doi.org/10.1109/ENIC.2014.14}
\showDOI{\tempurl}


\bibitem[\protect\citeauthoryear{{Follow That Page}}{{Follow That
  Page}}{2008}]%
        {FollowThatPage2008}
\bibfield{author}{\bibinfo{person}{{Follow That Page}}.}
  \bibinfo{year}{2008}\natexlab{}.
\newblock \bibinfo{booktitle}{\emph{Follow That Page - web monitor: we send you
  an email when your favorite page has changed}}.
\newblock Follow That Page.
\newblock
\urldef\tempurl%
\url{https://www.followthatpage.com}
\showURL{%
Retrieved February 8, 2017 from \tempurl}


\bibitem[\protect\citeauthoryear{Francisco-Revilla, Shipman, Furuta, Karadkar,
  and Arora}{Francisco-Revilla et~al\mbox{.}}{2001}]%
        {FranciscoRevilla2001}
\bibfield{author}{\bibinfo{person}{L. Francisco-Revilla}, \bibinfo{person}{F.
  Shipman}, \bibinfo{person}{R. Furuta}, \bibinfo{person}{U. Karadkar}, {and}
  \bibinfo{person}{A. Arora}.} \bibinfo{year}{2001}\natexlab{}.
\newblock \showarticletitle{Managing Change on the Web}. In
  \bibinfo{booktitle}{\emph{Proceedings of the 1st ACM/IEEE-CS Joint Conference
  on Digital Libraries}} \emph{(\bibinfo{series}{JCDL '01})}.
  \bibinfo{publisher}{ACM}, \bibinfo{address}{New York, NY, USA},
  \bibinfo{pages}{67--76}.
\newblock
\showISBNx{1-58113-345-6}
\urldef\tempurl%
\url{https://doi.org/10.1145/379437.379973}
\showDOI{\tempurl}


\bibitem[\protect\citeauthoryear{{Google}}{{Google}}{2003}]%
        {GoogleAlerts2003}
\bibfield{author}{\bibinfo{person}{{Google}}.} \bibinfo{year}{2003}\natexlab{}.
\newblock \bibinfo{booktitle}{\emph{Google Alerts - Monitor the Web for
  interesting new content}}.
\newblock Google LLC.
\newblock
\urldef\tempurl%
\url{https://www.google.com/alerts}
\showURL{%
Retrieved February 8, 2017 from \tempurl}


\bibitem[\protect\citeauthoryear{{Google}}{{Google}}{2013a}]%
        {GoogleBot2013}
\bibfield{author}{\bibinfo{person}{{Google}}.}
  \bibinfo{year}{2013}\natexlab{a}.
\newblock \bibinfo{booktitle}{\emph{Googlebot: Search Console Help}}.
\newblock Google LLC.
\newblock
\urldef\tempurl%
\url{https://support.google.com/webmasters/answer/182072?hl=en}
\showURL{%
Retrieved March 4, 2018 from \tempurl}


\bibitem[\protect\citeauthoryear{{Google}}{{Google}}{2013b}]%
        {OfficialGoogleBlog2013}
\bibfield{author}{\bibinfo{person}{{Google}}.}
  \bibinfo{year}{2013}\natexlab{b}.
\newblock \bibinfo{booktitle}{\emph{Official Google Blog: A second spring of
  cleaning.}}
\newblock Google LLC.
\newblock
\urldef\tempurl%
\url{https://googleblog.blogspot.com/2013/03/a-second-spring-of-cleaning.html}
\showURL{%
Retrieved November 8, 2019 from \tempurl}


\bibitem[\protect\citeauthoryear{{Google}}{{Google}}{2018}]%
        {SubmitURL2018}
\bibfield{author}{\bibinfo{person}{{Google}}.} \bibinfo{year}{2018}\natexlab{}.
\newblock \bibinfo{booktitle}{\emph{Submit URL}}.
\newblock Google LLC.
\newblock
\urldef\tempurl%
\url{https://www.google.com/webmasters/tools/submit-url}
\showURL{%
Retrieved March 2, 2018 from \tempurl}


\bibitem[\protect\citeauthoryear{Grimes and O’Brien}{Grimes and
  O’Brien}{2008}]%
        {Grimes2008}
\bibfield{author}{\bibinfo{person}{C. Grimes} {and} \bibinfo{person}{S.
  O’Brien}.} \bibinfo{year}{2008}\natexlab{}.
\newblock \showarticletitle{{Microscale Evolution of Web Pages}}. In
  \bibinfo{booktitle}{\emph{Proceedings of the 17th international conference on
  World Wide Web}} \emph{(\bibinfo{series}{WWW '08})}. \bibinfo{publisher}{ACM
  Press}, \bibinfo{address}{New York, NY, USA}, \bibinfo{pages}{1149--1150}.
\newblock


\bibitem[\protect\citeauthoryear{Hersovici, Jacovi, Maarek, Pelleg, Shtalhaim,
  and Ur}{Hersovici et~al\mbox{.}}{1998}]%
        {Hersovici1998}
\bibfield{author}{\bibinfo{person}{M. Hersovici}, \bibinfo{person}{M. Jacovi},
  \bibinfo{person}{Y.~S. Maarek}, \bibinfo{person}{D. Pelleg},
  \bibinfo{person}{M. Shtalhaim}, {and} \bibinfo{person}{S. Ur}.}
  \bibinfo{year}{1998}\natexlab{}.
\newblock \showarticletitle{The shark-search algorithm. An application:
  tailored Web site mapping}.
\newblock \bibinfo{journal}{\emph{Computer Networks and ISDN Systems}}
  \bibinfo{volume}{30}, \bibinfo{number}{1} (\bibinfo{year}{1998}),
  \bibinfo{pages}{317--326}.
\newblock
\showISSN{0169-7552}
\urldef\tempurl%
\url{https://doi.org/10.1016/S0169-7552(98)00038-5}
\showDOI{\tempurl}
\newblock
\shownote{Proceedings of the Seventh International World Wide Web Conference.}


\bibitem[\protect\citeauthoryear{Hmedeh, Vouzoukidou, Travers, Christophides,
  du~Mouza, and Scholl}{Hmedeh et~al\mbox{.}}{2011}]%
        {Hmedeh2011}
\bibfield{author}{\bibinfo{person}{Z. Hmedeh}, \bibinfo{person}{N.
  Vouzoukidou}, \bibinfo{person}{N. Travers}, \bibinfo{person}{V.
  Christophides}, \bibinfo{person}{C. du Mouza}, {and} \bibinfo{person}{M.
  Scholl}.} \bibinfo{year}{2011}\natexlab{}.
\newblock \showarticletitle{Characterizing Web Syndication Behavior and
  Content}. In \bibinfo{booktitle}{\emph{Web Information System Engineering --
  WISE 2011}}, \bibfield{editor}{\bibinfo{person}{A.~Bouguettaya},
  \bibinfo{person}{M.~Hauswirth}, {and} \bibinfo{person}{L.~Liu}} (Eds.).
  \bibinfo{publisher}{Springer Berlin Heidelberg}, \bibinfo{address}{Berlin,
  Heidelberg}, \bibinfo{pages}{29--42}.
\newblock
\showISBNx{978-3-642-24434-6}


\bibitem[\protect\citeauthoryear{Jacob, Sanka, Pandrangi, and
  Chakravarthy}{Jacob et~al\mbox{.}}{2004}]%
        {Jacob2004}
\bibfield{author}{\bibinfo{person}{J. Jacob}, \bibinfo{person}{A. Sanka},
  \bibinfo{person}{N. Pandrangi}, {and} \bibinfo{person}{S. Chakravarthy}.}
  \bibinfo{year}{2004}\natexlab{}.
\newblock \bibinfo{booktitle}{\emph{WebVigiL: An Approach to Just-In-Time
  Information Propagation in Large Network-Centric Environments}}.
\newblock \bibinfo{publisher}{Springer Berlin Heidelberg},
  \bibinfo{address}{Berlin, Heidelberg}, \bibinfo{pages}{301--318}.
\newblock
\showISBNx{978-3-662-10874-1}
\urldef\tempurl%
\url{https://doi.org/10.1007/978-3-662-10874-1_13}
\showDOI{\tempurl}


\bibitem[\protect\citeauthoryear{Jain and Khandagale}{Jain and
  Khandagale}{2014}]%
        {Jain2014}
\bibfield{author}{\bibinfo{person}{S. Jain} {and} \bibinfo{person}{H.
  Khandagale}.} \bibinfo{year}{2014}\natexlab{}.
\newblock \showarticletitle{{A Web Page Change Detection System For Selected
  Zone Using Tree Comparison Technique}}.
\newblock \bibinfo{journal}{\emph{International Journal of Computer
  Applications Technology and Research}} \bibinfo{volume}{3},
  \bibinfo{number}{4} (\bibinfo{date}{April} \bibinfo{year}{2014}),
  \bibinfo{pages}{254--262}.
\newblock


\bibitem[\protect\citeauthoryear{Jamali, Sayyadi, Hariri, and
  Abolhassani}{Jamali et~al\mbox{.}}{2006}]%
        {Jamali2006}
\bibfield{author}{\bibinfo{person}{M. Jamali}, \bibinfo{person}{H. Sayyadi},
  \bibinfo{person}{B.~B. Hariri}, {and} \bibinfo{person}{H. Abolhassani}.}
  \bibinfo{year}{2006}\natexlab{}.
\newblock \showarticletitle{A Method for Focused Crawling Using Combination of
  Link Structure and Content Similarity}. In
  \bibinfo{booktitle}{\emph{Proceedings of the 2006 IEEE/WIC/ACM International
  Conference on Web Intelligence}} \emph{(\bibinfo{series}{WI '06})}.
  \bibinfo{publisher}{IEEE Computer Society}, \bibinfo{address}{Washington, DC,
  USA}, \bibinfo{pages}{753--756}.
\newblock
\showISBNx{0-7695-2747-7}
\urldef\tempurl%
\url{https://doi.org/10.1109/WI.2006.19}
\showDOI{\tempurl}


\bibitem[\protect\citeauthoryear{{Jayarathna} and {Poursardar}}{{Jayarathna}
  and {Poursardar}}{2016}]%
        {Jayarathna2016}
\bibfield{author}{\bibinfo{person}{S. {Jayarathna}} {and} \bibinfo{person}{F.
  {Poursardar}}.} \bibinfo{year}{2016}\natexlab{}.
\newblock \showarticletitle{Change detection and classification of digital
  collections}. In \bibinfo{booktitle}{\emph{2016 IEEE International Conference
  on Big Data (Big Data)}}. \bibinfo{pages}{1750--1759}.
\newblock
\urldef\tempurl%
\url{https://doi.org/10.1109/BigData.2016.7840790}
\showDOI{\tempurl}


\bibitem[\protect\citeauthoryear{Johnson}{Johnson}{1977}]%
        {Johnson1977}
\bibfield{author}{\bibinfo{person}{D.~B. Johnson}.}
  \bibinfo{year}{1977}\natexlab{}.
\newblock \showarticletitle{Efficient Algorithms for Shortest Paths in Sparse
  Networks}.
\newblock \bibinfo{journal}{\emph{J. ACM}} \bibinfo{volume}{24},
  \bibinfo{number}{1} (\bibinfo{date}{Jan.} \bibinfo{year}{1977}),
  \bibinfo{pages}{1--13}.
\newblock
\showISSN{0004-5411}
\urldef\tempurl%
\url{https://doi.org/10.1145/321992.321993}
\showDOI{\tempurl}


\bibitem[\protect\citeauthoryear{Johnson and Tanimoto}{Johnson and
  Tanimoto}{1999}]%
        {Johnson1999}
\bibfield{author}{\bibinfo{person}{D.~B. Johnson} {and} \bibinfo{person}{S.~L.
  Tanimoto}.} \bibinfo{year}{1999}\natexlab{}.
\newblock \showarticletitle{Reusing Web Documents in Tutorials With the
  Current-Documents Assumption: Automatic Validation of Updates}. In
  \bibinfo{booktitle}{\emph{Proceedings of EdMedia + Innovate Learning 1999}},
  \bibfield{editor}{\bibinfo{person}{Betty Collis} {and} \bibinfo{person}{Ron
  Oliver}} (Eds.). \bibinfo{publisher}{Association for the Advancement of
  Computing in Education (AACE)}, \bibinfo{address}{Seattle, WA USA},
  \bibinfo{pages}{74--79}.
\newblock
\urldef\tempurl%
\url{https://www.learntechlib.org/p/17401}
\showURL{%
\tempurl}


\bibitem[\protect\citeauthoryear{Kausar, Dhaka, and Singh}{Kausar
  et~al\mbox{.}}{2013}]%
        {Kausar2013}
\bibfield{author}{\bibinfo{person}{M.~A. Kausar}, \bibinfo{person}{V.~S.
  Dhaka}, {and} \bibinfo{person}{S.~K. Singh}.}
  \bibinfo{year}{2013}\natexlab{}.
\newblock \showarticletitle{Web Crawler: A Review}.
\newblock \bibinfo{journal}{\emph{International Journal of Computer
  Applications}} \bibinfo{volume}{63}, \bibinfo{number}{2}
  (\bibinfo{year}{2013}), \bibinfo{pages}{31--36}.
\newblock


\bibitem[\protect\citeauthoryear{{Kc}, {Hagenbuchner}, and {Tsoi}}{{Kc}
  et~al\mbox{.}}{2008}]%
        {Kc2008}
\bibfield{author}{\bibinfo{person}{M. {Kc}}, \bibinfo{person}{M.
  {Hagenbuchner}}, {and} \bibinfo{person}{A.~C. {Tsoi}}.}
  \bibinfo{year}{2008}\natexlab{}.
\newblock \showarticletitle{A Scalable Lightweight Distributed Crawler for
  Crawling with Limited Resources}. In \bibinfo{booktitle}{\emph{2008
  IEEE/WIC/ACM International Conference on Web Intelligence and Intelligent
  Agent Technology}}, Vol.~\bibinfo{volume}{3}. \bibinfo{pages}{663--666}.
\newblock
\urldef\tempurl%
\url{https://doi.org/10.1109/WIIAT.2008.234}
\showDOI{\tempurl}


\bibitem[\protect\citeauthoryear{Keogh}{Keogh}{2005}]%
        {Keogh2005}
\bibfield{author}{\bibinfo{person}{J.~E. Keogh}.}
  \bibinfo{year}{2005}\natexlab{}.
\newblock \bibinfo{booktitle}{\emph{ASP.NET 2.0 Demystified}
  (\bibinfo{edition}{1st} ed.)}.
\newblock \bibinfo{publisher}{McGraw Hill Professional}, \bibinfo{address}{New
  York, USA}.
\newblock
\showISBNx{0072261412}


\bibitem[\protect\citeauthoryear{Klein, Sanderson, Van~de Sompel, Warner,
  Haslhofer, Lagoze, and Nelson}{Klein et~al\mbox{.}}{2013}]%
        {Klein2013}
\bibfield{author}{\bibinfo{person}{M. Klein}, \bibinfo{person}{R. Sanderson},
  \bibinfo{person}{H. Van~de Sompel}, \bibinfo{person}{S. Warner},
  \bibinfo{person}{B. Haslhofer}, \bibinfo{person}{C. Lagoze}, {and}
  \bibinfo{person}{M.~L. Nelson}.} \bibinfo{year}{2013}\natexlab{}.
\newblock \showarticletitle{A Technical Framework for Resource
  Synchronization}.
\newblock \bibinfo{journal}{\emph{D-Lib Magazine}} \bibinfo{volume}{19},
  \bibinfo{number}{1/2} (\bibinfo{date}{Jan} \bibinfo{year}{2013}).
\newblock


\bibitem[\protect\citeauthoryear{Koster}{Koster}{1993}]%
        {Koster1993}
\bibfield{author}{\bibinfo{person}{M. Koster}.}
  \bibinfo{year}{1993}\natexlab{}.
\newblock \bibinfo{booktitle}{\emph{{Guidelines for Robot Writers}}}.
\newblock robotstxt.org.
\newblock
\urldef\tempurl%
\url{http://www.robotstxt.org/guidelines.html}
\showURL{%
Retrieved May 28, 2019 from \tempurl}


\bibitem[\protect\citeauthoryear{Kumar and Neelima}{Kumar and Neelima}{2011}]%
        {Kumar2011}
\bibfield{author}{\bibinfo{person}{M. Kumar} {and} \bibinfo{person}{P.
  Neelima}.} \bibinfo{year}{2011}\natexlab{}.
\newblock \showarticletitle{Design and Implementation of Scalable, Fully
  Distributed Web Crawler for a Web Search Engine}.
\newblock \bibinfo{journal}{\emph{International Journal of Computer
  Applications}} \bibinfo{volume}{15}, \bibinfo{number}{7}
  (\bibinfo{date}{Feb.} \bibinfo{year}{2011}), \bibinfo{pages}{8--13}.
\newblock


\bibitem[\protect\citeauthoryear{Levene and Poulovassilis}{Levene and
  Poulovassilis}{2013}]%
        {Levene2004}
\bibfield{author}{\bibinfo{person}{M. Levene} {and} \bibinfo{person}{A.
  Poulovassilis}.} \bibinfo{year}{2013}\natexlab{}.
\newblock \bibinfo{booktitle}{\emph{{Web Dynamics: Adapting to Change in
  Content, Size, Topology and Use}} (\bibinfo{edition}{2013} ed.)}.
\newblock \bibinfo{publisher}{Springer Science \& Business Media},
  \bibinfo{address}{Berlin, Germany}.
\newblock
\showISBNx{3662108747}
\urldef\tempurl%
\url{https://doi.org/10.1007/978-3-662-10874-1}
\showDOI{\tempurl}


\bibitem[\protect\citeauthoryear{Li, Furuse, and Yamaguchi}{Li
  et~al\mbox{.}}{2005}]%
        {Li2005}
\bibfield{author}{\bibinfo{person}{J. Li}, \bibinfo{person}{K. Furuse}, {and}
  \bibinfo{person}{K. Yamaguchi}.} \bibinfo{year}{2005}\natexlab{}.
\newblock \showarticletitle{{Focused crawling by exploiting anchor text using
  decision tree}}. In \bibinfo{booktitle}{\emph{Special interest tracks and
  posters of the 14th international conference on World Wide Web}}
  \emph{(\bibinfo{series}{WWW '05})}. \bibinfo{publisher}{ACM Press},
  \bibinfo{address}{New York, NY, USA}, \bibinfo{pages}{1190--1191}.
\newblock


\bibitem[\protect\citeauthoryear{Liu, Pu, and Tang}{Liu et~al\mbox{.}}{2000}]%
        {Liu2000}
\bibfield{author}{\bibinfo{person}{L. Liu}, \bibinfo{person}{C. Pu}, {and}
  \bibinfo{person}{W. Tang}.} \bibinfo{year}{2000}\natexlab{}.
\newblock \showarticletitle{WebCQ-detecting and Delivering Information Changes
  on the Web}. In \bibinfo{booktitle}{\emph{Proceedings of the Ninth
  International Conference on Information and Knowledge Management}}
  \emph{(\bibinfo{series}{CIKM '00})}. \bibinfo{publisher}{ACM},
  \bibinfo{address}{New York, NY, USA}, \bibinfo{pages}{512--519}.
\newblock
\showISBNx{1-58113-320-0}
\urldef\tempurl%
\url{https://doi.org/10.1145/354756.354860}
\showDOI{\tempurl}


\bibitem[\protect\citeauthoryear{Mali and Meshram}{Mali and Meshram}{2011}]%
        {Mali2011}
\bibfield{author}{\bibinfo{person}{S. Mali} {and} \bibinfo{person}{B.~B.
  Meshram}.} \bibinfo{year}{2011}\natexlab{}.
\newblock \showarticletitle{{Focused Web Crawler with Page Change Detection
  Policy}}. In \bibinfo{booktitle}{\emph{Proceedings of the 2nd International
  Conference and workshop on Emerging Trends in Technology}}
  \emph{(\bibinfo{series}{ICWET 2011})}. \bibinfo{pages}{51--57}.
\newblock


\bibitem[\protect\citeauthoryear{McBryan}{McBryan}{1994}]%
        {McBryan1994}
\bibfield{author}{\bibinfo{person}{O. McBryan}.}
  \bibinfo{year}{1994}\natexlab{}.
\newblock \showarticletitle{{GENVL and WWWW: Tools for Taming the Web}}. In
  \bibinfo{booktitle}{\emph{Proceedings of the First International World Wide
  Web Conference}} \emph{(\bibinfo{series}{International World Wide Web
  Conference 1994})}.
\newblock


\bibitem[\protect\citeauthoryear{{Meegahapola}, {Alwis}, {Heshan},
  {Mallawaarachchi}, {Meedeniya}, and {Jayarathna}}{{Meegahapola}
  et~al\mbox{.}}{2017a}]%
        {Meegahapola2017d}
\bibfield{author}{\bibinfo{person}{L. {Meegahapola}}, \bibinfo{person}{R.
  {Alwis}}, \bibinfo{person}{E. {Heshan}}, \bibinfo{person}{V.
  {Mallawaarachchi}}, \bibinfo{person}{D. {Meedeniya}}, {and}
  \bibinfo{person}{S. {Jayarathna}}.} \bibinfo{year}{2017}\natexlab{a}.
\newblock \showarticletitle{Adaptive technique for web page change detection
  using multi-threaded crawlers}. In \bibinfo{booktitle}{\emph{2017 Seventh
  International Conference on Innovative Computing Technology (INTECH)}}.
  \bibinfo{pages}{120--125}.
\newblock
\urldef\tempurl%
\url{https://doi.org/10.1109/INTECH.2017.8102430}
\showDOI{\tempurl}


\bibitem[\protect\citeauthoryear{{Meegahapola}, {Alwis}, {Nimalarathna},
  {Mallawaarachchi}, {Meedeniya}, and {Jayarathna}}{{Meegahapola}
  et~al\mbox{.}}{2017b}]%
        {Meegahapola2017c}
\bibfield{author}{\bibinfo{person}{L. {Meegahapola}}, \bibinfo{person}{R.
  {Alwis}}, \bibinfo{person}{E. {Nimalarathna}}, \bibinfo{person}{V.
  {Mallawaarachchi}}, \bibinfo{person}{D. {Meedeniya}}, {and}
  \bibinfo{person}{S. {Jayarathna}}.} \bibinfo{year}{2017}\natexlab{b}.
\newblock \showarticletitle{Detection of change frequency in web pages to
  optimize server-based scheduling}. In \bibinfo{booktitle}{\emph{2017
  Seventeenth International Conference on Advances in ICT for Emerging Regions
  (ICTer)}}. \bibinfo{publisher}{IEEE}, \bibinfo{pages}{165--172}.
\newblock
\urldef\tempurl%
\url{https://doi.org/10.1109/ICTER.2017.8257791}
\showDOI{\tempurl}


\bibitem[\protect\citeauthoryear{{Meegahapola}, {Alwis}, {Nimalarathna},
  {Mallawaarachchi}, {Meedeniya}, and {Jayarathna}}{{Meegahapola}
  et~al\mbox{.}}{2017c}]%
        {Meegahapola2017a}
\bibfield{author}{\bibinfo{person}{L. {Meegahapola}}, \bibinfo{person}{R.
  {Alwis}}, \bibinfo{person}{E. {Nimalarathna}}, \bibinfo{person}{V.
  {Mallawaarachchi}}, \bibinfo{person}{D. {Meedeniya}}, {and}
  \bibinfo{person}{S. {Jayarathna}}.} \bibinfo{year}{2017}\natexlab{c}.
\newblock \showarticletitle{Optimizing change detection in distributed digital
  collections: An architectural perspective of change detection}. In
  \bibinfo{booktitle}{\emph{2017 18th IEEE/ACIS International Conference on
  Software Engineering, Artificial Intelligence, Networking and
  Parallel/Distributed Computing (SNPD)}}. \bibinfo{publisher}{IEEE},
  \bibinfo{pages}{277--282}.
\newblock
\urldef\tempurl%
\url{https://doi.org/10.1109/SNPD.2017.8022733}
\showDOI{\tempurl}


\bibitem[\protect\citeauthoryear{Meegahapola, Mallawaarachchi, Alwis,
  Nimalarathna, Meedeniya, and Jayarathna}{Meegahapola et~al\mbox{.}}{2018}]%
        {Meegahapola2018}
\bibfield{author}{\bibinfo{person}{L. Meegahapola}, \bibinfo{person}{V.
  Mallawaarachchi}, \bibinfo{person}{R. Alwis}, \bibinfo{person}{E.
  Nimalarathna}, \bibinfo{person}{D. Meedeniya}, {and} \bibinfo{person}{S.
  Jayarathna}.} \bibinfo{year}{2018}\natexlab{}.
\newblock \showarticletitle{Random Forest Classifier Based Scheduler
  Optimization for Search Engine Web Crawlers}. In
  \bibinfo{booktitle}{\emph{Proceedings of the 2018 7th International
  Conference on Software and Computer Applications}}
  \emph{(\bibinfo{series}{ICSCA 2018})}. \bibinfo{publisher}{ACM},
  \bibinfo{address}{New York, NY, USA}, \bibinfo{pages}{285--289}.
\newblock
\showISBNx{978-1-4503-5414-1}
\urldef\tempurl%
\url{https://doi.org/10.1145/3185089.3185103}
\showDOI{\tempurl}


\bibitem[\protect\citeauthoryear{{Meegahapola}, {Alwis}, {Nimalarathna},
  {Mallawaarachchi}, {Meedeniya}, and {Jayarathna}}{{Meegahapola}
  et~al\mbox{.}}{2017d}]%
        {Meegahapola2017b}
\bibfield{author}{\bibinfo{person}{L.~B. {Meegahapola}}, \bibinfo{person}{P.~K.
  D. R.~M. {Alwis}}, \bibinfo{person}{L.~B. E.~H. {Nimalarathna}},
  \bibinfo{person}{V.~G. {Mallawaarachchi}}, \bibinfo{person}{D.~A.
  {Meedeniya}}, {and} \bibinfo{person}{S. {Jayarathna}}.}
  \bibinfo{year}{2017}\natexlab{d}.
\newblock \showarticletitle{Change detection optimization in frequently
  changing web pages}. In \bibinfo{booktitle}{\emph{2017 Moratuwa Engineering
  Research Conference (MERCon)}}. \bibinfo{publisher}{IEEE},
  \bibinfo{pages}{111--116}.
\newblock
\urldef\tempurl%
\url{https://doi.org/10.1109/MERCon.2017.7980466}
\showDOI{\tempurl}


\bibitem[\protect\citeauthoryear{Menczer, Pant, Srinivasan, and Ruiz}{Menczer
  et~al\mbox{.}}{2001}]%
        {Menczer2001}
\bibfield{author}{\bibinfo{person}{F. Menczer}, \bibinfo{person}{G. Pant},
  \bibinfo{person}{P. Srinivasan}, {and} \bibinfo{person}{M.~E. Ruiz}.}
  \bibinfo{year}{2001}\natexlab{}.
\newblock \showarticletitle{Evaluating Topic-driven Web Crawlers}. In
  \bibinfo{booktitle}{\emph{Proceedings of the 24th Annual International ACM
  SIGIR Conference on Research and Development in Information Retrieval}}
  \emph{(\bibinfo{series}{SIGIR '01})}. \bibinfo{publisher}{ACM},
  \bibinfo{address}{New York, NY, USA}, \bibinfo{pages}{241--249}.
\newblock
\showISBNx{1-58113-331-6}
\urldef\tempurl%
\url{https://doi.org/10.1145/383952.383995}
\showDOI{\tempurl}


\bibitem[\protect\citeauthoryear{Mesbah, van Deursen, and Lenselink}{Mesbah
  et~al\mbox{.}}{2012}]%
        {Mesbah2012}
\bibfield{author}{\bibinfo{person}{A. Mesbah}, \bibinfo{person}{A. van
  Deursen}, {and} \bibinfo{person}{S. Lenselink}.}
  \bibinfo{year}{2012}\natexlab{}.
\newblock \showarticletitle{Crawling Ajax-Based Web Applications Through
  Dynamic Analysis of User Interface State Changes}.
\newblock \bibinfo{journal}{\emph{ACM Trans. Web}} \bibinfo{volume}{6},
  \bibinfo{number}{1}, Article \bibinfo{articleno}{3} (\bibinfo{date}{March}
  \bibinfo{year}{2012}), \bibinfo{numpages}{30}~pages.
\newblock
\showISSN{1559-1131}
\urldef\tempurl%
\url{https://doi.org/10.1145/2109205.2109208}
\showDOI{\tempurl}


\bibitem[\protect\citeauthoryear{Minu and Shetty}{Minu and Shetty}{2015}]%
        {Minu2015}
\bibfield{author}{\bibinfo{person}{S. Minu} {and} \bibinfo{person}{A. Shetty}.}
  \bibinfo{year}{2015}\natexlab{}.
\newblock \showarticletitle{A Comparative Study of Image Change Detection
  Algorithms in MATLAB}.
\newblock \bibinfo{journal}{\emph{Aquatic Procedia}}  \bibinfo{volume}{4}
  (\bibinfo{date}{March} \bibinfo{year}{2015}), \bibinfo{pages}{1366--1373.}
\newblock


\bibitem[\protect\citeauthoryear{{Mirtaheri}, {Zou}, {Bochmann}, {Jourdan}, and
  {Onut}}{{Mirtaheri} et~al\mbox{.}}{2013}]%
        {Mirtaheri2013}
\bibfield{author}{\bibinfo{person}{S.~M. {Mirtaheri}}, \bibinfo{person}{D.
  {Zou}}, \bibinfo{person}{G.~V. {Bochmann}}, \bibinfo{person}{G. {Jourdan}},
  {and} \bibinfo{person}{I.~V. {Onut}}.} \bibinfo{year}{2013}\natexlab{}.
\newblock \showarticletitle{Dist-RIA Crawler: A Distributed Crawler for Rich
  Internet Applications}. In \bibinfo{booktitle}{\emph{2013 Eighth
  International Conference on P2P, Parallel, Grid, Cloud and Internet
  Computing}}. \bibinfo{pages}{105--112}.
\newblock
\urldef\tempurl%
\url{https://doi.org/10.1109/3PGCIC.2013.22}
\showDOI{\tempurl}


\bibitem[\protect\citeauthoryear{{Misra} and {Sorenson}}{{Misra} and
  {Sorenson}}{1975}]%
        {Misra1975}
\bibfield{author}{\bibinfo{person}{P. {Misra}} {and} \bibinfo{person}{H.
  {Sorenson}}.} \bibinfo{year}{1975}\natexlab{}.
\newblock \showarticletitle{Parameter estimation in Poisson processes
  (Corresp.)}.
\newblock \bibinfo{journal}{\emph{IEEE Transactions on Information Theory}}
  \bibinfo{volume}{21}, \bibinfo{number}{1} (\bibinfo{date}{January}
  \bibinfo{year}{1975}), \bibinfo{pages}{87--90}.
\newblock
\urldef\tempurl%
\url{https://doi.org/10.1109/TIT.1975.1055324}
\showDOI{\tempurl}


\bibitem[\protect\citeauthoryear{{Moyeen}, {Ali}, {Chong}, and
  {Islam}}{{Moyeen} et~al\mbox{.}}{2016}]%
        {Moyeen2016}
\bibfield{author}{\bibinfo{person}{M.~A. {Moyeen}}, \bibinfo{person}{G.~G.
  M.~N. {Ali}}, \bibinfo{person}{P.~H.~J. {Chong}}, {and} \bibinfo{person}{N.
  {Islam}}.} \bibinfo{year}{2016}\natexlab{}.
\newblock \showarticletitle{An automatic layout faults detection technique in
  responsive web pages considering JavaScript defined dynamic layouts}. In
  \bibinfo{booktitle}{\emph{2016 3rd International Conference on Electrical
  Engineering and Information Communication Technology (ICEEICT)}}.
  \bibinfo{pages}{1--5}.
\newblock
\urldef\tempurl%
\url{https://doi.org/10.1109/CEEICT.2016.7873146}
\showDOI{\tempurl}


\bibitem[\protect\citeauthoryear{Muscovitch}{Muscovitch}{2012}]%
        {Muscovitch2012}
\bibfield{author}{\bibinfo{person}{Z. Muscovitch}.}
  \bibinfo{year}{2012}\natexlab{}.
\newblock \bibinfo{booktitle}{\emph{{Who owns your Craigslist advert?}}}
\newblock Intellectual Property magazine.
\newblock
\urldef\tempurl%
\url{http://www.dnattorney.com/056-057-IPM_October_2012.pdf}
\showURL{%
Retrieved May 20, 2019 from \tempurl}


\bibitem[\protect\citeauthoryear{Nadaraj}{Nadaraj}{2016}]%
        {Nadaraj2016}
\bibfield{author}{\bibinfo{person}{S. Nadaraj}.}
  \bibinfo{year}{2016}\natexlab{}.
\newblock \showarticletitle{{Distributed Content Aggregation \& Content Change
  Detection using Bloom Filters}}.
\newblock \bibinfo{journal}{\emph{{International Journal of Computer Science
  and Information Technologies (IJCSIT)}}} \bibinfo{volume}{7},
  \bibinfo{number}{2} (\bibinfo{date}{Mar} \bibinfo{year}{2016}),
  \bibinfo{pages}{745--748}.
\newblock


\bibitem[\protect\citeauthoryear{{NetMind}}{{NetMind}}{1996}]%
        {Mindit1996}
\bibfield{author}{\bibinfo{person}{{NetMind}}.}
  \bibinfo{year}{1996}\natexlab{}.
\newblock \bibinfo{booktitle}{\emph{NetMind Mind-it}}.
\newblock {NETMIND Srl}.
\newblock
\urldef\tempurl%
\url{http://www.netmind.com/}
\showURL{%
Retrieved May 12, 2018 from \tempurl}


\bibitem[\protect\citeauthoryear{Nottingham and Sayre}{Nottingham and
  Sayre}{2005}]%
        {RFC4287Atom2005}
\bibfield{author}{\bibinfo{person}{M. Nottingham} {and} \bibinfo{person}{R.
  Sayre}.} \bibinfo{year}{2005}\natexlab{}.
\newblock \bibinfo{booktitle}{\emph{RFC 4287 - The Atom Syndication Format.}}
\newblock
\urldef\tempurl%
\url{https://tools.ietf.org/html/rfc4287}
\showURL{%
Retrieved May 4, 2019 from \tempurl}


\bibitem[\protect\citeauthoryear{Oita and Senellart}{Oita and
  Senellart}{2011}]%
        {Oita2011}
\bibfield{author}{\bibinfo{person}{M. Oita} {and} \bibinfo{person}{P.
  Senellart}.} \bibinfo{year}{2011}\natexlab{}.
\newblock \showarticletitle{{Deriving Dynamics of Web Pages: A Survey}}. In
  \bibinfo{booktitle}{\emph{{TWAW (Temporal Workshop on Web Archiving)}}}.
  \bibinfo{publisher}{HAL-Inria}, \bibinfo{address}{Hyderabad, India}.
\newblock
\urldef\tempurl%
\url{https://hal.inria.fr/inria-00588715}
\showURL{%
\tempurl}


\bibitem[\protect\citeauthoryear{Olston and Najork}{Olston and Najork}{2010}]%
        {Olston2010}
\bibfield{author}{\bibinfo{person}{C. Olston} {and} \bibinfo{person}{M.
  Najork}.} \bibinfo{year}{2010}\natexlab{}.
\newblock \showarticletitle{Web Crawling}.
\newblock \bibinfo{journal}{\emph{Foundations and Trends® in Information
  Retrieval}} \bibinfo{volume}{4}, \bibinfo{number}{3} (\bibinfo{year}{2010}),
  \bibinfo{pages}{175--246}.
\newblock
\showISSN{1554-0669}
\urldef\tempurl%
\url{https://doi.org/10.1561/1500000017}
\showDOI{\tempurl}


\bibitem[\protect\citeauthoryear{Olston and Pandey}{Olston and Pandey}{2008}]%
        {Olston2008}
\bibfield{author}{\bibinfo{person}{C. Olston} {and} \bibinfo{person}{S.
  Pandey}.} \bibinfo{year}{2008}\natexlab{}.
\newblock \showarticletitle{Recrawl Scheduling Based on Information Longevity}.
  In \bibinfo{booktitle}{\emph{Proceedings of the 17th International Conference
  on World Wide Web}} \emph{(\bibinfo{series}{WWW '08})}.
  \bibinfo{publisher}{ACM}, \bibinfo{address}{New York, NY, USA},
  \bibinfo{pages}{437--446}.
\newblock
\showISBNx{978-1-60558-085-2}
\urldef\tempurl%
\url{https://doi.org/10.1145/1367497.1367557}
\showDOI{\tempurl}


\bibitem[\protect\citeauthoryear{{OnWebChange}}{{OnWebChange}}{2009}]%
        {OnWebChange2009}
\bibfield{author}{\bibinfo{person}{{OnWebChange}}.}
  \bibinfo{year}{2009}\natexlab{}.
\newblock \bibinfo{booktitle}{\emph{{OnWebChange - Track web page changes and
  get notified. Free Sign-up. }}}.
\newblock {OnWebChange.com}.
\newblock
\urldef\tempurl%
\url{https://onwebchange.com/}
\showURL{%
Retrieved April 3, 2018 from \tempurl}


\bibitem[\protect\citeauthoryear{{Open Archives Initiative}}{{Open Archives
  Initiative}}{2017}]%
        {OAIPMH2002}
\bibfield{author}{\bibinfo{person}{{Open Archives Initiative}}.}
  \bibinfo{year}{2017}\natexlab{}.
\newblock \bibinfo{booktitle}{\emph{{Open Archives Initiative Protocol for
  Metadata Harvesting}}}.
\newblock Open Archives Initiative.
\newblock
\urldef\tempurl%
\url{https://www.openarchives.org/pmh/}
\showURL{%
Retrieved May 2, 2019 from \tempurl}


\bibitem[\protect\citeauthoryear{Page, Brin, Motwani, and Winograd}{Page
  et~al\mbox{.}}{1999}]%
        {Page1999}
\bibfield{author}{\bibinfo{person}{L. Page}, \bibinfo{person}{S. Brin},
  \bibinfo{person}{R. Motwani}, {and} \bibinfo{person}{T. Winograd}.}
  \bibinfo{year}{1999}\natexlab{}.
\newblock \bibinfo{booktitle}{\emph{The PageRank Citation Ranking: Bringing
  Order to the Web.}}
\newblock \bibinfo{type}{Technical Report} 1999-66.
  \bibinfo{institution}{Stanford InfoLab}.
\newblock
\urldef\tempurl%
\url{http://ilpubs.stanford.edu:8090/422/}
\showURL{%
\tempurl}
\newblock
\shownote{Previous number = SIDL-WP-1999-0120.}


\bibitem[\protect\citeauthoryear{{Pagescreen}}{{Pagescreen}}{2018}]%
        {Pagescreen2018}
\bibfield{author}{\bibinfo{person}{{Pagescreen}}.}
  \bibinfo{year}{2018}\natexlab{}.
\newblock \bibinfo{booktitle}{\emph{{Monitor Website Changes : automated Alerts
  and Archives}}}.
\newblock {Pagescreen}.
\newblock
\urldef\tempurl%
\url{https://pagescreen.io/}
\showURL{%
Retrieved April 3, 2018 from \tempurl}


\bibitem[\protect\citeauthoryear{Pandey and Olston}{Pandey and Olston}{2005}]%
        {Pandey2005}
\bibfield{author}{\bibinfo{person}{S. Pandey} {and} \bibinfo{person}{C.
  Olston}.} \bibinfo{year}{2005}\natexlab{}.
\newblock \showarticletitle{User-centric Web Crawling}. In
  \bibinfo{booktitle}{\emph{Proceedings of the 14th International Conference on
  World Wide Web}} \emph{(\bibinfo{series}{WWW '05})}.
  \bibinfo{publisher}{ACM}, \bibinfo{address}{New York, NY, USA},
  \bibinfo{pages}{401--411}.
\newblock
\showISBNx{1-59593-046-9}
\urldef\tempurl%
\url{https://doi.org/10.1145/1060745.1060805}
\showDOI{\tempurl}


\bibitem[\protect\citeauthoryear{Pant and Menczer}{Pant and Menczer}{2003}]%
        {Pant2003}
\bibfield{author}{\bibinfo{person}{G. Pant} {and} \bibinfo{person}{F.
  Menczer}.} \bibinfo{year}{2003}\natexlab{}.
\newblock \showarticletitle{Topical Crawling for Business Intelligence}. In
  \bibinfo{booktitle}{\emph{Research and Advanced Technology for Digital
  Libraries}}, \bibfield{editor}{\bibinfo{person}{T.~Koch} {and}
  \bibinfo{person}{I.~T. S{\o}lvberg}} (Eds.). \bibinfo{publisher}{Springer
  Berlin Heidelberg}, \bibinfo{address}{Berlin, Heidelberg},
  \bibinfo{pages}{233--244}.
\newblock
\showISBNx{978-3-540-45175-4}


\bibitem[\protect\citeauthoryear{Pehlivan, Ben-Saad, and
  Gan{\c{c}}arski}{Pehlivan et~al\mbox{.}}{2010}]%
        {Pehlivan2010}
\bibfield{author}{\bibinfo{person}{Z. Pehlivan}, \bibinfo{person}{M. Ben-Saad},
  {and} \bibinfo{person}{S. Gan{\c{c}}arski}.} \bibinfo{year}{2010}\natexlab{}.
\newblock \showarticletitle{Vi-DIFF: Understanding Web Pages Changes}. In
  \bibinfo{booktitle}{\emph{Database and Expert Systems Applications}},
  \bibfield{editor}{\bibinfo{person}{P.~G. Bringas},
  \bibinfo{person}{A.~Hameurlain}, {and} \bibinfo{person}{G.~Quirchmayr}}
  (Eds.). \bibinfo{publisher}{Springer Berlin Heidelberg},
  \bibinfo{address}{Berlin, Heidelberg}, \bibinfo{pages}{1--15}.
\newblock
\showISBNx{978-3-642-15364-8}


\bibitem[\protect\citeauthoryear{Prieto, Álvarez, Carneiro, and
  Cacheda}{Prieto et~al\mbox{.}}{2015}]%
        {Prieto2015}
\bibfield{author}{\bibinfo{person}{V.~M. Prieto}, \bibinfo{person}{M.
  Álvarez}, \bibinfo{person}{V. Carneiro}, {and} \bibinfo{person}{F.
  Cacheda}.} \bibinfo{year}{2015}\natexlab{}.
\newblock \showarticletitle{Distributed and Collaborative Web Change Detection
  System}.
\newblock \bibinfo{journal}{\emph{Computer Science and Information Systems}}
  \bibinfo{volume}{12}, \bibinfo{number}{1} (\bibinfo{year}{2015}),
  \bibinfo{pages}{91--114}.
\newblock


\bibitem[\protect\citeauthoryear{Radlinski, Bennett, and Yilmaz}{Radlinski
  et~al\mbox{.}}{2011}]%
        {Radlinski2011}
\bibfield{author}{\bibinfo{person}{F. Radlinski}, \bibinfo{person}{P.~N.
  Bennett}, {and} \bibinfo{person}{E. Yilmaz}.}
  \bibinfo{year}{2011}\natexlab{}.
\newblock \showarticletitle{Detecting Duplicate Web Documents Using
  Clickthrough Data}. In \bibinfo{booktitle}{\emph{Proceedings of the Fourth
  ACM International Conference on Web Search and Data Mining}}
  \emph{(\bibinfo{series}{WSDM '11})}. \bibinfo{publisher}{ACM},
  \bibinfo{address}{New York, NY, USA}, \bibinfo{pages}{147--156}.
\newblock
\showISBNx{978-1-4503-0493-1}
\urldef\tempurl%
\url{https://doi.org/10.1145/1935826.1935859}
\showDOI{\tempurl}


\bibitem[\protect\citeauthoryear{Reis, Gribble, Kohno, and Weaver}{Reis
  et~al\mbox{.}}{2008}]%
        {Reis2008}
\bibfield{author}{\bibinfo{person}{C. Reis}, \bibinfo{person}{S.~D. Gribble},
  \bibinfo{person}{T. Kohno}, {and} \bibinfo{person}{N.~C. Weaver}.}
  \bibinfo{year}{2008}\natexlab{}.
\newblock \showarticletitle{{Detecting In-Flight Page Changes with Web
  Tripwires}}. In \bibinfo{booktitle}{\emph{Proceedings of the 5th USENIX
  Symposium on Networked Systems Design and Implementation}}.
  \bibinfo{publisher}{USENIX Association}, \bibinfo{address}{San Francisco,
  CA}, \bibinfo{pages}{31--44}.
\newblock


\bibitem[\protect\citeauthoryear{{RSS-DEV Working Group}}{{RSS-DEV Working
  Group}}{2000}]%
        {RDFSiteSummary2000}
\bibfield{author}{\bibinfo{person}{{RSS-DEV Working Group}}.}
  \bibinfo{year}{2000}\natexlab{}.
\newblock \bibinfo{booktitle}{\emph{RDF Site Summary (RSS) 1.0.}}
\newblock {web.resource.org}.
\newblock
\urldef\tempurl%
\url{http://web.resource.org/rss/1.0/spec}
\showURL{%
Retrieved March 4, 2018 from \tempurl}


\bibitem[\protect\citeauthoryear{{Ryou} and {Ryu}}{{Ryou} and {Ryu}}{2018}]%
        {Ryou2018}
\bibfield{author}{\bibinfo{person}{Y. {Ryou}} {and} \bibinfo{person}{S.
  {Ryu}}.} \bibinfo{year}{2018}\natexlab{}.
\newblock \showarticletitle{Automatic Detection of Visibility Faults by Layout
  Changes in HTML5 Web Pages}. In \bibinfo{booktitle}{\emph{2018 IEEE 11th
  International Conference on Software Testing, Verification and Validation
  (ICST)}}. \bibinfo{pages}{182--192}.
\newblock
\urldef\tempurl%
\url{https://doi.org/10.1109/ICST.2018.00027}
\showDOI{\tempurl}


\bibitem[\protect\citeauthoryear{Saad and Gan{\c{c}}arski}{Saad and
  Gan{\c{c}}arski}{2012}]%
        {Saad2012}
\bibfield{author}{\bibinfo{person}{M.~B. Saad} {and}
  \bibinfo{person}{St{\'e}phane Gan{\c{c}}arski}.}
  \bibinfo{year}{2012}\natexlab{}.
\newblock \showarticletitle{Archiving the web using page changes patterns: a
  case study}.
\newblock \bibinfo{journal}{\emph{International Journal on Digital Libraries}}
  \bibinfo{volume}{13}, \bibinfo{number}{1} (\bibinfo{date}{01 Dec}
  \bibinfo{year}{2012}), \bibinfo{pages}{33--49}.
\newblock
\showISSN{1432-1300}
\urldef\tempurl%
\url{https://doi.org/10.1007/s00799-012-0094-z}
\showDOI{\tempurl}


\bibitem[\protect\citeauthoryear{Santos, Ziviani, Almeida, Carvalho, de~Moura,
  and da~Silva}{Santos et~al\mbox{.}}{2013}]%
        {Santos2013}
\bibfield{author}{\bibinfo{person}{A.~S.~R. Santos}, \bibinfo{person}{N.
  Ziviani}, \bibinfo{person}{J. Almeida}, \bibinfo{person}{C.~R. Carvalho},
  \bibinfo{person}{E.~S. de Moura}, {and} \bibinfo{person}{A.~S. da Silva}.}
  \bibinfo{year}{2013}\natexlab{}.
\newblock \showarticletitle{Learning to Schedule Webpage Updates Using Genetic
  Programming}. In \bibinfo{booktitle}{\emph{String Processing and Information
  Retrieval}}, \bibfield{editor}{\bibinfo{person}{O.~Kurland},
  \bibinfo{person}{M.~Lewenstein}, {and} \bibinfo{person}{E.~Porat}} (Eds.).
  \bibinfo{publisher}{Springer International Publishing},
  \bibinfo{address}{Cham}, \bibinfo{pages}{271--278}.
\newblock
\showISBNx{978-3-319-02432-5}


\bibitem[\protect\citeauthoryear{Schonfeld and Shivakumar}{Schonfeld and
  Shivakumar}{2009}]%
        {Schonfeld2009}
\bibfield{author}{\bibinfo{person}{U. Schonfeld} {and} \bibinfo{person}{N.
  Shivakumar}.} \bibinfo{year}{2009}\natexlab{}.
\newblock \showarticletitle{Sitemaps: Above and Beyond the Crawl of Duty}. In
  \bibinfo{booktitle}{\emph{Proceedings of the 18th International Conference on
  World Wide Web}} \emph{(\bibinfo{series}{WWW '09})}.
  \bibinfo{publisher}{ACM}, \bibinfo{address}{New York, NY, USA},
  \bibinfo{pages}{991--1000}.
\newblock
\showISBNx{978-1-60558-487-4}
\urldef\tempurl%
\url{https://doi.org/10.1145/1526709.1526842}
\showDOI{\tempurl}


\bibitem[\protect\citeauthoryear{Sebesta}{Sebesta}{2011}]%
        {Sebesta2001}
\bibfield{author}{\bibinfo{person}{R.~W. Sebesta}.}
  \bibinfo{year}{2011}\natexlab{}.
\newblock \bibinfo{booktitle}{\emph{Programming the World Wide Web}
  (\bibinfo{edition}{4th} ed.)}.
\newblock \bibinfo{publisher}{Addison-Wesley Publishing Company},
  \bibinfo{address}{Boston, MA, USA}.
\newblock
\showISBNx{0132130815}


\bibitem[\protect\citeauthoryear{{Shkapenyuk} and {Suel}}{{Shkapenyuk} and
  {Suel}}{2002}]%
        {Shkapenyuk2002}
\bibfield{author}{\bibinfo{person}{V. {Shkapenyuk}} {and} \bibinfo{person}{T.
  {Suel}}.} \bibinfo{year}{2002}\natexlab{}.
\newblock \showarticletitle{Design and implementation of a high-performance
  distributed Web crawler}. In \bibinfo{booktitle}{\emph{Proceedings 18th
  International Conference on Data Engineering}}. \bibinfo{pages}{357--368}.
\newblock
\urldef\tempurl%
\url{https://doi.org/10.1109/ICDE.2002.994750}
\showDOI{\tempurl}


\bibitem[\protect\citeauthoryear{Shobhna and Chaudhary}{Shobhna and
  Chaudhary}{2013}]%
        {Shobhna2013}
\bibfield{author}{\bibinfo{person}{Shobhna} {and} \bibinfo{person}{M.~C.
  Chaudhary}.} \bibinfo{year}{2013}\natexlab{}.
\newblock \showarticletitle{A Survey on Web Page Change Detection System Using
  Different Approaches}.
\newblock \bibinfo{journal}{\emph{International Journal of Computer Science and
  Mobile Computing (IJCSMC)}} \bibinfo{volume}{2}, \bibinfo{number}{6}
  (\bibinfo{date}{June} \bibinfo{year}{2013}), \bibinfo{pages}{294--299}.
\newblock


\bibitem[\protect\citeauthoryear{Silverstein, Marais, Henzinger, and
  Moricz}{Silverstein et~al\mbox{.}}{1999}]%
        {Silverstein1999}
\bibfield{author}{\bibinfo{person}{C. Silverstein}, \bibinfo{person}{H.
  Marais}, \bibinfo{person}{M. Henzinger}, {and} \bibinfo{person}{M. Moricz}.}
  \bibinfo{year}{1999}\natexlab{}.
\newblock \showarticletitle{Analysis of a Very Large Web Search Engine Query
  Log}.
\newblock \bibinfo{journal}{\emph{SIGIR Forum}} \bibinfo{volume}{33},
  \bibinfo{number}{1} (\bibinfo{date}{Sept.} \bibinfo{year}{1999}),
  \bibinfo{pages}{6--12}.
\newblock
\showISSN{0163-5840}
\urldef\tempurl%
\url{https://doi.org/10.1145/331403.331405}
\showDOI{\tempurl}


\bibitem[\protect\citeauthoryear{{Sitemaps.org}}{{Sitemaps.org}}{2008}]%
        {sitemaps2008}
\bibfield{author}{\bibinfo{person}{{Sitemaps.org}}.}
  \bibinfo{year}{2008}\natexlab{}.
\newblock \bibinfo{booktitle}{\emph{sitemaps.org – Home}}.
\newblock {Sitemaps.org}.
\newblock
\urldef\tempurl%
\url{https://www.sitemaps.org/index.html}
\showURL{%
Retrieved March 4, 2018 from \tempurl}


\bibitem[\protect\citeauthoryear{Smith}{Smith}{2008}]%
        {Smith2008}
\bibfield{author}{\bibinfo{person}{B.~E. Smith}.}
  \bibinfo{year}{2008}\natexlab{}.
\newblock \bibinfo{booktitle}{\emph{Creating Web Pages For Dummies}
  (\bibinfo{edition}{9th} ed.)}.
\newblock \bibinfo{publisher}{John Wiley \& Sons Inc}, \bibinfo{address}{New
  Jersey, USA}.
\newblock
\showISBNx{0470385359}


\bibitem[\protect\citeauthoryear{Sproul}{Sproul}{2009}]%
        {Sproul2008}
\bibfield{author}{\bibinfo{person}{K. Sproul}.}
  \bibinfo{year}{2009}\natexlab{}.
\newblock \bibinfo{booktitle}{\emph{The Dao of SEO} (\bibinfo{edition}{2009}
  ed.)}.
\newblock \bibinfo{publisher}{Lulu.com}, \bibinfo{address}{Morrisville, North
  Carolina, USA}.
\newblock
\showISBNx{1435714199}


\bibitem[\protect\citeauthoryear{Srinivasan, Menczer, and Pant}{Srinivasan
  et~al\mbox{.}}{2005}]%
        {Srinivasan2005}
\bibfield{author}{\bibinfo{person}{P. Srinivasan}, \bibinfo{person}{F.
  Menczer}, {and} \bibinfo{person}{G. Pant}.} \bibinfo{year}{2005}\natexlab{}.
\newblock \showarticletitle{A General Evaluation Framework for Topical
  Crawlers}.
\newblock \bibinfo{journal}{\emph{Information Retrieval}} \bibinfo{volume}{8},
  \bibinfo{number}{3} (\bibinfo{date}{01 Jan} \bibinfo{year}{2005}),
  \bibinfo{pages}{417--447}.
\newblock
\showISSN{1573-7659}
\urldef\tempurl%
\url{https://doi.org/10.1007/s10791-005-6993-5}
\showDOI{\tempurl}


\bibitem[\protect\citeauthoryear{{Sun}, {Councill}, and {Giles}}{{Sun}
  et~al\mbox{.}}{2010}]%
        {Sun2010}
\bibfield{author}{\bibinfo{person}{Y. {Sun}}, \bibinfo{person}{I.~G.
  {Councill}}, {and} \bibinfo{person}{C.~L. {Giles}}.}
  \bibinfo{year}{2010}\natexlab{}.
\newblock \showarticletitle{The Ethicality of Web Crawlers}. In
  \bibinfo{booktitle}{\emph{2010 IEEE/WIC/ACM International Conference on Web
  Intelligence and Intelligent Agent Technology}}, Vol.~\bibinfo{volume}{1}.
  \bibinfo{pages}{668--675}.
\newblock
\urldef\tempurl%
\url{https://doi.org/10.1109/WI-IAT.2010.316}
\showDOI{\tempurl}


\bibitem[\protect\citeauthoryear{Sun, Zhuang, and Giles}{Sun
  et~al\mbox{.}}{2007}]%
        {Sun2007}
\bibfield{author}{\bibinfo{person}{Y. Sun}, \bibinfo{person}{Z. Zhuang}, {and}
  \bibinfo{person}{C.~L. Giles}.} \bibinfo{year}{2007}\natexlab{}.
\newblock \showarticletitle{A Large-scale Study of Robots.Txt}. In
  \bibinfo{booktitle}{\emph{Proceedings of the 16th International Conference on
  World Wide Web}} \emph{(\bibinfo{series}{WWW '07})}.
  \bibinfo{publisher}{ACM}, \bibinfo{address}{New York, NY, USA},
  \bibinfo{pages}{1123--1124}.
\newblock
\showISBNx{978-1-59593-654-7}
\urldef\tempurl%
\url{https://doi.org/10.1145/1242572.1242726}
\showDOI{\tempurl}


\bibitem[\protect\citeauthoryear{Thelwall and Stuart}{Thelwall and
  Stuart}{2006}]%
        {Thelwall2006}
\bibfield{author}{\bibinfo{person}{M. Thelwall} {and} \bibinfo{person}{D.
  Stuart}.} \bibinfo{year}{2006}\natexlab{}.
\newblock \showarticletitle{Web crawling ethics revisited: Cost, privacy, and
  denial of service}.
\newblock \bibinfo{journal}{\emph{Journal of the American Society for
  Information Science and Technology}} \bibinfo{volume}{57},
  \bibinfo{number}{13} (\bibinfo{year}{2006}), \bibinfo{pages}{1771--1779}.
\newblock
\urldef\tempurl%
\url{https://doi.org/10.1002/asi.20388}
\showDOI{\tempurl}
\showeprint{https://onlinelibrary.wiley.com/doi/pdf/10.1002/asi.20388}


\bibitem[\protect\citeauthoryear{{Trackly}}{{Trackly}}{2016}]%
        {Trackly2016}
\bibfield{author}{\bibinfo{person}{{Trackly}}.}
  \bibinfo{year}{2016}\natexlab{}.
\newblock \bibinfo{booktitle}{\emph{{Trackly | website change detection}}}.
\newblock {Trackly}.
\newblock
\urldef\tempurl%
\url{https://trackly.io/}
\showURL{%
Retrieved November 9, 2019 from \tempurl}


\bibitem[\protect\citeauthoryear{Umbrich, Hausenblas, Hogan, Polleres, and
  Decker}{Umbrich et~al\mbox{.}}{2010}]%
        {Umbrich2010}
\bibfield{author}{\bibinfo{person}{J. Umbrich}, \bibinfo{person}{M.
  Hausenblas}, \bibinfo{person}{A. Hogan}, \bibinfo{person}{A. Polleres}, {and}
  \bibinfo{person}{S. Decker}.} \bibinfo{year}{2010}\natexlab{}.
\newblock \showarticletitle{{Towards Dataset Dynamics: Change Frequency of
  Linked Open Data Sources}}. In \bibinfo{booktitle}{\emph{Proceedings of the
  3rd International Workshop on Linked Data on the Web (LDOW2010), in
  conjunction with 19th International World Wide Web Conference, CEUR, 2010}}
  \emph{(\bibinfo{series}{LDOW 2010})}. \bibinfo{publisher}{CEUR}.
\newblock


\bibitem[\protect\citeauthoryear{Van~de Sompel, Nelson, Lagoze, and
  Warner}{Van~de Sompel et~al\mbox{.}}{2004}]%
        {VandeSompel2004}
\bibfield{author}{\bibinfo{person}{H. Van~de Sompel}, \bibinfo{person}{M.~L.
  Nelson}, \bibinfo{person}{C. Lagoze}, {and} \bibinfo{person}{S. Warner}.}
  \bibinfo{year}{2004}\natexlab{}.
\newblock \showarticletitle{Resource Harvesting within the OAI-PMH Framework}.
\newblock \bibinfo{journal}{\emph{D-Lib Magazine}} \bibinfo{volume}{10},
  \bibinfo{number}{12} (\bibinfo{year}{2004}).
\newblock


\bibitem[\protect\citeauthoryear{Van~de Sompel, Sanderson, Klein, Nelson,
  Haslhofer, Warner, and Lagoze}{Van~de Sompel et~al\mbox{.}}{2012}]%
        {VandeSompel2012}
\bibfield{author}{\bibinfo{person}{H. Van~de Sompel}, \bibinfo{person}{R.
  Sanderson}, \bibinfo{person}{M. Klein}, \bibinfo{person}{M.~L. Nelson},
  \bibinfo{person}{B. Haslhofer}, \bibinfo{person}{S. Warner}, {and}
  \bibinfo{person}{C. Lagoze}.} \bibinfo{year}{2012}\natexlab{}.
\newblock \showarticletitle{A Perspective on Resource Synchronization}.
\newblock \bibinfo{journal}{\emph{D-Lib Magazine}} \bibinfo{volume}{18},
  \bibinfo{number}{9/10} (\bibinfo{date}{Sep} \bibinfo{year}{2012}).
\newblock


\bibitem[\protect\citeauthoryear{{Versionista}}{{Versionista}}{2007}]%
        {Versionista2007}
\bibfield{author}{\bibinfo{person}{{Versionista}}.}
  \bibinfo{year}{2007}\natexlab{}.
\newblock \bibinfo{booktitle}{\emph{{Versionista: Monitor Website Changes}}}.
\newblock {Versionista}.
\newblock
\urldef\tempurl%
\url{https://versionista.com/}
\showURL{%
Retrieved April 3, 2018 from \tempurl}


\bibitem[\protect\citeauthoryear{{Visualping}}{{Visualping}}{2017}]%
        {VisualPing2017}
\bibfield{author}{\bibinfo{person}{{Visualping}}.}
  \bibinfo{year}{2017}\natexlab{}.
\newblock \bibinfo{booktitle}{\emph{Visualping: \#1 Website change detection,
  monitoring and alerts}}.
\newblock {VisualPing}.
\newblock
\urldef\tempurl%
\url{https://visualping.io/}
\showURL{%
Retrieved November 9, 2019 from \tempurl}


\bibitem[\protect\citeauthoryear{{w3computing.com}}{{w3computing.com}}{2017}]%
        {ThreeDWeb2017}
\bibfield{author}{\bibinfo{person}{{w3computing.com}}.}
  \bibinfo{year}{2017}\natexlab{}.
\newblock \bibinfo{booktitle}{\emph{{Dynamic and Three-Dimensional Web
  Pages}}}.
\newblock w3computing.com.
\newblock
\urldef\tempurl%
\url{https://www.w3computing.com/systemsanalysis/dynamic-three-dimensional-web-pages/}
\showURL{%
Retrieved May 28, 2019 from \tempurl}


\bibitem[\protect\citeauthoryear{{Wachete}}{{Wachete}}{2014}]%
        {Wachete2014}
\bibfield{author}{\bibinfo{person}{{Wachete}}.}
  \bibinfo{year}{2014}\natexlab{}.
\newblock \bibinfo{booktitle}{\emph{Wachete - Monitor web changes.}}
\newblock {Wachete s.r.o.}
\newblock
\urldef\tempurl%
\url{https://www.wachete.com}
\showURL{%
Retrieved April 3, 2018 from \tempurl}


\bibitem[\protect\citeauthoryear{Walsh, Kapfhammer, and McMinn}{Walsh
  et~al\mbox{.}}{2017}]%
        {Walsh2017}
\bibfield{author}{\bibinfo{person}{T.~A. Walsh}, \bibinfo{person}{G.~M.
  Kapfhammer}, {and} \bibinfo{person}{P. McMinn}.}
  \bibinfo{year}{2017}\natexlab{}.
\newblock \showarticletitle{ReDeCheck: An Automatic Layout Failure Checking
  Tool for Responsively Designed Web Pages}. In
  \bibinfo{booktitle}{\emph{Proceedings of the 26th ACM SIGSOFT International
  Symposium on Software Testing and Analysis}} \emph{(\bibinfo{series}{ISSTA
  2017})}. \bibinfo{publisher}{ACM}, \bibinfo{address}{New York, NY, USA},
  \bibinfo{pages}{360--363}.
\newblock
\showISBNx{978-1-4503-5076-1}
\urldef\tempurl%
\url{https://doi.org/10.1145/3092703.3098221}
\showDOI{\tempurl}


\bibitem[\protect\citeauthoryear{{Wang}, {DeWitt}, and {Cai}}{{Wang}
  et~al\mbox{.}}{2003}]%
        {Wang2003}
\bibfield{author}{\bibinfo{person}{Y. {Wang}}, \bibinfo{person}{D.~J.
  {DeWitt}}, {and} \bibinfo{person}{J.~. {Cai}}.}
  \bibinfo{year}{2003}\natexlab{}.
\newblock \showarticletitle{X-Diff: an effective change detection algorithm for
  XML documents}. In \bibinfo{booktitle}{\emph{Proceedings 19th International
  Conference on Data Engineering (Cat. No.03CH37405)}}.
  \bibinfo{pages}{519--530}.
\newblock
\urldef\tempurl%
\url{https://doi.org/10.1109/ICDE.2003.1260818}
\showDOI{\tempurl}


\bibitem[\protect\citeauthoryear{Wolf, Squillante, Yu, Sethuraman, and
  Ozsen}{Wolf et~al\mbox{.}}{2002}]%
        {Wolf2002}
\bibfield{author}{\bibinfo{person}{J.~L. Wolf}, \bibinfo{person}{M.~S.
  Squillante}, \bibinfo{person}{P.~S. Yu}, \bibinfo{person}{J. Sethuraman},
  {and} \bibinfo{person}{L. Ozsen}.} \bibinfo{year}{2002}\natexlab{}.
\newblock \showarticletitle{Optimal Crawling Strategies for Web Search
  Engines}. In \bibinfo{booktitle}{\emph{Proceedings of the 11th International
  Conference on World Wide Web}} \emph{(\bibinfo{series}{WWW '02})}.
  \bibinfo{publisher}{ACM}, \bibinfo{address}{New York, NY, USA},
  \bibinfo{pages}{136--147}.
\newblock
\showISBNx{1-58113-449-5}
\urldef\tempurl%
\url{https://doi.org/10.1145/511446.511465}
\showDOI{\tempurl}


\bibitem[\protect\citeauthoryear{{WordPress}}{{WordPress}}{2003}]%
        {WordPress2003}
\bibfield{author}{\bibinfo{person}{{WordPress}}.}
  \bibinfo{year}{2003}\natexlab{}.
\newblock \bibinfo{booktitle}{\emph{{WordPress.com: Create a Free Website or
  Blog}}}.
\newblock {Automattic Inc.}
\newblock
\urldef\tempurl%
\url{https://wordpress.com/}
\showURL{%
Retrieved November 8, 2019 from \tempurl}


\bibitem[\protect\citeauthoryear{{Yadav}, {Sharma}, and {Gupta}}{{Yadav}
  et~al\mbox{.}}{2007}]%
        {Yadav2007}
\bibfield{author}{\bibinfo{person}{D. {Yadav}}, \bibinfo{person}{A.~K.
  {Sharma}}, {and} \bibinfo{person}{J.~P. {Gupta}}.}
  \bibinfo{year}{2007}\natexlab{}.
\newblock \showarticletitle{Change Detection in Web Pages}. In
  \bibinfo{booktitle}{\emph{10th International Conference on Information
  Technology (ICIT 2007)}}. \bibinfo{pages}{265--270}.
\newblock
\urldef\tempurl%
\url{https://doi.org/10.1109/ICIT.2007.37}
\showDOI{\tempurl}


\bibitem[\protect\citeauthoryear{Yohanes, Handoko, and Wardana}{Yohanes
  et~al\mbox{.}}{2011}]%
        {Yohanes2011}
\bibfield{author}{\bibinfo{person}{B.~W. Yohanes}, \bibinfo{person}{Handoko},
  {and} \bibinfo{person}{H.~K. Wardana}.} \bibinfo{year}{2011}\natexlab{}.
\newblock \showarticletitle{Focused Crawler Optimization Using Genetic
  Algorithm}.
\newblock \bibinfo{journal}{\emph{TELKOMNIKA}} \bibinfo{volume}{9},
  \bibinfo{number}{3} (\bibinfo{date}{Dec.} \bibinfo{year}{2011}),
  \bibinfo{pages}{403--410}.
\newblock


\bibitem[\protect\citeauthoryear{{Zheng}}{{Zheng}}{2011}]%
        {Zheng2011}
\bibfield{author}{\bibinfo{person}{S. {Zheng}}.}
  \bibinfo{year}{2011}\natexlab{}.
\newblock \showarticletitle{Genetic and Ant Algorithms Based Focused Crawler
  Design}. In \bibinfo{booktitle}{\emph{2011 Second International Conference on
  Innovations in Bio-inspired Computing and Applications}}.
  \bibinfo{pages}{374--378}.
\newblock
\urldef\tempurl%
\url{https://doi.org/10.1109/IBICA.2011.98}
\showDOI{\tempurl}


\bibitem[\protect\citeauthoryear{Zilberman}{Zilberman}{2013}]%
        {Check4Change2006}
\bibfield{author}{\bibinfo{person}{R. Zilberman}.}
  \bibinfo{year}{2013}\natexlab{}.
\newblock \bibinfo{booktitle}{\emph{{Check4Change}}}.
\newblock {Check4Change}.
\newblock
\urldef\tempurl%
\url{https://check4change.com}
\showURL{%
Retrieved April 3, 2018 from \tempurl}


\end{thebibliography}

\end{document}